\documentclass[11pt]{ncsuthesis4}

\usepackage{amsfonts}
\usepackage{amssymb}
\usepackage{amsmath}
\usepackage[numbers,sort&compress]{natbib}
\usepackage{latexsym}
\usepackage{tocbibind}
\usepackage[pdftex]{graphicx}      
\DeclareGraphicsExtensions{.pdf}   
\usepackage{subfigure}
\usepackage{dcolumn}
\usepackage{bm}

\begin{document}
\newcommand{\puttitle}[0]{Generalized Pairing Wave Functions
                          and Nodal Properties \\
                          for Electronic Structure 
                          Quantum Monte Carlo}

\title{\puttitle}

\author{Michal Bajdich}
\degreeyear{2007}
\degree{Doctor of Philosophy}
\chair{Prof. Lubos Mitas}
\memberII{Prof. Jerry L. Whitten}
\memberIII{Prof. Dean Lee}
\memberIV{Prof. Christopher Roland}
\numberofmembers{4}

\newpage
\pagestyle{empty}
\begin{center}
\begin{large}
\textbf{ABSTRACT}
\end{large}
\end{center}
{\ssp {\bf BAJDICH, MICHAL}. 
\puttitle .
(Under the direction of Prof. Lubos Mitas.)

}

The quantum Monte Carlo (QMC) is one of the most promising 
many-body electronic structure approaches.
It employs stochastic techniques for solving the stationary 
Schr\" odinger equation and for evaluation of expectation values. 
The key advantage of QMC is its capability to use the explicitly correlated 
wave functions, which allow the study of many-body effects 
beyond the reach of mean-field methods. The most important 
limit on QMC accuracy is the fixed-node approximation, 
which comes from necessity to circumvent the fermion sign problem.  
The size of resulting fixed-node errors depends on the quality of the 
nodes (the subset of position space where the wave function vanishes) 
of a used wave function.
In this dissertation, we analyze the nodal properties of the existing fermionic wave functions
and offer new types of variational wave functions with improved nodal structure.

In the first part of this dissertation, we study the fermion nodes 
for spin-polarized states of a few-electron
ions and molecules with $s$, $p$, $d$ and $f$ one-particle orbitals. 
We find exact nodes for some cases of two electron atomic and 
molecular states and also the first exact node for the three-electron
atomic system in  $^4S(p^3)$ state using appropriate coordinate
maps and wave function symmetries. We analyze the cases of nodes 
for larger number of electrons in the Hartree-Fock approximation
and for some cases we find transformations for
projecting the high-dimensional nodal manifolds into
3D space. The nodal topologies and other properties are studied
using these projections. Finally, for two specific cases of 
spin-unpolarized states, we show how correlations reduce 
the nodal structure to only two maximal nodal cells.

In the second part, we investigate several types of trial 
wave functions with pairing orbitals and their nodal properties in the fixed-node quantum Monte Carlo. 
Using a set of first row atoms and molecules we find that the 
wave functions in the form of single Pfaffian provide very consistent and systematic
behavior in recovering the correlation energies on the 
level of 95\%. In order to get beyond this limit we explore
the possibilities of expanding 
the wave function in linear combinations of Pfaffians. 
We observe that molecular systems require much larger expansions 
than atomic systems and that the linear combinations of a few Pfaffians 
lead to rather small gains in correlation energy.
Further, we test the wave function based on fully-antisymmetrized product
of independent pair orbitals. Despite its seemingly large
variational potential, we do not observe significant gains in correlation energy.
Finally, we combine these developments with the recently 
proposed inhomogeneous backflow transformations. 

\newpage
\pagestyle{plain}
\renewcommand{\baselinestretch}{2}
\vspace{0.2in}

\field{Physics}
\campus{Raleigh, NC}

\maketitle

\begin{frontmatter}
\thispagestyle{plain}
\begin{dedication}
\begin{center}
\begin{large}
\textbf{Dedication}
\end{large}
\end{center}

\begin{center}
\null\vfil
Dedicated to my parents.
\vfil\null
\end{center}
\end{dedication}

\begin{center}
\begin{large}
\thispagestyle{plain}
\textbf{Biography}
\end{large}
\end{center}
I was born on August 16 1978, 
exactly 58 years after the great Henry Charles Bukowski. 


\begin{acknowledgements}
\thispagestyle{plain}
First of all, I would like to thank my advisor 
Prof. Lubos Mitas for his long-time encouragement and support.
It was a pleasure for me to be a part of our stimulating 
discussions. His elucidating approach to physical problems is exemplary 
and I hope it will be reflected in my future scientific career. 

I am also thankful to all current and former members of
our NCSU research group, namely Jind\v rich Koloren\v c, 
Lucas K. Wagner, Hiori Kino, Gabriel Drobn\'y, Ji-Woo Lee, Prasenjit Sen, 
David Sulock and Zack Helms. Jind\v rich is especially 
deserving my deep thanks for his many helpful suggestions 
throughout the preparation of this thesis. 
Furthermore, I want to thank my friend Cheng Wang, who made these five years at graduate school 
so much more entertaining.
I would like to thank the Physics Department at large: faculty, staff and fellow students, 
especially to Prof. Michael A. Paesler, Cecilia C. Upchurch and Jennifer T. Allen.  

I would also like to thank Jeffrey Grossman from UC Berkeley, for his invitations to Bay Area 
and Prof. Kevin E. Schmidt from Arizona State University 
for grateful sharing of his ideas about Pfaffian pairing wave functions.
Further, let me thank my undergraduate advisor Richard Hlubina and to my former graduate advisor Martin Mo\v sko.

Special thanks goes to Lauren A. Griffin for her careful reading and many corrections of the manuscript. 
Last but certainly not least, thanks to my parents Silvia and Milan for 
their unyielding support of my education throughout my life and their encouragement and 
guidance which has proven to be invaluable. 

I am grateful for the support by NSF Grants No. DMR-0121361, and  EAR-0530110 
and the computer time allocations at PAMS NCSU and NCSA facilities. 
\end{acknowledgements}

\tableofcontents
\listoffigures
\listoftables

\def\listsymbolname{List of Abbreviations and Symbols}
\chapter*{\listsymbolname\markboth{\uppercase{\listsymbolname}}%
{\uppercase{\listsymbolname}}}
\addcontentsline{toc}{chapter}{\listsymbolname}
\begin{tabular}{p{2cm}r}
a.u.    & Hartree Atomic Units ($\hbar=m_e=e=4\pi/\epsilon_0=1$)\\
AIP     & Antisymmetric Independent Pairs wave function\\
ADA     & Averaged Density Approximation \\
AGP     & Antisymetrized Geminal Product \\
BCS     & Bardeen-Cooper-Schrieffer wave function\\
BF      & Back-Flow transformation \\
CC      & Coupled Cluster technique \\
CCSD(T) & Coupled Cluster with Singles, Doubles, and perturbative Triples \\
CI      & Configuration Interaction \\
CISD    & Configuration Interaction with Singles and Doubles \\
DFT     & Density Functional Theory \\
DMC     & Diffusion Monte Carlo \\
GGA     & Generalized Gradient Approximation of DFT \\
H       & Hartree atomic energy ($1$ H$=27.21138$ eV)\\
HF      & Hartree-Fock theory\\
IPFSE   & Independent-Particle Finite-Size Errors \\
LDA     & Local Density approximation of DFT \\
LM      & Levenberg-Marquardt method \\
MC      & Monte Carlo \\
MCSCF   & Multi-Configurational Self-Consistent Field \\
MPF     & Multi-Pfaffian wave function\\
NP      & Nondeterministic Polynomial \\
PIMC    & Path Integral Monte Carlo \\
\end{tabular}
\newpage
\thispagestyle{plain}
\begin{tabular}{p{2cm}r}
PF      & Pfaffian functional form \\
QMC     & Quantum Monte Carlo \\
SCF     & Self-Consistent Field \\
SIC     & Self-Interaction Corrections \\
STU     & Singlet-Triplet-Unpaired wave function\\ 
VMC     & Variational Monte Carlo \\
WDA     & Weighted Density Approximation \\
\\
${\mathcal A}$                & Antisymmetrization operator \\
$A({\bf R}\to{\bf R}')$       & Acceptance probability of step from ${\bf R}$ to ${\bf R}'$\\ 
$\alpha_i(\sigma_j)$          & Spin orbital\\
$\{c\}$                       & Set of variational parameters \\ 
$\chi(i,j)$                   & Triplet pair orbital \\
$E_L({\bf R})$                & Local energy at ${\bf R}$\\
$E_n$                         & Energy eigenvalue of a Hamiltonian \\
$f({\bf R},\tau)$             & Importance function in DMC\\
$\Phi_n$                      & Eigenfunction of a Hamiltonian \\
$\boldsymbol \Phi^{\uparrow\downarrow}$  & Singlet pairing matrix \\
$\tilde{\phi}(i,j)$           & Two-particle (or pair) orbital \\
$\phi(i,j)$                   & Singlet pair orbital\\
${\bf g}$                     & Gradient vector in parameter space\\
\mbox{$G({\bf R}\to{\bf R}',\tau)$}  & Green's function \\
$I$                           & Identity matrix \\
${\mathcal H}$                & Hamiltonian operator \\
$H$                           & Hessian matrix \\
$\mu$                         & Dumping parameter \\
${\bf M}_i({\bf R})$          & Metric tensor of many-body coordinates\\
$N_e$ or $N$                  & Number of electrons \\
$N_I$                         & Number of nuclei \\
$\Omega$                      & Three Euler angles \\
$\Omega({\bf R})$             & Nodal cell of ${\bf R}$ \\
${\mathcal P}({\bf R})$       & Probability distribution \\
$\Psi_T$                      & Trial wave function \\                 
\end{tabular}
\newpage
\thispagestyle{plain}
\begin{tabular}{p{2cm}r}
$\varphi_i^\sigma(j)$         & One-particle spartial orbital \\
$\tilde\varphi_i(j)$          & One-particle spin-orbital\\
$P$                           & Permutation operator \\
$P_I$                         & Parity operator \\
$P_l$                         & Legendre polynomial of the $l$th degree\\
$P_c(a_{1, j})$               & Pfaffian cofactor of $a_{1, j}$\\
$\mathfrak{R}$                & Complete configuration many-body  space \\
$R(\pi z)$                    & Operator of rotation by $\pi$ degrees around $z$-axis\\ 
${\bf R}$                     & Point in many-body space \\
${\bf r}$                     & Position of a single electron \\
$\rho({\bf r})$               & One-particle electronic density at point ${\bf r}$\\
$\rho$                        & Gain ratio\\
$\sigma$                      & Spin projection of single electron\\
${\boldsymbol \Sigma}$        & Vector of spin projections\\
$\sigma^2$                    & Variance of local energy \\
$T({\bf R}\to{\bf R}')$       & Sampling distribution for step from ${\bf R}$ to ${\bf R}'$\\
$\tau$                        & Imaginary time\\
${\bf t}_i({\bf R})$          & Spatial offset in a back-flow transformation\\
${\bf v}_D({\bf R})$          & Drift velocity in a many-body space\\
$w_i$                         & Weight of a walker\\
$W_{\{c_{new}\}}({\bf R})$    & Re-weighting factor \\

${\boldsymbol \xi}^{\uparrow\uparrow}$  & Triplet spin-up spin-up pairing matrix \\
${\boldsymbol \xi}_i({\bf R})$        & Backflow displacement\\
$Z$                           & Nuclear charge \\
\end{tabular}
\end{frontmatter}
\pagestyle{nofancy}
\chapter{Introduction}\label{ch:intro}
The properties of quantum chemical and condensed matter systems of our everyday world
are determined by the laws of quantum physics, which have been known since the 1930s.  
The distribution and motion of electrons surrounding 
the atomic nucleus are described by time-dependent 
Schr\"odinger equation\footnote{Quite often it is necessary to solve the Dirac 
equation for core electrons and use relativistic (spin-orbit) corrections elsewhere.},
but solving it for many electrons is extremely hard. The difficulty 
comes from presence of electron-electron interaction term, which makes it 
impossible to separate this {\em many-body problem} into set of one-electron problems. 
In the past, the methods for solving the Schr\"odinger equation were
based on replacing the difficult interaction term by some effective one, 
designed ``to capture'' the essential physics. 
The great success of these theories is the proof of the genius of these approximations.
In this dissertation, we present a quantum Monte Carlo (QMC) method, which 
incorporates the electron-electron term directly and is able to solve the 
many-body Schr\"odinger equation almost exactly. Currently, it is the only method, 
which treats the full many-body problem and scales up to large systems. 

The term ``quantum Monte Carlo'' covers several different stochastic techniques 
adapted to determine either the ground state or finite-temperature equilibrium properties. 
Further, we restrict our discussion only to the ground state electronic properties.
The simplest method explained is the variational Monte Carlo (VMC), which uses 
the stochastic integration for evaluation of expectation values for chosen 
{\em trial wave function}. Its accuracy sorely depends on the quality of a used trial wave function. 
This drawback is partially removed by diffusion Monte Carlo (DMC), which 
projects out the ground-state component of the starting trial wave function.
The only inaccuracy of DMC comes from the fermion-sign problem, 
the inability to directly sample the fermionic wave function. 
It can be circumvent by the fixed-node approximation, 
which enforces the {\em nodes} (the subset of position space where the wave function vanishes) of 
a trial wave function on the projected wave function, but introduces a
small {\em fixed-node error}. 

Fixed-node QMC simulations are typically more computationally demanding than traditional 
independent-particle quantum chemistry techniques, 
but have been very effective in providing high accuracy results 
for many real systems such as molecules, clusters and solids. 
Typically, for cohesive and binding energies, band gaps, and other energy differences 
the agreement with experiments is within 1-3\%~\cite{jeff_benchmark,qmcrev}.
The computational cost of QMC increases as the cube with system size, making
calculations with hundreds of electrons tractable; 
the large clusters~\cite{williamson01} and solids~\cite{williamson_fse}
up to 1000 electrons have already been studied. 

Today, the fundamental problem of accuracy improvement of QMC simulations 
lies in the elimination or at least in the control of the fixed-node errors.  
To achieve this, we focus in this thesis on studying the structure and properties of fermionic wave functions,
as well as on finding better approximations to their nodes.
This is quite challenging, because the fermion nodes are complicated high-dimensional manifolds determined 
by the many-body effects. Despite this difficulty, 
we were able to discover the exact nodes for several high-symmetry cases and 
to analyze the nodal structure of many spin-polarized and unpolarized systems~\cite{michal_prb,lubosmc28}. 

In the second part of the thesis, we provide the partial answer 
to the search for better approximations to the nodes of fermionic wave functions. 
We propose a generalized pairing trial wave function in the {\em Pfaffian functional form}, 
which leads us to accurate and compact description of the nodes and results in 
overall improvement in the accuracy of QMC~\cite{pfaffianprl,pfaffianprb}. 

The last part of the thesis deals with yet another way how to improve 
the nodes of wave functions. It proposes the fermion coordinate 
transformation of a {\em backflow type}~\cite{feynman,schmidt_bf,panoff,moskowitz,kwon1,kwon2,kwon3}, which was recently demonstrated 
to work also for inhomogeneous systems~\cite{drummond_bf,rios_bf}. We generalize its application to
Pfaffian pairing wave functions and perform the first tests of this approach.

\newpage
\noindent{\bf This thesis is organized as follows:}
\begin{itemize}
\item The remainder of this chapter gives an overview of the techniques 
used to study the quantum mechanics of many-body and chemical systems.
\item The Chapter~\ref{ch:qmc} follows with the description of the methodology 
behind the VMC and DMC methods. 
\item Our study of the exact and 
approximate nodes of fermionic wave functions is summarized in the Ch.~\ref{ch:nodes}.
\item Chapter~\ref{ch:pfaffians} discusses our calculations with 
generalized pairing wave functions. 
\item Some preliminary results on the 
backflow corrected trial wave functions are the content of the 
Ch.~\ref{ch:bf}. 
\item Finally, the last chapter concludes this dissertation.
\end{itemize}

\section{The Many-Body Schr\"odinger Equation}
One of the main challenges of condensed-matter physics and quantum
chemistry is the accurate solution of the Schr\"odinger equation.  
Since the nuclei are about thousand times heavier
than the electrons, the most common approach is to decouple 
the electronic and ionic degrees of freedom that is known as the {\em Born-Oppenheimer approximation}.
The electronic part of non-relativistic Born-Oppenheimer Hamiltonian for quantum system
of $N_e$ electrons in the presence of $N_I$ nuclei in Hartree atomic units
($\hbar=m_e=e=4\pi/\epsilon_0=1$) is then given by
\begin{equation}\label{eq:hamiltonian}
{\mathcal H}=- \frac 1 2 \sum_i \nabla^2_i-\sum_i\sum_I^{N_I}
        \frac{Z_I}{|r_i-R_I|} + \frac 1 2 \sum_i\sum_{j\neq
        i}\frac{1}{|r_i-r_j|} ,
\end{equation}
where $i$ and $j$ indexes are summing over all electrons and $I$ index
is summing over all nuclei.  A spectrum of states $\{\Phi_n\}$ which
diagonalizes the stationary Schr\"odinger equation
\begin{equation}\label{eq:sch}
{\mathcal H}\Phi_n=E_n\Phi_n
\end{equation}
with Hamiltonian~(\ref{eq:hamiltonian}) is the solution of above
many-body problem.

The history of {\it ab initio} methods\footnote{For more details of
the history of electronic structure, refer to a book by
R. M. Martin~\cite{martin}.}  for electronic structure started soon
after the invention of quantum mechanics by pioneering work of Heitler
and London~\cite{london}, who calculated the binding energy of H$_2$.
The first quantitative calculations of multi-electron systems where
accomplished by Hartree~\cite{hartree}, and
Hylleraas~\cite{hylleraas,node1s2s_1}. Fock~\cite{fock} was initial in
using the properly antisymmetrized wave function, the
Slater~\cite{slater} determinant, which established the Hartree-Fock
(HF) theory.  The HF theory replaces the hard problem of many
interacting electrons with system of independent electrons in
self-consistent field (SCF).

In electronic structure of solids, the discoveries of the effective
free electron theory together with Pauli {\em exclusion
principle}~\cite{pauli} and the {\em band theory} of Bloch~\cite{bloch}
were the first critical steps toward understanding crystals. In
1930s, the foundations for the basic classification of solids into
metals, semiconductors and insulators were laid.  Soon after, Wigner
and Seitz~\cite{wigner1,wigner2} performed the first quantitative
calculation of electronic states of sodium metal.

Today, the density functional theory (DFT) invented by Hohenberg and
Kohn~\cite{kohn_dft} and applied by Kohn and Sham~\cite{kohn_sham} is
the principal method for calculations of solids. Together with HF and
post HF methods, they are very relevant to our discussion of quantum
Monte Carlo (QMC), which relies on these methods mainly for its
input. What follows is their brief overview.

\section{Hartree-Fock Theory}\label{sec:intro:HF}
Due to the Pauli exclusion principle, any solution to the stationary
Schr\"odinger equation with Hamiltonian~(\ref{eq:hamiltonian}) has to
be antisymmetric under exchange of any two electrons with the same
spin as
\begin{align}
\Psi(\ldots,i,\ldots,j,\ldots)=-\Psi(\ldots,j,\ldots,i,\ldots).
\end{align} 
The Hartree-Fock theory ~\cite{fock,slater} uses the simplest
antisymmetric wave function, the Slater determinant,
\begin{align}\label{slaterdet}
\Psi(1,2,\ldots,N)&= \frac{1}{\sqrt{N!}} \sum_P (-1)^P \tilde
\varphi_{i_1}(1)\tilde \varphi_{i_2}(2)\ldots\tilde \varphi_{i_N}(N)\\
\nonumber
&=\frac{1}{\sqrt{N!}}\begin{vmatrix} \tilde\varphi_1(1) &
\tilde\varphi_1(2) & \ldots & \tilde\varphi_1(N)\\ \tilde\varphi_2(1)
& \tilde\varphi_2(2) & \ldots & \tilde\varphi_2(N)\\ \vdots & \vdots &
\vdots & \vdots \\ \tilde\varphi_N(1) & \tilde\varphi_N(2) & \ldots &
\tilde\varphi_N(N)\\
\end{vmatrix}
\equiv {\rm det}[\tilde\varphi_1(1),\ldots, \tilde\varphi_N(N)],
\end{align}
to approximate the state of $N$ electron system.  The $\tilde
\varphi_i(j)$ are one particle spin-orbitals, each of which is a product 
of spatial $\varphi_i^\sigma(j)$ and spin $\alpha_i(\sigma_j)$ orbitals.
(Note that $\varphi_i^\sigma(j)$ is independent of spin $\sigma$ 
in closed-shell cases. In open-shell systems, this assumption
corresponds to {\em spin-restricted Hartree-Fock approximation}).
If the Hamiltonian  is independent of spin or is diagonal 
in spin basis $\sigma=\{|\uparrow\rangle,|\downarrow\rangle\}$, the
expectation value of  Hamiltonian~(\ref{eq:hamiltonian}) with the wave function~(\ref{slaterdet}) is given by
\begin{align}
E_{HF}=&\sum_{i,\sigma}\!\int \varphi^{\sigma*}_i({\bf r})\left[
-\frac{1}{2} \nabla^2 + V_{ext} \right] \varphi^{\sigma}_i({\bf r})\,{\rm
d}{\bf r} \nonumber \\ &+\frac{1}{2}\sum_{i,j,\sigma_i,\sigma_j}\!\int
\!\!\!\int \frac{\varphi_i^{\sigma_i*}({\bf r}) \varphi_j^{\sigma_j*}({\bf r'})
\varphi_i^{\sigma_i}({\bf r}) \varphi_j^{\sigma_j}({\bf r'})}{|{\bf r}-{\bf r'}|} \,{\rm
d}{\bf r} \,{\rm d}{\bf r'}\nonumber \\
&-\frac{1}{2}\sum_{i,j,\sigma}\!\int \!\!\!\int
\frac{\varphi_i^{\sigma*}({\bf r}) \varphi_j^{\sigma*}({\bf r'}) \varphi_j^{\sigma}({\bf
r}) \varphi_i^{\sigma}({\bf r'})}{|{\bf r}-{\bf r'}|} \,{\rm d}{\bf r} \,{\rm
d}{\bf r'}.
\end{align}
Above expression for  HF energy contains three different terms, the first
being just sum of independent one-particle energies in the external
potential of nuclei $V_{ext}$, the second is the direct contribution
to Coulomb interaction also called the Hartree term and the last one
is the {\em exchange term} (or Fock term). Note, that in the
case of $i=j$ ({\em self-interaction contribution}) the last two terms
explicitly cancel each other.

\section{Post Hartree-Fock Methods}\label{sec:intro:postHF}
The single-determinant HF theory contains proper antisymmetry,
which introduces the effects related to the exchange. However, the
full electron-electron Coulomb repulsion is only approximated by Hartree
term. What is left out is referred to as {\em electronic
correlation}. The correlation energy is then defined as a difference
of energies of exact wave function and the best HF wave function of the
same state. Typically, the correlation energy constitutes only a
fraction of total energy, but it accounts for large portion of
cohesion and excitation energies. Obtaining the missing correlation is
therefore the principal challenge of all modern electronic structure
methods. The high accuracy is needed to access such problems as
description of magnetism or superconductivity.

The missing correlation in HF wave function can be accounted for by
additional Slater determinants. There exist several methods, which can
produce these multi-determinantal wave functions. The most frequently used are
configuration interaction (CI), multi-configurational self-consistent
field (MCSCF), and coupled cluster (CC) methods.  The idea behind CI
is to diagonalize the $N$-electron Hamiltonian on the space of
orthogonal Slater determinants.  These determinants are typically
constructed as multi-particle excitations from reference
determinant (usually HF solution).  If all the determinants are
included, the full CI is in principle exact. However, the number of
terms grows exponentially with $N$ and therefore, we have
to do limited expansions in practice.  
The most used is the configuration
interaction with all the single and double excitations (CISD). The
disadvantage of this approach is that the expansion converges very
slowly in correlation energy (as $\sqrt{N}$) and creates the 
{\em size-consistency problem} (i.e., wrong scaling of total energy with $N$).

The MCSCF is in some way a modification of truncated CI expansion,
when the orbitals used for construction of determinants are also
optimized together with determinantal weights.  The optimization of
all parameters is a difficult task and limits the number of
determinants in the expansion.

The size-consistency problem can be overcome by coupled cluster expansion~\cite{cizek}. All the excitations 
from the reference determinant are in principle included, but the coefficients of expansion are approximated and method is 
non-variational. The computational cost for CCSD (with singles and doubles) scales as 
$N^6$ and therefore constitutes a formidable challenge for larger systems. The commonly used 
CCSD(T) (CC with singles, doubles and approximated triples) produces the most accurate energies for small 
and intermediate systems available at the present time,  which in many cases serve as benchmarks for all other methods including QMC.
For excellent treatment and overview of above quantum chemical methods we refer the reader to book by A. Szabo and N. S. Ostlund~\cite{szabo}.


\section{Density Functional Theory}\label{sec:intro:DFT}
Previous methods were examples of the wave function based theories. 
For density functional theory the primary object is the {\em one-particle electron density}. 
It is formally an exact method based on Hohenberg and Kohn theorem~\cite{kohn_dft} 
that states that the ground state properties of a many-electron system can be obtained by 
minimizing the total energy functional 
\begin{equation}
E[\rho({\bf r})]=\int V_{ext}({\bf r}) \rho({\bf r})\,{\rm d}{\bf r} + F[\rho({\bf r})],
\end{equation}
where the $F[\rho({\bf r})]$ is some unknown, universal functional of electronic density $\rho({\bf r})$. 
The total energy $E$ has a minimum when the electronic density  $\rho({\bf r})$ is equal 
to the exact electronic density in some external potential $V_{ext}({\bf r})$. 
 
Kohn and Sham~\cite{kohn_sham} wrote the {\it ansatz} for the density in the
terms of one-electron orbitals of an auxiliary non-interacting system as
\begin{equation}
\rho({\bf r})=\sum_i^N |\varphi_i({\bf r})|^2.
\end{equation}
 The total energy of electronic system can be then expressed as  
\begin{equation}
E[\rho({\bf r})]=-\frac{1}{2} \sum_i \int \varphi_i^{*}({\bf r}) \nabla^2\varphi_i({\bf r})\,{\rm d}{\bf r}
+\int \rho({\bf r})V_{ext}({\bf r})\,{\rm d}{\bf r}
+ \frac{1}{2}\int\!\!\!\int \frac{\rho({\bf r})\rho({\bf r'})}{|{\bf r}-{\bf r'}|}\,{\rm d}{\bf r}{\rm d}{\bf r'}
+E_{xc}[\rho({\bf r})],
\end{equation}
where the first 2 terms represent the familiar energy of non-interacting system in the external potential, 
the third term is just the Hartree term and the rest is an unknown universal {\em exchange-correlation 
functional} of density. If the $E_{xc}[\rho({\bf r})]$ would be precisely known, 
we would have one to one mapping between difficult many-electron system and this 
one-particle problem. 

The simplest approximation to exchange-correlation functional is the local density approximation (LDA),
\begin{equation}
E_{xc}[\rho({\bf r})]=\int \epsilon_{xc}^{heg}(\rho({\bf r}))\rho({\bf r})\,{\rm d}{\bf r},
\end{equation}
where the $\epsilon_{xc}^{heg}$ is the exchange-correlation energy per electron in 
homogeneous electron gas. It is interesting to know that for practical application of the LDA, the correlation portion 
of this energy was taken from high-precision QMC calculation~\cite{ceperley_adler}. Even today, the LDA is widely used in solids. 

In cases where the LDA is not accurate enough, it seem natural to 
express the $\epsilon_{xc}(\rho({\bf r}), \nabla \rho({\bf r}))$ as a function of the local density and its gradient. 
This was the essential idea behind the generalized gradient approximation (GGA)~\cite{becke_3parm,becke_exchange,LYP}, 
which increased precision enabled wide-spread use in quantum chemical systems. 

There exist many flavors of density functional theory, 
e.g., hybrid-functionals~\cite{PBE,becke_3parm,LYP}, time-dependent DFT (TD-DFT)~\cite{tddft1,tddft2}, 
extensions to non-local density functionals (averaged density approximation (ADA)~\cite{ADA_WDA} and weighted density approximation (WDA)~\cite{ADA_WDA}),
or orbital-dependent functionals (e.g. self-interaction corrections (SIC)~\cite{self_interaction} and LDA+U~\cite{lda_plus_u}).   

The best known  density functional results are typically an order of magnitude less accurate than good QMC results.
However, the computational cost is much more favorable for DFT and consequently DFT 
is much more widely applied to 
variety of interesting applications in many fields of science and technology.

\chapter{Quantum Monte Carlo Methods}\label{ch:qmc}

\section{Introduction}
There are several Quantum Monte Carlo methods. They all apply a stochastic approach to find a solutions of 
a stationary Schr\"odinger equation of quantum systems.
However, in this dissertation we will be restricted to description of only the variational and diffusion Monte Carlo methods. 
In the variational Monte Carlo the expectation values are evaluated via stochastic integration over
$3N$ dimensional space. A variational theorem ensures that the expectation value of Hamiltonian with respect to given 
trial wave function is a true upper bound to the exact ground-state energy and 
sorely depends on the accuracy of the trial wave function. 
The second method, the diffusion Monte Carlo, removes some deficiency in the accuracy of trial wave function 
by employing imaginary-time projection of trial wave function onto the ground state. The nodal structure of 
trial wave function is enforced on the projected wave function by fixed node approximation.  
Hence, if the nodes of trial wave function were identical to the nodes of an exact one, 
we would know the ground state energy of a quantum system in polynomial time.


\subsection{Metropolis Sampling}
Most of quantum Monte Carlo calculations work with the Metropolis rejection algorithm~\cite{metropolis},
in which a Markov process is constructed to generate a random walk through the state space.
(In VMC and DMC methods we walk in $3N$ dimensional spatial space $\{{\bf R}\}$ and one fixed point ${\bf R}$
is also commonly referred to as {\em walker}.) The algorithm enables us to sample desired multi-dimensional 
probability distribution ${\mathcal P}({\bf R})$ without any prior knowledge of its normalization. 
Given some transition rule $P({\bf R}\to{\bf R}')$ from ${\bf R}$ to ${\bf R}'$, 
which satisfies {\em ergodicity} and {\em detailed balance},
\begin{align}
P({\bf R}\to{\bf R}'){\mathcal P}({\bf R})=P({\bf R}'\to{\bf R}){\mathcal P}({\bf R}'),
\end{align}
the desired probability density ${\mathcal P}({\bf R})$ will converge to an equilibrium state
given by
\begin{align}
\int P({\bf R}\to{\bf R}'){\mathcal P}({\bf R})\,{\rm d}{\bf R}={\mathcal P}({\bf R}').
\end{align}
The transition rule $P({\bf R}\to{\bf R}')$ can be further written as a product of the sampling distribution $T({\bf R}\to{\bf R}')$ 
(typically Gaussian distribution centered around {\bf R}) and the probability of an acceptance $A({\bf R}\to{\bf R}')$ of the proposed step
\begin{align}
P({\bf R}\to{\bf R}')=T({\bf R}\to{\bf R}')A({\bf R}\to{\bf R}').
\end{align}
The conditions of detailed balance are satisfied by choosing the $A({\bf R}\to{\bf R}')$ to be
\begin{equation}
A({\bf R}\to{\bf R}')=\min\left[1,\frac{T({\bf R}'\to{\bf R}){\mathcal P}({\bf R}')}{T({\bf R}\to{\bf R}'){\mathcal P}({\bf R})}\right].
\end{equation}
A Monte Carlo (MC) simulation initialized at some random state evolves toward an equilibrium. After the stabilization period we reach the point, where 
we can start to collect the statistics. The usual estimators for the mean, variance and error bars of the mean of some operator 
${\mathcal O}({\bf R})$ are given as  
\begin{align}
\langle {\mathcal O} \rangle^M &=\frac{1}{M}\sum_m^M {\mathcal O}({\bf R}_m),\\ \nonumber
\langle \sigma_{\mathcal O}^2\rangle^M&=\frac{1}{M-1}\sum_m^M ({\mathcal O}({\bf R}_m)-\langle {\mathcal O} \rangle_M)^2, \\ \nonumber
\epsilon_{\mathcal O}^M&=\frac{\sigma_{\mathcal O}^M}{\sqrt{M}},
\end{align}
where $M$ represents a number of sampling points. Furthermore, 
the well known {\em central limit theorem} then ensures that 
\begin{align}
\lim_{M \to \infty} \langle {\mathcal O} \rangle^M = \langle {\mathcal O} \rangle
\end{align}
with statistical error bars going to zero as $1/\sqrt{M}$.   

\section{Variational Monte Carlo}
Variational Monte Carlo (VMC) performs stochastic evaluation of an expectation value of Hamiltonian ${\mathcal H}$ over 
{\em trial wave function} $\Psi_T({\bf R})$, which is a reasonably good approximation of the true ground state. 
The $\Psi_T$ is subject to several necessary conditions, which are addressed in Sec.~\ref{sec:twf} of this chapter.

As follows from variational theorem, this expectation value provides 
rigorous upper bound on ground state energy $E_0$,
\begin{equation}
E_{VMC}=\frac{\int \Psi_T^*({\bf R}){\mathcal H}\Psi_T({\bf R})\,{\rm d}{\bf R}}{\int \Psi_T^*({\bf R})\Psi_T({\bf R})\,{\rm d}{\bf R}}
=\frac{\int |\Psi_T({\bf R})|^2 E_L({\bf R})\,{\rm d}{\bf R}}{\int  |\Psi_T({\bf R})|^2\,{\rm d}{\bf R}}\geq E_0,
\end{equation}
where in the second step we have introduced the term called local energy 
\begin{align}\label{eq:localenergy}
E_L({\bf R})=\frac{{\mathcal H}\Psi_T({\bf R})}{\Psi_T({\bf R})}
\end{align}
evaluated over probability density (also called the {\em importance function})
\begin{align}\label{eq:vmcP}
{\mathcal P}({\bf R})=\frac{|\Psi_T({\bf R})|^2}{\int  |\Psi_T({\bf R})|^2\,{\rm d}{\bf R}}.
\end{align}
The estimator of $E_{VMC}$ is then given as 
\begin{equation}
E_{VMC}^M=\frac{1}{M}\sum_m^M E_L({\bf R}_m),
\end{equation}
where $M$ configurations are distributed according  ${\mathcal P}({\bf R})$ (\ref{eq:vmcP}) via Metropolis algorithm.

Trial wave function typically depends on a set of variational parameters, which can be optimized in order to achieve 
the minimum of $E_{VMC}$ or the minimum of variance of local energy 
\begin{align}
\sigma_{VMC}^2=\frac{\int |\Psi_T({\bf R})|^2 (E_L({\bf R})-E_{VMC})^2\,{\rm d}{\bf R}}{\int  |\Psi_T({\bf R})|^2\,{\rm d}{\bf R}}.
\end{align}
The variance $\sigma_{VMC}^2$ is especially good function to minimize, since it is always 
positive and bounded from bellow ($\sigma_{VMC}^2 \to 0$ as $\Psi_T \to \Phi_0$). 
In practice, there are other possible combinations of $E_{VMC}$, $\sigma_{VMC}^2$ and some 
other functions of $\Psi_T$ which serve as good minimizers. We devote the Sec.~\ref{sec:qmc:opt} 
of this chapter to optimization of variational trial wave functions.

\subsection{Correlated Sampling}
The technique of a correlated sampling exploits the fact that the 
variance of difference of some correlated random variables $X$ and $Y$
decreases as their correlation increases. In QMC, the standard 
application of correlated sampling is to the VMC energy differences of two close (i.e., correlated) trial-wave functions, 
$\Psi_T^{(1)}$ and $\Psi_T^{(2)}$. The energy difference can be written as
\begin{align}
E_1-E_2&=\frac{\int |\Psi_T^{(1)}({\bf R})|^2 E_L^{(1)}({\bf R})\,{\rm d}{\bf R}}{\int |\Psi_T^{(1)}({\bf R})|^2\,{\rm d}{\bf R}} 
-\frac{\int |\Psi_T^{(2)}({\bf R})|^2 E_L^{(2)}({\bf R})\,{\rm d}{\bf R}}{\int |\Psi_T^{(2)} ({\bf R})|^2\,{\rm d}{\bf R}} \nonumber \\
&=\int |\Psi_T^{(1)}({\bf R})|^2 \left[\frac{E_L^{(1)} ({\bf R}) }{\int |\Psi_T^{(1)}({\bf R})|^2\,{\rm d}{\bf R}}
      -\frac{w({\bf R})E_L^{(2)}}{\int w({\bf R})|\Psi_T^{(1)}({\bf R})|^2\,{\rm d}{\bf R}}\right]\,{\rm d}{\bf R},
\end{align} 
where in the second line we used re-weighting factor $w({\bf R})=\frac{|\Psi_T^{(2)}({\bf R})|^2}{|\Psi_T^{(1)}({\bf R})|^2}$. 
Therefore, we can obtain the above energy difference as a sum over weighted local energy differences. 
Correlated sampling in DMC, however, requires further modification to the Green's function
(best known is the  Filippi and Umrigar's approximation~\cite{filippi_force}).

\section{Diffusion Monte Carlo}
\subsection{Imaginary Time Schr\" odinger Equation}
Diffusion Monte Carlo method belongs to a larger class of projection and Green's function MC methods. 
It exploits the imaginary time many-body Schr\" odinger equation 
\begin{equation}\label{eq:timesch}
\frac{{\rm d}\Psi({\bf R},\tau)}{{\rm d}\tau}=-({\mathcal H}-E_T)\Psi({\bf R},\tau),
\end{equation}
where $\Psi({\bf R},\tau)$ is the projected wave function with real variable $\tau$ 
measuring the progress in imaginary time and $E_T$ is an energy offset.
The equation~(\ref{eq:timesch}) is formally similar to 
a diffusion equation.
Its effect is to 
converge the initial wave function to the ground state. This can be 
easily seen by expanding $\Psi({\bf R},\tau)$ in the eigenstates $\{\Phi_n\}$ of ${\mathcal H}$ and projecting time $\tau \to \infty$
\begin{align}
\lim_{\tau \to \infty}\Psi({\bf R},\tau)&=\lim_{\tau \to \infty}\sum_n  \langle \Phi_n| \Psi(0) \rangle \,\,{\rm e}^{-\tau(E_n-E_T)} \,\Phi_n({\bf R}) \\ \nonumber
&=\lim_{\tau \to \infty}\langle \Phi_0| \Psi(0) \rangle \,\,{\rm e}^{-\tau(E_0-E_T)} \Phi_0({\bf R}).
\end{align}
Therefore, as far as there is non-zero overlap of the 
initial wave function with a ground state, we can always project out the later one. The parameter $E_T$ is adjusted during simulation 
to keep the amplitude of $\Psi$ constant. The initial wave function is typically analytically known [e.g. $\Psi({\bf R},0)=\Psi_T({\bf R})$] and 
the final projected wave function is represented via ensemble of walkers. Therefore, in this case the $\Psi$ not $\Psi^2$
is considered to be a population density of walkers. 

The diffusion algorithm is performed via Green's function 
\begin{align}
G({\bf R}\to{\bf R}',\tau)&=\langle {\bf R}| \,{\rm e}^{-\tau({\mathcal H}-E_T)}|{\bf R}'\rangle,
\end{align}
which determines the wave function after projection time $\tau$ as 
\begin{equation}
\Psi({\bf R},\tau)=\int G({\bf R}\to{\bf R}',\tau)\Psi({\bf R}',0)\,{\rm d}{\bf R}'. 
\end{equation}
The Green's function's short-time approximation~\cite{reynolds82} can be written using Trotter-Suzuki formula as
\begin{align}\label{eq:GFshortsimple}
G({\bf R}\to{\bf R}',\tau)&=\langle {\bf R}| \,{\rm e}^{-\tau(\hat{T}+\hat{V}-E_T)}|{\bf R}'\rangle\\ \nonumber
&\approx\,{\rm e}^{-\tau(V({\bf R})-E)/2}\langle {\bf R}| \,{\rm e}^{-\tau\hat{T}}|{\bf R}'\rangle \,{\rm e}^{-\tau(V({\bf R}')-E_T)/2}\\ \nonumber
&\approx\, (2\pi\tau)^{-3N/2} \exp\left[-\frac{({\bf R}-{\bf R}')^2}{2\tau}\right]
\exp\left[-\tau\left(\frac{V({\bf R})+V({\bf R}')}{2}-E_T\right)\right],
\end{align}
where in the last line we used the solution for kinetic term as a Gaussian expanding in time. 
We can interpret the action of the short-time Green's function as a {\em diffusion process} (kinetic energy term) and 
{\em branching} or {\em re-weighting process} (potential-energy term) on the population of walkers. 

\subsection{Importance Sampling}
Recent DMC methods treat diffusion problem by introducing the {\em importance sampling}~\cite{Grimm,qmchistory1,reynolds82} with the mixed distribution 
$f({\bf R},\tau)=\Psi_T({\bf R})\Psi({\bf R},\tau)$, where $\Psi_T({\bf R})$ is some trial wave function.
The modified diffusion equation for above mixed distribution $f$ is then 
\begin{align}
\frac{{\rm d}f({\bf R},\tau)}{{\rm d}\tau}=& -\frac{1}{2}\nabla^2 f({\bf R},\tau)
+\nabla \cdot [{\bf v}_D({\bf R})f({\bf R},\tau)] \\ \nonumber
&+(E_L({\bf R})-E_T)f({\bf R},\tau),
\end{align}
where ${\bf v}_D({\bf R})={\bf \nabla} \ln|\Psi_T({\bf R})|$ 
represents a drift velocity term and $E_L({\bf R})$ is the local energy of $\Psi_T({\bf R})$ from Eq.~(\ref{eq:localenergy}). 
The short-time approximation to the Green's function is then 
correspondingly
\begin{align}\label{eq:GFshort}
\tilde G({\bf R}\to{\bf R}',\tau)\approx &(2\pi\tau)^{-3N/2} \exp\left[-\frac{({\bf R}-{\bf R}'-\tau{\bf v}_D({\bf R}'))^2}{2\tau}\right]\\ \nonumber
&\times \exp\left[-\tau\left(\frac{E_L({\bf R})+E_L({\bf R}')}{2}-E_T\right)\right]\\ \nonumber
&=\tilde G_D({\bf R}\to{\bf R}',\tau)\times \tilde G_B({\bf R}\to{\bf R}',\tau).
\end{align}
Here, we can again interpret the action of the short-time Green's function as a diffusion process $\tilde G_D$ (kinetic energy term)
and branching/re-weighting process $\tilde G_B$ controlled by difference of averaged local energies at ${\bf R}$ and ${\bf R}'$ to $E_T$. 
This transformation has several consequences. 
First, the density of walkers is enhanced in the regions, where $\Psi_T({\bf R})$ is large due to drift velocity ${\bf v}_D({\bf R})$. 
Second, the re-weighting term contains the local energy $E_L({\bf R})$ in the exponent, which for a good $\Psi_T$ is 
close to a constant and therefore much better behaved than unbounded potential $V({\bf R})$. 

In the actual calculation, we start with walkers distributed according to $f({\bf R},0)=|\Psi_T({\bf R})|^2$
with the weights $\{w\}$ of all walkers set to one.
Then we perform MC step sampled from the kinetic part $\tilde G_D$ of Eq.~(\ref{eq:GFshort}). The step is accepted with Metropolis probability
\begin{equation}
A({\bf R}\to{\bf R}')=\min\left(1,\frac{\tilde G_D({\bf R}'\to{\bf R})|\Psi_T({\bf R}')|^2}{\tilde G_D({\bf R}\to{\bf R}')|\Psi_T({\bf R})|^2}\right).
\end{equation} 
Consequently, we assign to each walker a new weight according to 
\begin{equation}\label{eq:reweiht}
w'_i=\tilde G_B({\bf R}_i\to{\bf R}'_i,\tau)\,w_i.
\end{equation}
The advantage of this approach is that the DMC algorithm has essentially the VMC dynamics with a small time step with 
additional re-weighting [Eq.~(\ref{eq:reweiht})]. Over the simulation, some weights will start to dominate over others. It is therefore necessary to 
control their population. One way is to take two walkers, one with a large weight $w_1$ and another with a
small weight $w_2$. The first walker is then branched with probability $\frac{w_1}{w_1+w_2}$ to two, both having 
an average weight of $\frac{w_1+w_2}{2}$. Hence the walker with a small weight gets killed. 
After an equilibration period, we can start to collect statistics needed for the 
calculation of projected ground state energy
\begin{align}
E_{DMC}&=\lim_{\tau \to \infty}\frac{\int \Psi^*({\bf R},\tau){\mathcal H}\Psi_T({\bf R})\,{\rm d}{\bf R}}{\int \Psi^*({\bf R},\tau)\Psi_T({\bf R})\,{\rm d}{\bf R}} 
\\ \nonumber
&=\lim_{\tau \to \infty} \frac{\int f({\bf R},\tau) E_L({\bf R})\,{\rm d}{\bf R}}{\int  f({\bf R},\tau)\,{\rm d}{\bf R}}\\ \nonumber
&\approx\ \frac{\sum_i w_i E_L({\bf R}_i)}{\sum_i w_i}.
\end{align} 

\subsection{Fixed-Node Approximation}\label{subsec:qmc:fixednode}
\begin{figure}[!t]
\begin{center}
\includegraphics[width=0.80\columnwidth]{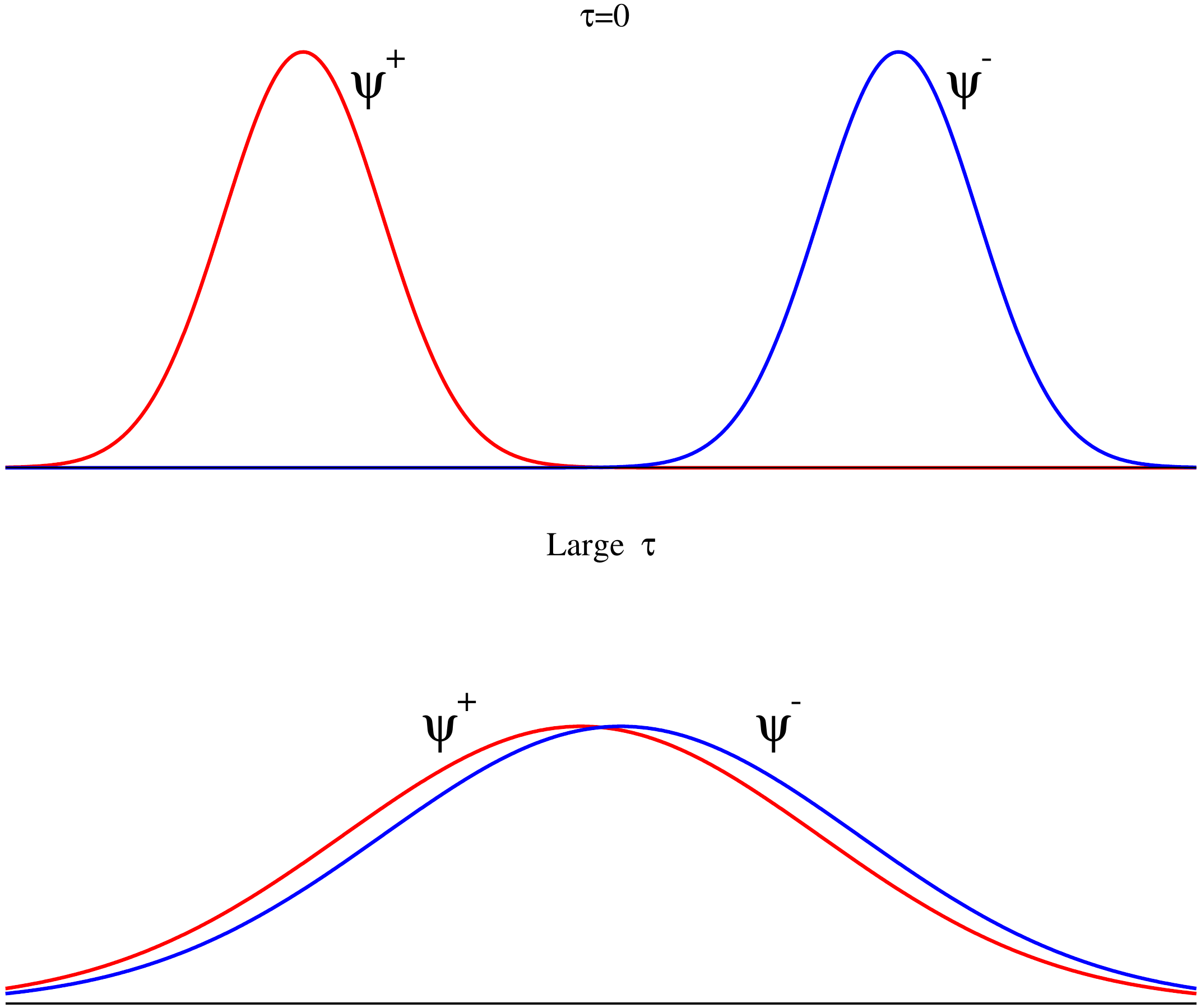}
\end{center}
\caption{Imaginary time behavior of walker distributions $\Psi^+$ (red) and $\Psi^-$ (blue). Upper figure: at the beginning of the diffusion ($\tau=0$).
Lower figure: After large enough imaginary-time evolution. The fermionic signal decays as ($\Psi^+ - \Psi^-) \propto \exp[-\tau(E_0^F-E_0^B)]$, where $E_0^F-E_0^B$ is a 
difference of energies of fermionic to bosonic ground states.}
\label{fig:sign_problem}
\end{figure}

For the bosonic systems, the ground state wave function (also $\Psi({\bf R},\tau)$ for that matter) 
has no nodes and can be considered to be positive everywhere. However, the fermionic wave function is naturally antisymmetric, 
and therefore its amplitude can no longer represent the sampling distribution for the walkers, which has to be positive everywhere. 
We could try to overcome this problem in naive way by decomposing $\Psi({\bf R},\tau)$ into 
positive functions $\Psi^+({\bf R},\tau)$ and $\Psi^-({\bf R},\tau)$ such that 
\begin{align}
\Psi({\bf R},\tau)=\Psi^+({\bf R},\tau)-\Psi^-({\bf R},\tau)
\end{align}
However, since our diffusion equation is linear in $\Psi({\bf R},\tau)$, both $\Psi^+({\bf R},\tau)$ and $\Psi^-({\bf R},\tau)$ 
will converge to the same bosonic distribution, hence making the averaging of fermionic observables difficult (See Fig.~\ref{fig:sign_problem}).
This is a well-known {\em fermion sign problem}. Up to day, no 
satisfactory 
solution
has been found, 
however it can be circumvented by the {\em fixed-node approximation}~\cite{jbanderson75,jbanderson76,moskowitz82,reynolds82}, 
which restricts the evolution of positive and negative 
walkers into  separate regions defined by sign of trial wave function $\Psi_T({\bf R})$.
This is achieved by introducing a mixed distribution $f({\bf R},\tau)=\Psi_T({\bf R})\Psi({\bf R},\tau)$ in the importance sampling, 
so that we can 
extend the stochastic sampling of the walkers for fermionic wave function to the regions where 
\begin{align}\label{eq:fixednode}
f({\bf R},\tau)\ge 0
\end{align} 
and prohibiting it everywhere else. In other words, by restricting the $\Psi({\bf R},\tau)$ to have the same sign as $\Psi_T({\bf R})$ for all $\tau$ 
we can readily satisfy Eq.~(\ref{eq:fixednode}). 
Therefore, the fixed-node approximation reduces the nonlocal antisymmetric condition for fermionic wave function to local boundary 
condition in Eq.~(\ref{eq:fixednode}). 
Furthermore, the drift-velocity term diverges at a node of $\Psi_T({\bf R})=0$ and acts like a repulsive potential, 
which stabilizes the simulation. The fixed-node DMC energy is still a rigorous upper bound to the true ground state energy~\cite{reynolds82,moskowitz82}. 
The resulting fixed-node error is proportional to the square of the nodal displacement error. 

\subsection{Beyond Fixed-Node Approximation}
The fixed-node approximation was historically the first and the simplest way how to overcome the spurious fermion sign problem. 
Besides improving the nodes with better trial wave functions (see Chs.~\ref{ch:pfaffians} and~\ref{ch:bf}) 
or attempts to understand them (see Ch.~\ref{ch:nodes}), there have been several efforts to go beyond the fixed-node approximation. 
One of the first were ideas based on the {\em cancellation of paired walkers} with opposite signs~\cite{kalos_pairing1}. 
This method was subsequently improved to work with fully interacting ensemble of walkers~\cite{kalos_pairing2}. 
Although the method suffered from unfavorable scaling caused by computation of inter-walker distances, 
it was applied (among others) to Li atom ground state with good agreement with estimated exact energy.

Ceperley and Alder~\cite{ceperley_node_release} on the other hand developed the {\em released-node method}.
Their method starts from an initial fixed-node distribution, but walkers are allowed to cross the nodes of 
a trial wave function. Each time a walker crosses the node it caries the additional weight in the form of $\pm$ sign, which 
contributes to the released-node energy. This algorithm is in principle transient (the decay of fermion first excited 
state has to be faster then the decay of fermion ground state to bosonic state), the authors were able to 
successfully calculate the total energies of a number of small atoms and molecules. 

Subsequently, Anderson and Traynor~\cite{anderson91} combined the best features of fixed-node, released-node and 
cancellation methods to algorithm, which employed improvements as relocation after node crossing, self-cancellations 
and multiple cancellations, and maximum use of symmetry in cancellations together with rigorous energy evaluation 
using importance sampling with trial wave functions. Among the most interesting systems, the authors calculated 
the excited state of H$_2$ $^3\Sigma_u^+$ and the barrier height for the reaction ${\rm H}+{\rm H}_2\to{\rm H}_2+{\rm H}$~\cite{anderson92}. 
However, the main obstacle to the application of this method to large systems is the computational cost associated with 
efficient annihilation of walkers. 

Among other alternative methods, let us also mention the method of {\em adaptive nodes} (see, e.g., Ref.~\cite{adaptive_nodes}).
The principle is based on representing the approximate nodal function directly from ensemble of walkers 
(e.g. Gaussian centered on each walker). This results in the adaptive description of the nodes
that does not depend upon {\it a priori} knowledge of the wave function. The main difficulty 
of the method however comes from scaling at higher dimensions $N$, where the antisymmetry condition results 
in the calculation of $N!$ permutations.   

The critical component of any method, which would solve the now famous fermion sign problem, 
is the polynomial scaling with the number of particles. There might exist approaches, which scale polynomially 
for some cases~\cite{kalos_pairing3}, but in general will scale exponentially. 
Indeed, recently has been shown~\cite{troyer_nphard} that the problem of quantum Monte Carlo simulation of 
a random coupling Ising model with fermion sign problem belongs into the class of {\em nondeterministic polynomial} (NP) {\em hard} problems.
It is widely excepted that no NP problem can be solved in the polynomial (P) time, hence $\rm NP\neq P$. 
This finding thus implies that the general solution to the fermion sign problem 
in polynomial time is not possible. 
   
\section{Variational Trial Wave Functions}\label{sec:twf}
The great advantage of QMC methods is the capability 
to use explicitly correlated trial wave functions. 
Their quality is the most important factor on final accuracy and 
efficiency of QMC. On the other hand, the repeated evaluation of $\Psi_T$ (and $\nabla \Psi_T$, $\nabla^2 \Psi_T$) 
is  the most costly part of QMC calculation. 
Therefore, it is desirable to seek both {\em highly accurate} and {\em efficient} representations of $\Psi_T$. 

\subsection{Basic Properties}
Any reasonably accurate trial wave function for QMC has to obey several basic properties. 
Since $\Psi_T$ is the approximate solution to the bound electronic system,
we demand the $\int \Psi_T^* \Psi_T $, $\int \Psi_T^* {\mathcal H} \Psi_T$
and $\int \Psi_T^* {\mathcal H}^2 \Psi_T$ to exist. 
This is obvious, since the VMC expectation value and its variance would not be properly defined otherwise.
In addition, if Hamiltonian does not explicitly depend on
the magnetic field (has time-reversal symmetry), $\Psi_T$ can be made real.

Further, $\Psi_T$ and $\nabla \Psi_T$ has to be continuous, wherever the potential is finite. 
In addition, the continuity of the potential implies also the continuity of $\nabla^2 \Psi_T$.
Important corollary is as $\Psi_T$ approaches 
an exact solution, local energy becomes constant everywhere. Therefore, any singularity from 
the potential has to be properly canceled by an opposite singularity in the kinetic energy, which 
gives rise to Kato {\em cusp conditions}~\cite{Kato57,Pack66}.
As the distance of electron to nucleus goes $r_{I,i} \to 0$, the potential energy divergence  
gets properly canceled  when
\begin{align}\label{eq:necusp}
{\frac{1}{\Psi_T}{\frac{\partial \Psi_T}{\partial r_{Ii}} }\bigg\arrowvert}_{r_{Ii}=0}=-Z_I.
\end{align}
The electron-nucleus cusp condition [Eq.~(\ref{eq:necusp})] is typically satisfied by the choice of proper orbitals or 
removed by the use of pseudopotentials (see Sec.~\ref{sec:pseudo}).
Similarly, as the electron-electron distance goes $r_{ij} \to 0$, there is a divergence in electron-electron Coulomb potential. 
Let us for now assume that $i$ and $j$ electrons bear the same spin. 
If we write trial wave function as a product of an antisymmetric function $\Psi_A(r_{ij})$ and a symmetric function
$\Psi_S(r_{i,j})$ with respect to $r_{i,j}$, then the divergent terms cancel if 
\begin{align}\label{eq:cupsupup}
{\frac{1}{\Psi_S}{\frac{\partial \Psi_S}{\partial r_{ij}} }\bigg\arrowvert}_{r_{ij}=0}=\frac{1}{4}.
\end{align}
For two electrons with unlike spins, the situation is different, since the antisymmetric part $\Psi_A(r_{ij}=0)$ is generally non-zero.
The unlike spins cusp condition therefore modifies to
\begin{align}\label{eq:cupsupdown}
{\frac{1}{\Psi_S}{\frac{\partial \Psi_S}{\partial r_{ij}} }\bigg\arrowvert}_{r_{ij}=0}=\frac{1}{2}.
\end{align}
Finally, $\Psi_T$ has to be antisymmetric with respect 
to exchange of both spin and spatial 
coordinates
\begin{align}
\Psi_T(P{\bf R},P{\boldsymbol \Sigma})=(-1)^P \Psi_T({\bf R},{\boldsymbol \Sigma}),
\end{align}
where ${\boldsymbol \Sigma}=(\sigma_1,\ldots,\sigma_N)$ are discrete spin variables with values $\pm \frac{1}{2}$
and $P$ is an arbitrary permutation with sign equal to $(-1)^P$. However, if the 
Hamiltonian does not contain spin-dependent terms, we will impose antisymmetry only with respect to interchanges between the 
electrons of the same spin. That is, the spatial antisymmetry is only required when 
\begin{align}\label{eq:PSigma}
P{\boldsymbol \Sigma}={\boldsymbol \Sigma}.
\end{align}
Then the spins of all electrons are being fixed with total spin projection 
equal to $\frac{1}{2}(N_\uparrow-N_\downarrow)$. We can therefore 
label the first $N_{\uparrow}$ particles as spin-up and the rest  $N-N_\downarrow$
as spin-down. Throughout this dissertation we will always assume that Eq.~(\ref{eq:PSigma}) 
holds and that the spin variable is factored out by the above labeling scheme.

\subsection{Trial Wave Functions Forms}\label{subsec:twf}
Almost all trial wave functions used in the QMC today are expressed as
product of some antisymmetric part $\Psi_A({\bf X})$ and symmetric exponential of 
Jastrow correlation factor $U_{corr}({\bf R})$ 
\begin{align}
\Psi_T({\bf R})=\Psi_A({\bf X})\times \exp[U_{corr}({\bf R})],
\end{align}
where ${\bf X}=({\bf x}_1,\ldots,{\bf x}_N)$ represents some general quasi-particle coordinates 
dependent on all $N$ electron positions ${\bf R}$. For simplicity, let us assume 
${\bf X}\equiv {\bf R}$. We will come back to a more general case when we will 
be discussing the {\em backflow transformation}.

Bellow, we will describe the particular forms for $\Psi_A({\bf R})$, $U_{corr}({\bf R})$ and 
${\bf X}({\bf R})$ as implemented in our QMC code QWALK~\cite{qwalk}.
Jastrow part~\cite{jastrow,reynolds82,moskowitz,cyrus1}, explicitly dependent on electron-nucleus and 
electron-electron distances, 
\begin{align}
U_{corr}(\{r_{ij}\},\{r_{iI}\},\{r_{jI}\})=\sum_{iI} \chi(r_{iI})+ \sum_{i\neq j} u(r_{ij}) 
+ \sum_{i\neq j,I} w(r_{ij},r_{iI},r_{jI}) 
\end{align}
is written as the sum of one-body (electron-nucleus), two-body (electron-electron) and three-body (electron-electron-nucleus) terms.
The function of these terms is twofold. First is to describe the electron-electron cusp conditions [Eqs.~(\ref{eq:cupsupup}) and~(\ref{eq:cupsupdown})]. 
These are satisfied if 
\begin{align}
{{\frac{\partial u_{\sigma_i \sigma_j}}{\partial r_{ij}} }\bigg\arrowvert}_{r_{ij}=0}=
\left\{
\begin{array}{cc}
\frac{1}{4} & \sigma_i=\sigma_j,\\
\frac{1}{2} & \sigma_i=-\sigma_j,
\end{array} \right.
\end{align}
and all the remaining functions have zero derivative.
Second purpose is to incorporate the correlation effects not present in $\Psi_A({\bf R})$.
Expanded in the basis of one dimensional functions the components of the Jastrow factor take form
\begin{align}
\chi(r)&=\sum_k c_k^{en} a_k(r),  \\
u(r)&=\sum_k c_k^{ee} b_k(r), \\
w(r_{ij},r_{iI},r_{jI})&=\sum_{klm} c_{klm}^{een} [a_k(r_{iI})a_l(r_{jI})+a_k(r_{jI})a_l(r_{iI})]b_m(r_{ij}).
\end{align}
The above Jastrow factor has proved to be very efficient in describing correlation effects with minimal number of variational 
parameters. More details about the 
implemented basis functions can be found in App.~\ref{appendix:functions} or in the QWalk documentation~\cite{qwalk}.

To propose a general fully antisymmetric wave function is a formidable challenge
given the scarcity of antisymmetric algebraic forms with known evaluation schemes. 
This dissertation deals with two of them, determinant and {\em Pfaffian}. 
The great success of quantum-chemical theories is based on the wave function constructed from
Slater determinants written in spin-factorized form as 
\begin{align}
\Psi_A^{Slater}({\bf R})=\sum_i w_i \, D^{\uparrow}_i(1,\ldots, N_\uparrow)\, D^{\downarrow}_i(N_\uparrow+1,\ldots, N).
\end{align}
Each spin-up and spin-down Slater determinant
\begin{align}
D^{\uparrow}_i(1,\ldots, N_\uparrow)&={\rm det}[\varphi_1^{i\uparrow}(1),\ldots, \varphi_{N_\uparrow}^{i\uparrow}(N_\uparrow)], \\
D^{\downarrow}_i(N_\uparrow+1,\ldots, N)&={\rm det}[\varphi_1^{i\downarrow}(N_\uparrow+1),\ldots, \varphi_{N_\downarrow}^{i\downarrow}(N)] 
\end{align}
is a function of one-particle orbitals taken from (post-)HF or DFT methods. In the introductory chapter we have touched upon some of the 
properties of $\Psi_A^{Slater}$. The most important one is that the Slater determinants can form a complete set of antisymmetric functions 
and are therefore in principle capable of describing the exact wave function. 
In practice, however, we are limited to finite expansions. The missing correlation is then added by Jastrow correlation factor. 
The resulting Slater-Jastrow wave function $\Psi^{S-J}_T=\Psi_A^{Slater}\exp[U_{corr}]$ 
is most commonly used type of trial wave function in QMC.
Although very successful, the $\Psi^{S-J}_T$ has its limitations summarized bellow:
\begin{itemize}
\item[(a)] The large CI-like expansions needed for high accuracy results are computationally expensive to evaluate. We are thus 
bound to small or intermediate chemical systems. The expansion is nonexistent in solids.
\item[(b)] The single-determinant Slater-Jastrow wave function $\Psi^{S-J}_T$ provides in many cases good results 
for energy differences such as cohesion and binding due to large cancellation of errors. 
However, any single-determinant $\Psi^{S-J}_T$, with exception of fully-polarized state, divides space into 4 (2 for each spin) nodal cells, 
while in many cases the exact ground state wave function has only two nodal cells (for more on {\em conjecture of minimal nodal cell division}, see 
Ch.~\ref{ch:nodes}).
\item[(c)] The accuracy of Slater determinant based wave function is strongly dependent on the quality of one-particle orbitals. 
These come from methods which do not incorporate the explicit correlation such as Jastrow correlation factor and therefore are usually not optimal. 
However, to optimize many orbital coefficients is still a formidable challenge.
\end{itemize}
The above limitations forced us and many others to think about novel wave functions with 
capabilities beyond $\Psi^{S-J}_T$. Since determinant is the simplest antisymmetric object constructed 
from one particle orbitals, what would this object look like if we would build it from two, three or more particle orbitals?
For two particle orbitals this object is called {\em Pfaffian} and the resulting wave function a {\em Pfaffian pairing wave function}.

Similarly to multi-determinantal case, we can construct the Pfaffian pairing wave function as a 
linear combination of several Pfaffians
\begin{align}\label{eq:mpf1}
\Psi_A^{PF}({\bf R})=\sum_i w_i \, {P\!f}_i(1,\ldots, N).
\end{align} 
Each $i$th Pfaffian in the expansion~(\ref{eq:mpf1})
\begin{align}
{P\!f}_i(1,\ldots, N)&={\mathcal A}[\tilde{\phi}_i(1,2), \tilde{\phi}_i(3,4), \ldots, 
                            \tilde{\phi}_i(N-1,N)]\\
&={\rm pf}[\tilde{\phi}_i(1,2), \tilde{\phi}_i(3,4), \ldots, 
                            \tilde{\phi}_i(N-1,N)]
\end{align} 
is an antisymmetrized product of one type of $N/2$ pairing orbitals $\tilde{\phi}_i$.
The description of Pfaffian functional form together with application of Pfaffian pairing wave functions 
to several systems and possible extensions is discussed in great detail in Ch.~\ref{ch:pfaffians}.

Lets us go back to a generalization mentioned at the beginning of this section. 
For several decades, it was known that it is possible to improve the nodal 
accuracy of trial wave functions for homogeneous systems by introducing 
a quasi-particle coordinates~${\bf X}$~\cite{feynman,schmidt_bf,panoff,moskowitz,kwon1,kwon2,kwon3}, i.e.,
\begin{align}\label{eq:bf_trans}
{\bf R}\stackrel{{T}_{BF}}{\longrightarrow}{\bf X},
\end{align}
where ${T}_{BF}$ represents the {\em backflow transformation}. As a consequence, the 
nodes of $\Psi_A({\bf X})$ will be in general different from those of $\Psi_A({\bf R})$.
Recently, some progress was also reported with chemical (inhomogeneous) systems~\cite{drummond_bf,rios_bf}. 
In this dissertation we put these new developments into test on several systems.  
The quasi-coordinate of $i$th electron at position ${\bf r}_i$ is given as 
\begin{align}
{\bf x}_i&={\bf r}_i+{\boldsymbol \xi}_i({\bf R}) \nonumber \\
&={\bf r}_i+{\boldsymbol \xi}_i^{en}({\bf R})+{\boldsymbol \xi}_i^{ee}({\bf R})+{\boldsymbol \xi}_i^{een}({\bf R}),
\end{align}
where we have again divided the contributions to one-body (electron-nucleus), two-body (electron-electron) 
and three-body (electron-electron-nucleus) backflow terms.
They can be further expressed as 
\begin{align}
{\boldsymbol \xi}_i^{en}({\bf R})&=\sum_I \chi(r_{iI}) {\bf r}_{iI} \\
{\boldsymbol \xi}_i^{ee}({\bf R})&=\sum_{j\ne i} u(r_{ij}) {\bf r}_{ij} \\
{\boldsymbol \xi}_i^{een}({\bf R})&=\sum_I \sum_{j\ne i} [w_1(r_{ij},r_{iI},r_{jI}) {\bf r}_{ij} + w_2(r_{ij},r_{iI},r_{jI}) {\bf r}_{iI}],
\end{align}
where ${\bf r}_{ij}={\bf r}_i-{\bf r}_j$  and $\chi$, $u$ and $w_1$ with $w_2$  are similar to one, two and three-body Jastrow terms. The calculational 
and implementation details together with some results are further discussed in Ch.~\ref{ch:bf}.

\section{Pseudopotentials}\label{sec:pseudo}
Many properties of interest in electronic structure methods are well 
determined by the behavior of only the valence electrons. 
On the other hand, the computational cost for QMC methods increases very rapidly with the atomic number $Z$
($\propto Z^{5.5}$~\cite{ceperley86} to $Z^{6.5}$~\cite{hammond87}).
It is therefore physically desirable 
and computationally necessary to approximate the effect of core electrons by {\em pseudopotentials}.

The presence of core electrons of atoms complicates the calculation for two reasons.
The shorter length scales associated with variation of wave functions near the 
atomic core with large $Z$ require the decrease of a time step in QMC simulations.  
The other problem is the increase of fluctuations of local energy as $Z$ gets bigger 
caused by improper cancellation of divergent kinetic and potential terms.

Analogously to independent particle theories such as Hartree-Fock and DFT, 
in QMC we remove the core electrons of atoms from calculation by introducing
effective core potentials, also called pseudopotentials. The action 
of pseudopotential is typically different for electrons with different angular momenta. 
It is conventional to divide the pseudopotential operator to local (common to all angular momenta) and
nonlocal (independent for each angular momentum) parts as 
\begin{align}
V_{ps}({\bf R})&=V_{loc}({\bf R})+\hat{V}_{nloc}\nonumber \\
&=\sum_i V_{loc}^{ps}({\bf r}_i) + \sum_i \hat{V}_{nloc}^{ps}({\bf r_i}),
\end{align}
where 
\begin{align}
\hat{V}_{nloc}^{ps}({\bf r}_i)=\sum_l V_{l}^{ps}({\bf r_i}) |l\rangle \langle l|
\end{align}
sums the contributions to nonlocal pseudopotential over angular momenta $l$ for $i$th electron.
The portion of the local energy $E_L({\bf R})$ from action of $\hat{V}_{nloc}$ on $\Psi_T({\bf R})$ is then  
\begin{align}
\frac{\hat{V}_{nloc}\Psi_T({\bf R})}{\Psi_T({\bf R})}
=\sum_i \frac{\hat{V}_{nloc}^{ps}({\bf r}_i) \Psi_T({\bf R})}{\Psi_T({\bf R})}.
\end{align}
Each $i$th argument in the above sum is evaluated as angular integration over 
the surface of the sphere passing through the $i$th electron and centered on the origin. 
If we choose the $z$-axis parallel to ${\bf r}_i$, the contribution for the $i$th electrons gets
\begin{align}\label{eq:ppint}
\frac{\hat{V}_{nloc}^{ps}({\bf r}_i) \Psi_T({\bf R})}{\Psi_T({\bf R})}=&\sum_l V^{ps}_l({\bf r}_i) \frac{2l+1}{4\pi} \int P_l[\cos(\theta_i')]
\times \frac{\Psi_T({\bf r}_1,\ldots,{\bf r}_i',\ldots,{\bf r}_N)}{\Psi_T({\bf r}_1,\ldots,{\bf r}_i,\ldots,{\bf r}_N)} {\rm d}\Omega_{{\bf r}_i'},
\end{align}
where $P_l$ denotes the $l$th Legendre polynomial. The integral is evaluated numerically 
on a spherical grid of Gaussian quadrature providing exact results up to certain maximum value of $l_{max}$. 
When the orientation of axes of a given quadrature are chosen randomly,  
the result is an unbiased Monte Carlo estimation of integral (\ref{eq:ppint}). 
The quadratures implemented in QWALK~\cite{qwalk} use $l_{max}=3$ (Octahedron - 6 points on the sphere) or 
higher precision $l_{max}=5$ (Icosahedron - 12 points). For more details, see original Ref.~\cite{lubos_ps}.

In the DMC method, the application of  nonlocal operator $\hat{V}_{nloc}$ causes the approximate 
propagator $\langle {\bf R}| \,\exp[-\tau \hat{V}_{nloc}]|{\bf R}'\rangle$ to have a fluctuating sign. 
Consequently, the imaginary time Schr\" odinger equation with importance function $f({\bf R},\tau)=\Psi_T({\bf R})\Psi({\bf R},\tau)$ will 
have modified form 
\begin{align}\label{eq:locapprox}
\frac{{\rm d}f({\bf R},\tau)}{{\rm d}\tau}=& -\frac{1}{2}\nabla^2 f({\bf R},\tau)
+\nabla \cdot [{\bf v}_D({\bf R})f({\bf R},\tau)] +[E_L({\bf R})-E_T]f({\bf R},\tau) \nonumber \\
&+\Bigg\{ \frac{\hat{V}_{nloc} \Psi_T({\bf R})}{\Psi_T({\bf R})} -\frac{\hat{V}_{nloc} \Psi({\bf R},\tau)}{\Psi({\bf R},\tau)} \Bigg\},
\end{align}
where the last term is unknown.
We can therefore introduce the {\em localization approximation} by neglecting the last term in Eq.~(\ref{eq:locapprox}). 
It has been shown  by Mitas {\it et. al.,}~\cite{lubos_ps}, that the error in energy from
localization approximation $\propto (\Psi_T -\Phi_0)^2$. Fortunately, 
accurate Jastrow-based trial wave functions are available in most cases, 
so the localization error is small. Comparison
of calculations with experiments for small systems had shown that the 
error from localization approximation is typically smaller then from fixed-node approximation.
Further tests  on transition metal atoms~\cite{lubosFe,flad&dolg} revealed some dependence of errors from localization approximation 
on quality of trial wave functions.
Recently, a new algorithm based on lattice-regularization by Causula~\cite{casula_lrdmc}
was demonstrated to eliminate the localization approximation.

\section{Optimization of Variational Wave Functions}\label{sec:qmc:opt}
As we have argued earlier in this chapter, the quality of trial wave functions controls the efficiency of VMC and DMC methods 
as well as final accuracy of their results. Given the freedom of choice for a wave function form in Monte Carlo, 
the accurate wave function may depend on many linear and non-linear parameters. 
The challenge is then the effective optimization of these variational parameters.

\subsection{Variance Minimization}
Let us denote the set of variational parameters $\{c\}$ of some
real valued trial variational wave function $\Psi_{\{c\}}$. 
The mean value of local energy (or VMC energy) with respect to $\Psi_{\{c\}}$ 
is given as
\begin{align}\label{eq:energymin}
E_{VMC}^{\{c\}}=\frac{\int \Psi_{\{c\}}^2 E_L^{\{c\}}\,{\rm d}{\bf R}}{\int \Psi^2_{\{c\}}\,{\rm d}{\bf R}}
=\langle E_L^{\{c\}} \rangle\equiv \bar{E}^{\{c\}}.
\end{align}
The variance of local energy is given as
\begin{align}\label{eq:varmin}
\sigma^2_{\{c\}}=\frac{\int \Psi^2_{\{c\}} (E_L^{\{c\}}-\bar{E}^{\{c\}})^2\,{\rm d}{\bf R}}{\int  \Psi^2_{\{c\}}\,{\rm d}{\bf R}}
=\langle(E_L^{\{c\}}-\bar{E}^{\{c\}})^2\rangle,
\end{align}
where we introduced a notation of the form:
\begin{align}
\langle {\mathcal O} \rangle\equiv \frac{\int {\mathcal O}({\bf R}) \Psi^2  \,{\rm d}{\bf R}}{\int  \Psi^2\,{\rm d}{\bf R}}.
\end{align}

The simplest of the wave function optimization methods is to minimize the variance~\cite{cyrus1} in Eq.~(\ref{eq:varmin}) on the
set of fixed, finite MC configurations, where the walkers are distributed according to $\Psi^2_{\{c_0\}}$.
The set of starting variational parameters is denoted as $\{c_0\}$. The variance is generally a good 
function to minimize, since it is always positive and bounded from bellow by zero. In practice, 
the  mean value of local energy at the end of optimization is not {\it a priori} known,
so we replace $\bar{E}^{\{c\}}$ in Eq.~(\ref{eq:varmin}) by a reference energy $E_{ref}$. 

There are several modifications to variance minimization.
One standard improvement is to use weights, which account for wave function change after each step 
of optimization. The {\em re-weightened variance} with the new set of parameters $\{c_{new}\}$ is then given as 
\begin{align}\label{eq:varmin2}
\sigma^2_{\{c_{new}\}}=\frac{\int \Psi^2_{\{c_0\}}W_{\{c_{new}\}}({\bf R})(E_L^{\{c_{new}\}}-E_{ref})^2\,{\rm d}{\bf R}}
{\int  \Psi^2_{\{c_0\}}W_{\{c_{new}\}}({\bf R})\,{\rm d}{\bf R}},
\end{align}
where we have introduced weights
\begin{align}\label{eq:weights}
W_{\{c_{new}\}}({\bf R})=\frac{\Psi^2_{\{c_{new}\}}({\bf R})}{\Psi^2_{\{c_0\}}({\bf R})}.
\end{align}
The advantage of the re-weighting scheme is a more accurate value of variance at each step of optimization. 
However, for optimization on small MC samples, sooner or later one weight will start to dominate over the others thus
biasing the variance estimate.

Given the tools above, it is now possible to employ a suitable minimization algorithm for 
the search of a minimum. We use two different methods. First one is 
a modified version of a {\em quasi-Newton method}~\cite{va10a}. 
It uses only the value of a function to be minimized and builds up the
information about the curvature of parameter space throughout the minimization.
Second algorithm is based on the general Levenberg-Marquardt (LM) method discussed bellow (see Sec.~\ref{sec:LM}).
It has a smooth transition between {\em Newton method} and {\em steepest descent method} with build-in stabilization parameter. 
However, it requires the information about gradient and Hessian of a minimized function and therefore it is more costly to calculate. 
The gradient of variance with respect to $i$th parameter is readily obtained by differentiating Eq.~(\ref{eq:varmin}) as 
\begin{align}
\sigma^2_i=2\bigg[ \langle E_{L,i}(E_L-\bar{E})\rangle +\Big \langle \frac{\Psi_i}{\Psi}E_L^2\Big\rangle 
-\Big \langle \frac{\Psi_i}{\Psi} \Big\rangle \Big \langle  E_L^2 \Big\rangle 
-2\bar{E}-\Big \langle \frac{\Psi_i}{\Psi} (E_L-\bar{E}) \Big\rangle
\bigg], 
\end{align}
where subscript $i$ denotes $\frac{\partial}{\partial c_i}$.
Since the variance minimization method can be viewed
as a fit of the local energy on a 
fixed MC configurations~\cite{cyrus1}, an alternative expression 
for variance gradient follows from ignoring the change of the wave function
\begin{align}\label{eq:vargrad}
\sigma^2_i=2\langle E_{L,i}(E_L-\bar{E})\rangle. 
\end{align}
The Hessian derived from gradient (\ref{eq:vargrad}) is then given as 
\begin{align}\label{eq:varhess}
\sigma^2_{ij}=2\langle (E_{L,i}-\bar{E})(E_{L,j}-\bar{E})\rangle. 
\end{align}
Important characteristic of Hessian (\ref{eq:varhess}) is that it is 
symmetric in $i$ and $j$ and positive definite. 
    
\subsection{Energy Minimization}
A straightforward  minimization of mean energy Eq.~(\ref{eq:energymin}) is in general 
quite unstable. The reason is that for sufficiently flexible variational wave function 
it is possible to lower the minimum of energy for a finite set of MC configurations,  
while in fact raising the true expectation value. As we have mentioned earlier, the 
variance is bounded from bellow and therefore this problem is far less severe~\cite{cyrus1}.
The simple implementation 
therefore requires 
to minimize the energy on large MC configurations and possibly adjust for parameter changes by re-weighting procedure. 
Even then, as the initial and final wave function start to diverge, it is necessary to re-sample the MC configurations.

Similarly to variance minimization, we can improve the convergence of the optimization by employing LM method
and use information about gradient and Hessian of mean of local energy. 
The gradient of the mean of local energy can be readily obtained from Eq.~(\ref{eq:energymin}) as
\begin{align}\label{eq:engrad}
\bar{E}_i&=\Big \langle  \frac{\Psi_i}{\Psi}E_L +  \frac{H\Psi_i}{\Psi} -2\bar{E}\frac{\Psi_i}{\Psi}\Big\rangle \nonumber \\
&=2\Big \langle \frac{\Psi_i}{\Psi}(E_L-\bar{E})\Big\rangle\qquad{\rm \mbox{(by Hermicity)}}. 
\end{align}
Note, that the expression in the last step of Eq.~(\ref{eq:engrad}) has a favorable property of 
zero fluctuations as $\Psi\to\Phi_0$.  
Taking direct derivative of  Eq.~(\ref{eq:engrad}), the Hessian is 
\begin{align}\label{eq:hess1}
\bar{E}_{ij}=&2\bigg[ \Big \langle \Big( \frac{\Psi_{ij}}{\Psi}+\frac{\Psi_i \Psi_j}{\Psi^2}\Big) (E_L-\bar{E}) \Big\rangle \nonumber \\
&-\Big \langle \frac{\Psi_i}{\Psi}\Big\rangle\bar{E}_j 
-\Big \langle \frac{\Psi_j}{\Psi}\Big\rangle\bar{E}_i + \Big\langle \frac{\Psi_i}{\Psi}E_{L,j}\Big\rangle \bigg ].
\end{align}
It is clear that the above Hessian is not symmetric in $i$ and $j$ when approximated by finite sample.
However, Umrigar and Filippi~\cite{cyrus2} recently demonstrated that the fully symmetric Hessian 
written entirely in the terms of covariances ($\langle ab \rangle - \langle a \rangle \langle b \rangle$) 
has much smaller 
fluctuations then Hessian~(\ref{eq:hess1}). 
Following their approach, modified Hessian from Ref.~\cite{cyrus2} is then given as
\begin{align}\label{eq:hess2}
\bar{E}_{ij}=&2\bigg[ \Big \langle \Big( \frac{\Psi_{ij}}{\Psi}+\frac{\Psi_i \Psi_j}{\Psi^2}\Big) (E_L-\bar{E}) \Big\rangle
-\Big \langle \frac{\Psi_i}{\Psi}\Big\rangle\bar{E}_j 
-\Big \langle \frac{\Psi_j}{\Psi}\Big\rangle\bar{E}_i 
\bigg ]\nonumber \\
&+\Big[ \Big \langle \frac{\Psi_i}{\Psi}E_{L,j}\Big\rangle
 -\Big \langle \frac{\Psi_i}{\Psi}\Big\rangle \langle  E_{L,j} \rangle \Big]
 +\Big[ \Big \langle \frac{\Psi_j}{\Psi}E_{L,i}\Big\rangle 
 -\Big \langle \frac{\Psi_j}{\Psi}\Big\rangle \langle  E_{L,i} \rangle \Big],
\end{align}
where we added three additional terms $\Big\langle \frac{\Psi_j}{\Psi}E_{L,i}\Big\rangle$, 
$\langle \frac{\Psi_i}{\Psi} E_{L,j} \rangle$ and $\langle \frac{\Psi_j}{\Psi} E_{L,i} \rangle$
of zero expectation value (for proof, see e.g. Ref.~\cite{lin}). 
The first term makes the Hessian symmetric, while  the remaining two terms put Hessian into covariance 
form. Hence, the addition of terms of zero expectation value for infinite sample has the effect of cancellation of 
most of the fluctuations in finite sample making the minimization method vastly more efficient.

Another useful rearrangement of the Hessian~(\ref{eq:hess2}) is
\begin{align}\label{eq:hess3}
\bar{E}_{ij}=&2\bigg[ \Big \langle \Big( \frac{\Psi_{ij}}{\Psi}-\frac{\Psi_i \Psi_j}{\Psi^2}\Big) (E_L-\bar{E}) \Big\rangle 
 \nonumber \\
&+2 \Big\langle \Big( \frac{\Psi_i}{\Psi} -\Big\langle \frac{\Psi_i}{\Psi} \Big\rangle \Big) 
                \Big( \frac{\Psi_j}{\Psi} -\Big\langle \frac{\Psi_j}{\Psi} \Big\rangle \Big)
(E_L-\bar{E}) \Big\rangle \bigg ]\nonumber \\
&+\Big[ \Big \langle \frac{\Psi_i}{\Psi}E_{L,j}\Big\rangle
 -\Big \langle \frac{\Psi_i}{\Psi}\Big\rangle \langle  E_{L,j} \Big\rangle \Big] 
 +\Big[ \Big \langle \frac{\Psi_j}{\Psi}E_{L,i}\Big\rangle 
 -\Big \langle \frac{\Psi_j}{\Psi}\Big\rangle \langle  E_{L,i} \Big\rangle \Big].
\end{align}
If our minimization procedure involves only the linear parameters of the form $\exp[-c(i)f] $ (e.g. linear coefficients in Jastrow factor),
the two terms on the first line of Eq.~(\ref{eq:hess3}) cancel out, thus removing the need for expensive calculation of the Hessian of a wave function. 

In analogy to variance minimization, once the gradient and Hessian are computed, 
we search for a new set of parameters with lower value of energy using the Levenberg-Marquardt method discussed bellow (see Sec.~\ref{sec:LM}). 

\subsection{Levenberg-Marquardt Method}\label{sec:LM}
Levenberg~\cite{levenberg} and Marquardt~\cite{marquardt}
suggested to use a damped Newton method where the new step $\Delta {\bf c}$ in the parameter space is given by 
\begin{align}
({\bf H}+\mu{\bf I})\Delta {\bf c}= -{\bf g},
\end{align}
where ${\bf H}$ and ${\bf I}$ are  Hessian and identity matrices, ${\bf g}$ denotes a gradient 
and $\mu$ is the positive damping parameter. This scheme has several 
favorable properties.
\begin{itemize}
\item[(a)] Positive $\mu$ makes ${\bf H}+\mu{\bf I}$ positive definite, which ensures that $\Delta {\bf c}$ is in descent direction.
\item[(b)] For large values of $\mu$ we get 
\begin{align}
\Delta {\bf c}\simeq -\frac{{\bf g}}{\mu},
\end{align}
which is a short step in steepest decent direction. This approach is favorable,
if we are far from minimum and the Newton approximation is not valid.
\item[(c)] If $\mu$ is very small, then the new step $\Delta {\bf c}$
is in the Newton direction, which if we are close to minimum leads to 
(almost) quadratic convergence. 
\end{itemize}

What is left to determine is the optimal value for the damping parameter $\mu$. The 
choice of $\mu$ is related to the values of Hessian ${\bf H}$. If the 
Hessian is not positive definite (which can happen for small samples or if we are far from minimum), 
we add to Hessian diagonal an absolute value of its lowest negative eigenvalue. 
Further, we start with $\mu_0=\tau \max({\bf H}_{ii})$, where the choice of $\tau$ 
is determined by the distance from minimum (from $\tau=10^{-6}$ if we are very close, up to even $\tau=1$).
Then we perform a correlated MC run (VMC or DMC) with three different damping
parameters (e.g. $\mu_1=\mu_0$, $\mu_2=10 \mu_0$ and $\mu_3=100 \mu_0$). A new 
damping $\mu_{new}$ is then chosen at the point where the parabola fitted to 
the three new energy values attains minimum ($\mu_{new}\in\langle\mu_1,\mu_3 \rangle$). 
On the other hand, if the correlated MC run is too expensive, we adjust the $\mu$ based on the value of the {\em gain ratio} 
\begin{align}
\rho=\frac{F({\bf c})-F({\bf c}+\Delta{\bf c})}{\frac{1}{2}\Delta {\bf c}(\Delta {\bf c}-{\bf g})},
\end{align}
where $F({\bf c})$ is the function of parameters ${\bf c}$ to be minimized. 
The value of a new damping $\mu_{new}$ is determined in the following way.
\begin{itemize}
\item If $\rho>0$ 
\begin{align}
\mu_{new}=\mu_0 \max[\frac{1}{3}, 1-(2\rho-1)^3],
\end{align}
\item else 
\begin{align}
\mu_{new}=2^n\mu_0,
\end{align}
\end{itemize}
where $n$ denotes a number of consequent times $\rho\leq 0$. 
For more details on Levenberg-Marquardt method and other least squares methods see for example Ref.~\cite{madsen}.

\subsection{Implementation of Optimization Methods in QWALK}\label{subsec:opt3}
Historically, the first implementation of minimization method in QWALK code was the OPTIMIZE method, 
which uses only the value of a function and a quasi-Newton minimizer. The optimization is bound to 
a fixed MC sample. It has proved effective for variance minimization of linear and nonlinear Jastrow parameters, 
however, the energy minimization required large MC samples. 

This deficiency was partially removed, when we have added the OPTIMIZE2 method, 
which  uses modified Hessians and gradients of variance and energy of Ref.~\cite{cyrus2}
together with the Levenberg-Marquardt minimization algorithm.
While the choice of modified Hessian and gradient decreases significantly their 
fluctuations, the additional information (they provide) consequently improves  
the convergence of the method. The down-side to this approach is an
additional cost of their calculation. For this reason, the analytical gradients and Hessians 
of wave function with respect to selected variational parameters were implemented 
(determinantal weights and orbital coefficients, Pfaffian weights and pair orbital coefficients).  
OPTIMIZE2 performs on a fixed MC sample, so the damping parameter is determined via 
the gain ratio $\rho$ as described in Levenberg-Marquardt method, Sec.~\ref{sec:LM}. 
OPTIMIZE2 has proved to be very effective method and was used for most of the
optimizations in this dissertation and in other references~\cite{lucas_dissertation}. 

As an attempt to remove the dependence of OPTIMIZE2 method on the fixed MC sample, 
I have also implemented the NEWTON\_OPT method. Its principal advantage is that the
energy and variance are obtained from a small MC run performed after each step of optimization. 
In addition, the optimal dumping can be chosen by an additional correlated MC run
as described in Levenberg-Marquardt method, Sec.~\ref{sec:LM}. Increased stability, however, comes 
at the price of additional MC calculations. 

Besides the well-known variance and energy minimization, it is possible
to minimize several other so-called {\em cost functions}.
Their incomplete list can be found in the Table~\ref{table:cost functions}.
\begin{table}[!t]
\caption{List of some useful cost functions for a wave function optimization
with denoted implementation in QWALK~\cite{qwalk}.}
\begin{center}
\begin{tabular}{l r c c c}
\hline
\hline
\multicolumn{1}{l}{Cost Function} & \multicolumn{1}{c}{Minimized quantity} & \multicolumn{3}{c}{QWALK implementation} \\
\multicolumn{1}{l}{} & \multicolumn{1}{c}{} & \multicolumn{1}{c}{{\small OPTIMIZE}} & \multicolumn{1}{c}{{\small OPTIMIZE2}} & \multicolumn{1}{c}{{\small NEWTON\_OPT}}\\
\hline
Variance  & $\langle(E_L-\bar{E})^2\rangle$ & $\surd$ & $\surd$ & $\surd$\\
Energy & $\bar{E}$  & $\surd$ & $\surd$ & $\surd$\\
Mixed & $x\bar{E} +(1-x)\langle(E_L-\bar{E})^2\rangle$  & $\surd$ & $\surd$ & $\surd$\\
Absolute value & $ \langle |E_L-\bar{E}| \rangle$  & $\surd$ & & \\
Lorentz & $\langle \ln(1+(E_L-\bar{E})^2/2) \rangle$ & $\surd$ & & \\
Ratio  & $\frac{\langle(E_L-\bar{E})^2\rangle}{\bar{E}}$ & & &  \\
Overlap & $\frac{\int \Psi_T\Psi(\tau\to\infty)}{\int \Psi_T^2}$ & & & \\
\hline
\hline
\end{tabular}
\end{center}
\label{table:cost functions}
\end{table}
As was pointed recently by Umrigar and Filippi~\cite{cyrus2}, 
the wave functions optimized by minimization of a mixture of energy and variance 
have almost as low energy as energy optimized wave functions 
but a lower variance. 

\begin{figure}[!h]
\begin{minipage}{\columnwidth}
\centering
\includegraphics[width=\columnwidth]{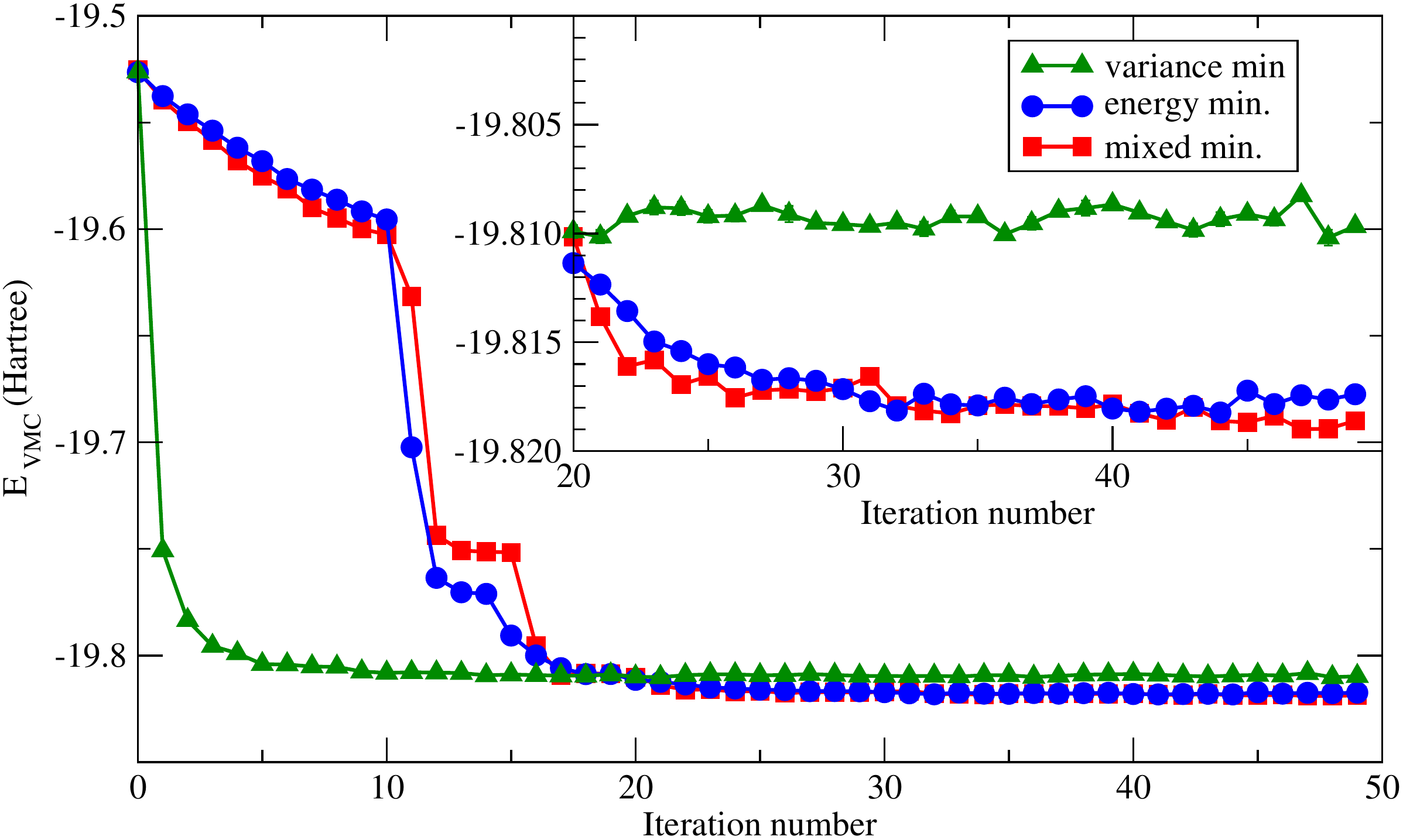}

\includegraphics[width=\columnwidth]{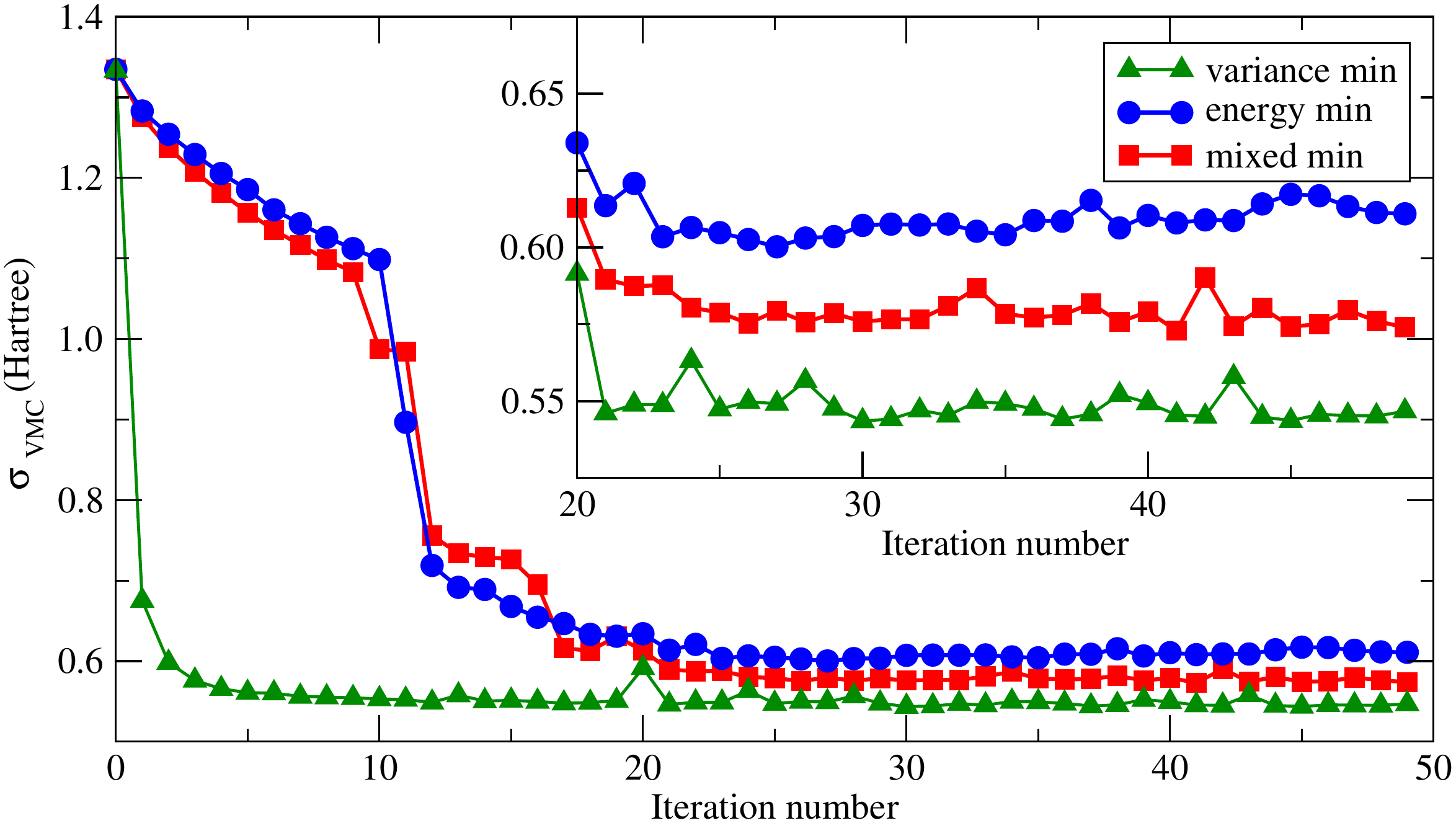}
\end{minipage}
\caption{
Energy $E$ and dispersion of local energy $\sigma$ of N$_2$ molecule versus iteration number
of minimization. Notation: minimization of variance (green triangles), energy (blue circles) and 95\% mixture of energy and 5\% of variance (red squares).
Upper figure: Energy versus iteration number. The error-bars are of the size of symbol or smaller. 
Inset: the later iterations on expanded scale. 
The mixture minimized energy is almost as good as energy minimized one while variance minimized energy is higher by almost 10 mH. 
Lower figure: Same as the upper figure but for the dispersion of local energy $\sigma$ rather than energy.
Inset shows that the mixture minimized $\sigma$ falls in between $\sigma$ from energy minimization and variance minimization. 
}
\label{fig:min_n2}
\end{figure}

We demonstrate this for the ground state wave function of N$_2$ molecule (see Fig~\ref{fig:min_n2}).
The employed NEWTON\_OPT method minimized 23 Jastrow parameters 
(the curvatures of basis functions and all linear coefficients)
of single determinant Slater-Jastrow wave function on 56000 walkers.
The mixture minimized energy is almost as good as energy minimized energy, 
while the mixture minimized dispersion $\sigma$ falls in between $\sigma$ from energy minimization and variance minimization.
Also note rather slower converge of energy and mixture minimization when compared to variance minimization. 
This behavior may have two reasons. One is the more complex structure of energy function. Alternatively, 
we just happen to be much further from minimum for energy function than for variance.

The mixture minimized wave functions are therefore the most efficient for DMC calculations.
In the same time, small addition of variance (typically 5\%) to energy in minimization
greatly decreases the fluctuations of Hessian and enables us to use smaller MC samples.


\section{Summary}
In this chapter, we tried to plot an overview of two principal Quantum Monte Carlo methods,
the variational and diffusion Monte Carlo. Further, we mentioned some essential sampling 
MC techniques such as Metropolis, importance and correlated sampling. 
A more technical details of QMC, like the use of non-local pseudopotentials are discussed. 
The only two uncontrolled approximations of DMC, 
the fixed node and localization approximations, are introduced to circumvent 
the fermion sign problem. We analyzed some of the properties of trial wave functions
and presented two very general and accurate wave functions, the Slater-Jastrow
and the Pfaffian-Jastrow trial wave functions. Both can be teamed-up with backflow
transformation which can further lead to improved nodes. Finally, 
we talked about the optimization of variational trial wave functions and 
acquainted the reader with its three implementations in QWALK code.

\chapter{Nodal Properties of Fermionic Wave Functions}\label{ch:nodes}
\begin{center}
\begin{minipage}{\columnwidth}
Sections of this chapter also appeared in:
\begin{center}
{\bf Approximate and exact nodes of fermionic wave functions:\\
  Coordinate transformations and topologies},\\
M.~Bajdich, L.~Mitas, G.~Drobn\'y, and L.~K. Wagner,\\
Physical Review B \textbf{72}(7), 075131 (pages~8) (2005),\\
(Received 14 September 2004; revised 23 May 2005; published 23 August 2005)\\
{\it \copyright2005 The American Physical Society}
\end{center}

\begin{center}
{\bf Investigation of nodes of fermionic wave functions},\\
L.~Mitas, G.~Drobn\'y, M.~Bajdich, and L.~K. Wagner,\\
In \emph{Condensed Matter Theories}, vol.~20 (Nova Science Publishers, 2006).\\
{\it \copyright2006 Nova Science Publishers}
\end{center}
\end{minipage}
\end{center}
\newpage
\section{Introduction}\label{nodessec:level0}
Nodes of fermionic wave functions and their related 
objects, density matrices, are of great interest to physicist 
for several reasons. From QMC point of view, the knowledge 
of exact nodes of fermionic wave functions would enable us  
to obtain exact ground state energies by means of fixed-node DMC.
Analogously, the knowledge of exact nodes of temperature density 
matrices would lead to the solution of the {\em thermal fermion problem}\footnote{The thermal fermion problem 
is the construction of a polynomial-time algorithm of the exact thermodynamics of many-fermion system at positive temperature~\cite{davidnode}.}
in path integral Monte Carlo (PIMC).
The nodal properties of wave functions 
dramatically change with increased correlations among electrons and 
are relevant to transport and many-body phases such as {\em superconductivity}. 

The general properties of fermion nodes were first analyzed in an extensive study by Ceperley~\cite{davidnode},
which included a proof of the tiling property and generalizations of the fermion nodes to density matrices. 
In addition, for some free particle systems, it was numerically shown~\cite{davidnode} that there are only {\em two  nodal cells}.
The fermion nodes for degenerate and excited states were further studied by Foulkes and co-workers~\cite{foulkes}. 
Recently, Mitas used the property of connectivity from Ref.~\cite{davidnode} to show that a number of spin-polarized noninteracting and mean-field systems 
(homogeneous electron gas, atomic states, fermions in the box, in the harmonic well and on the sphere)
has ground state wave functions (given as Slater determinants) with {\em minimal number of nodal cells}~\cite{lubos_nodeprl,lubos_nodeprb}. 
Further, he demonstrated that for spin-unpolarized systems an arbitrarily weak interaction introduced by Bardeen-Cooper-Schrieffer (BCS) wave function
reduces the four non-interacting nodal cells to just two. 
Finally, he has also shown that the minimal number of nodal cells property extends to the temperature density matrices.

The fermion nodes of small systems, mostly atoms, were investigated in several
previously published  papers~\cite{node1s2s_1,node1s2s_2,Breit30,jbanderson75,jbanderson76,andersp,lesternode}.
Interesting work by Bressanini, Reynolds and Ceperley revealed differences in the nodal surface 
topology between Hartree-Fock and correlated wave functions for the Be atom explaining the large impact of the $2s,2p$ 
near-degeneracy on the fixed-node DMC energy~\cite{dariobe}.
More recently, improvement in fixed-node DMC energies of small systems using CI expansions~\cite{Bressanininew,umrigarC2}, 
and pairing wave functions~\cite{sorellabcs1,sorellabcs2,pfaffianprl,pfaffianprb} were also reported. 

This chapter is organized as follows. In Sec.~\ref{nodessec:level1},
we summarize the general properties of fermion nodes. In Sec.~\ref{nodessec:level2},
we discover new {\em exact} fermion nodes for two and three-electron spin polarized systems.
In Sec.~\ref{nodessec:level3}, we categorize the nodal surfaces for the several half-filled 
subshells relevant for atomic and molecular states. 
In Sec.~\ref{nodessec:level5},
we show how opposite spin correlations eliminate a redundant nodal structure of HF wave functions 
for two specific cases of spin-unpolarized states.
Finally, in the last section we present our conclusions 
and suggestions for future work.

\section{General Properties}\label{nodessec:level1}
Let us assume a system of spin-polarized electrons  
described by a real wave function $\Psi({\bf R})$.
Then the $\Psi({\bf R})$ is antisymmetric with respect to any 
particle exchange as described in Sec.~\ref{sec:twf}.
Consequently, there exists a subset of electron configurations $\{ {\bf R}_{node}\}$, 
called a fermion node\footnote{The first question about the node structure of fermion wave functions was raised by J. B. Anderson in 1976~\cite{jbanderson76}.}, 
for which the wave function is zero, i.e., 
\begin{align}\label{eq:node1}
\Psi({\bf R}_{node})=0.
\end{align}
Naturally, we eliminate from the definition the regions, where the wave function vanishes because of other
reasons (e.g., external potential) than antisymmetry. In general, the fermion node is 
a $(ND-1)$-dimensional manifold (hypersurface) defined by implicit Eq.~(\ref{eq:node1}) assuming that we have
$N$ fermions in a $D-$dimensional space. When the positions of any two electrons are equal (i.e., ${\bf r}_i={\bf r}_j$), 
the antisymmetry ensures that the $\Psi({\bf R})=0$. However, that does not fully specify the nodes, 
but only defines the $(ND-D)$-dimensional subspace of {\em coincidence planes}, where wave function vanishes. 
Note that when we talk about the nodes, 
we always mean the nodes of a many-body wave functions and not the nodes of one-particle orbitals. 

Let us now introduce the basic properties of fermion nodes as they were 
studied by Ceperley\cite{davidnode} some time ago.
\begin{itemize}
\item[(a)] {\em Tiling property for the non-degenerate ground state}---Let us
define a nodal cell $\Omega({\bf R}_t)$ as a subset of configurations, which can be reached
from the point ${\bf R}_t$ by a continuous path without crossing the node. 
The tiling property 
says that by applying all possible particle permutations to an arbitrary 
nodal cell of a ground state wave function one covers the complete configuration
space $\mathfrak{R}$ (i.e., $\sum_P\Omega(P{\bf R}_t)+\Omega({\bf R}_t)=\mathfrak{R}$). 
Note that this does not specify how many nodal cells are there.
\item[(b)] {\em Nodal crossings}---If two nodal surfaces cross each other, they are orthogonal
at the crossing. If $n$ nodal surfaces cross each other, the crossing 
angles are all equal to $\pi/n$. 
\item[(c)] {\em Symmetry of the node}---Symmetry of the state is also
symmetry of the node.
\item[(d)] {\em Connectivity} $\iff$ {\em two maximal nodal cells}---It 
is possible to show that there are only two nodal cells 
using the argument based on connectivity of particles by triple exchanges. 
The three particles $i$, $j$, $k$ are called {\em connected} if there exist a cyclic 
exchange path $i \to j$, $j \to k$, $k \to i$, which does not cross the node.  
In addition, if all particles of some point ${\bf R}_t$ are 
connected together by triple exchanges, the  $\Psi({\bf R})$ has only 
two nodal cells. In other words, the whole configuration space is covered 
by only one positive and one negative nodal cell (i.e., the nodal cells are {\em maximal}). 
This can be better understood if we realize the following.
First, any triple exchange is just a two pair positive permutation.
The existence of point ${\bf R}_t$ with all particles connected is however 
equivalent to have all positive (not flipping the sign) permutations $P^+{\bf R}_t$ belonging 
to the same nodal cell. Second, the tilling property implies, that once
all particles are connected for ${\bf R}_t$ it is true for entire 
cell $\Omega({\bf R}_t)$. Therefore, there will be only 
one maximal cell per each sign. More details on this 
property can be found in Ref.~\cite{davidnode}.
\end{itemize}

\section{Exact Nodal Surfaces}\label{nodessec:level2} 
We assume the usual electron-ion Hamiltonian and we first 
investigate a few-electron ions focusing on fermion nodes for
subshells of one-particle states with $s,p,d,f ... $ symmetries using 
variable transformations, symmetry operations and explicit
expressions for the nodes.

\subsection{Three-Electron Quartet $^4S (p^3)$ State}\label{nodessec:level21}
\begin{figure}[!t]
\begin{center}
\includegraphics[width=\columnwidth]{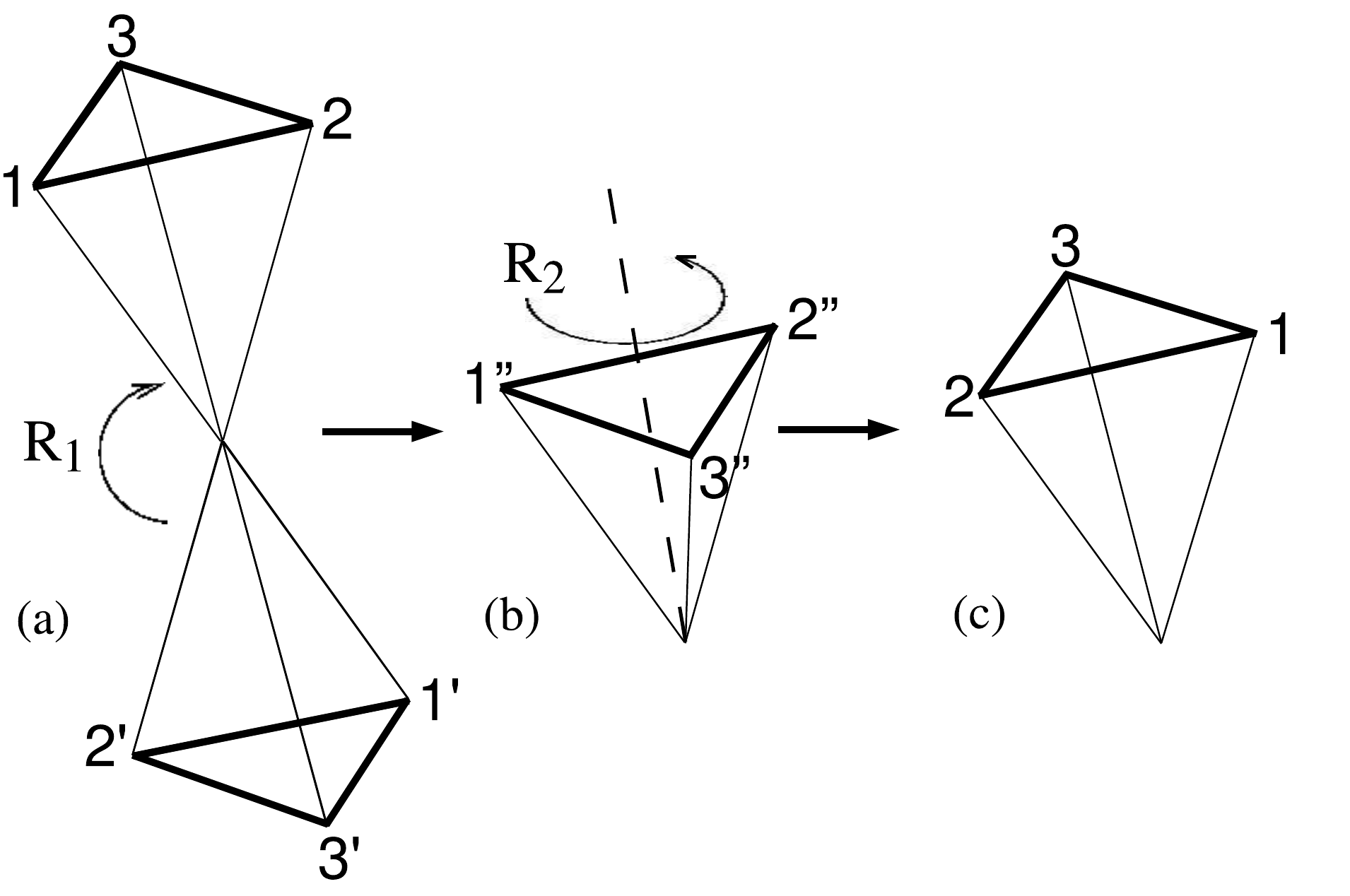}
\end{center}
\caption{Inversion and two subsequent rotations of three particles:
(a) Original and inverted (primed) positions;
(b) Positions after the rotation $R_1$  
in the plane given by the particles 1, 2 and the origin;
(c) Positions after the second rotation $R_2$
around the ${\bf r}_{1}+{\bf r}_2$ axis. Note that
the original positions of
the  particles 1 and 2 are exchanged.}\label{fig:0}
\end{figure}

Let us first analyze a special case with $r_1=r_2$ and 
$r_{23}=r_{31}$. It is then easy to see that the 
inversion around the origin with subsequent rotations is equivalent
to the exchange of two particles, say, 1 and 2 (Fig.~\ref{fig:0}). 
Therefore, for this particular configuration of particles,
the combination of parity and rotations 
is closely related to the exchange symmetry.
The illustration also shows that the six distances
do not specify the relative positions of the three electrons
unambiguously. For a given set of the distances there 
are two distinct positions, say, of the electron 
3, relative to the fixed positions
of electrons 1 and 2  (see Fig.~\ref{fig:0}) and compare positions
3 and 3'' of the third electron.

\begin{figure}[!t]
\centering
\begin{center}
\includegraphics[width=\columnwidth]{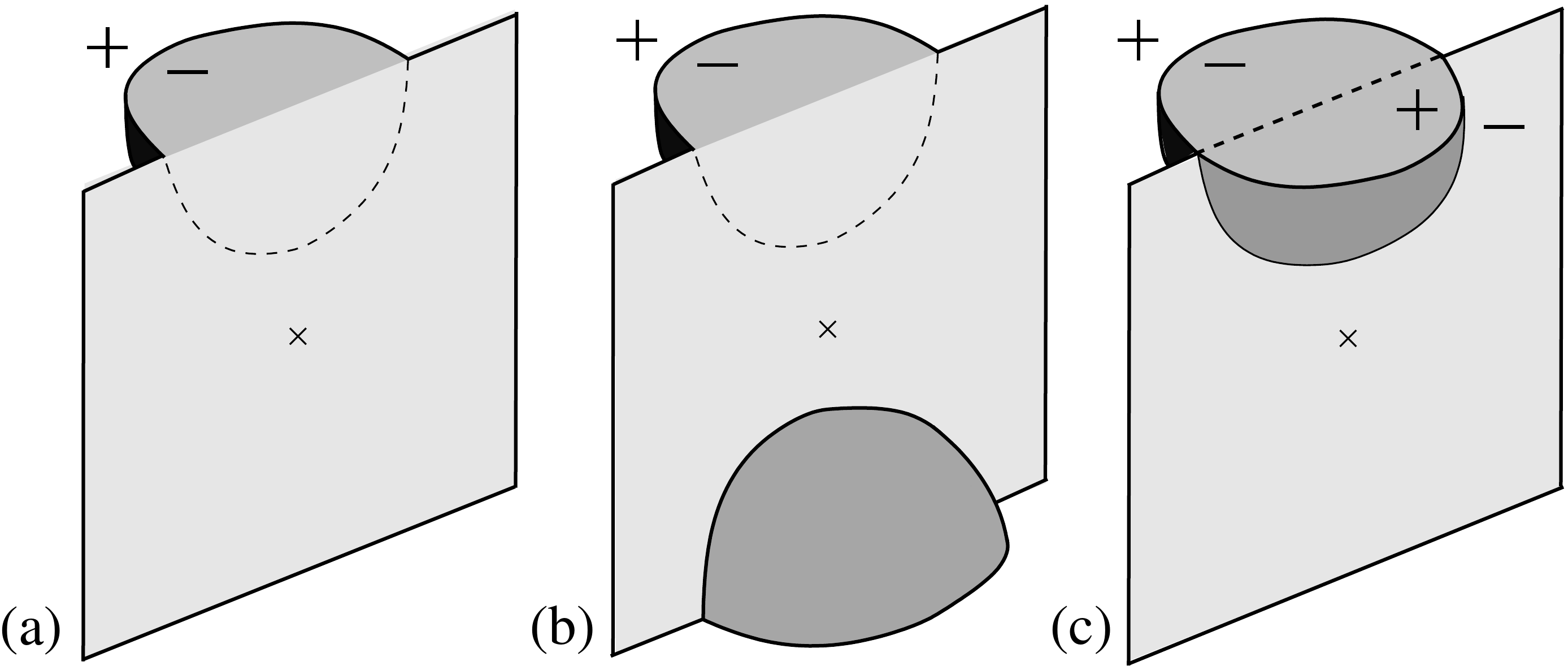}
\end{center}
\caption{ (a) An illustration of an artificial
  distortion of the planar ground state node
for the quartet state (see text);
(b) The original and parity transformed distorted node; 
(c) Finally, a subsequent 
rotation of the inverted distortion necessarily leads to  
a new nodal pocket which is artificial for the ground state. 
In fact, nodes with similar topologies are present
in excited states (see bellow).}\label{fig:blob}
\end{figure}

\begin{figure}[!t]
\centering
\begin{center}
\includegraphics[width=\columnwidth]{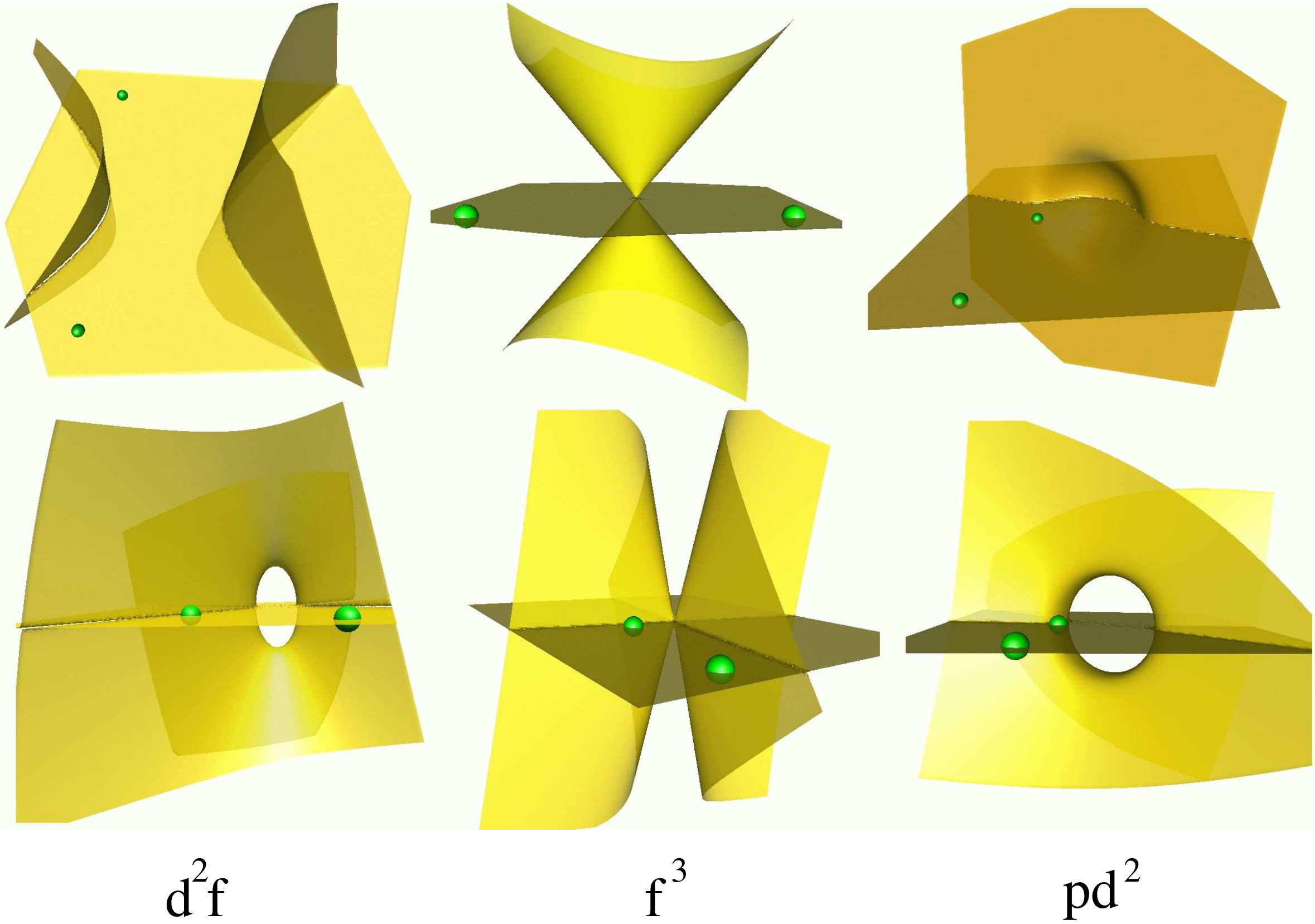}
\end{center}
\caption{The 3D projected nodes of a few selected excitations 
for the symmetry adopted CI expansion of the $^4S(p^3)$ ground state. 
The exact planar node of the quartet ground state is also possessed 
by all the excitations.
The small spheres show the fixed positions of two electrons
while the third one is scanning the nodal surface. 
Labels indicate the orbitals involved.}
\label{fig:p3_excit}
\end{figure}

In order to analyze the wave function in an unambiguous manner 
it is convenient to define new coordinates.
Let us denote
${\bf r}_{12}^+ ={\bf r}_1+{\bf r}_2, r_{12}^+=|{\bf r}_{12}^+| $,
together with the customary 
${\bf r}_{12} ={\bf r}_1-{\bf r}_2, r_{12}=|{\bf r}_{12}| $ . 
We can now introduce the following 
map of the Cartesian coordinates 
\begin{equation}
({\bf r}_1, {\bf r}_2, {\bf r}_3) \to
(r_{12}^+, r_{12}, r_3, \cos\alpha,\cos\beta,\gamma,\Omega)
\end{equation}
with definitions: $\cos\alpha=
{\bf r}_3\cdot({\bf r}_1\times {\bf r}_2)/(r_3|{\bf r}_1\times {\bf r}_2|)$,
$\cos\beta={\bf r}_{12}^+\cdot {\bf r}_{12}/(r_{12}^+ r_{12})$ and
$\gamma$ being an azimuthal angle of ${\bf r}_3$ in the
relative coordinate system with unit vectors
${\bf e}_x= {\bf r}_{12}^+/r_{12}^+$, 
${\bf e}_z={\bf r}_1\times {\bf r}_2/|{\bf r}_1\times {\bf r}_2|$, 
${\bf e}_y={\bf e}_z\times {\bf e}_x$.
For completeness, 
$\Omega$ denotes three Euler angles, which fix the orientation of the 
three-particle system in the original coordinates  
(e.g., two spherical angles of ${\bf r}_1\times {\bf r}_2$
and an azimuthal angle of ${\bf r}_{12}^+$). Since the angles $\Omega$
are irrelevant in $S$ symmetry, the
 first six variables fully specify the relative positions
of the three particles and
the wave function dependence simplifies to
$\Psi(r_{12}^+, r_{12}, r_3, \cos\alpha,\cos\beta,\gamma)$.
Consider now two symmetry operations which change the sign of the wave function and
keep the distances unchanged: parity $P_I$
and exchange $P_{12}$ between particles 1 and 2. 
The exchange flips the sign of all three $\cos\alpha,\cos\beta,\gamma$ 
while the parity changes only the sign of $\cos\alpha$.
The action of $P_I P_{12}$ on $\Psi$ leads to
\begin{equation}
\Psi(...,\cos\alpha,-\cos\beta,-\gamma)=
\Psi(...,\cos\alpha,\cos\beta,\gamma)
\end{equation}
showing that the wave function is even in the simultaneous sign flip
$(\cos\beta,\gamma) \to (-\cos\beta,-\gamma)$.
Applying the exchange operator $P_{12}$ to the wave function 
and taking advantage of the previous property gives us
\begin{equation}
\Psi(...,-\cos\alpha,\cos\beta,\gamma)=
-\Psi(...,\cos\alpha,\cos\beta,\gamma)
\end{equation}
suggesting that there is a node determined by the condition $\cos\alpha=0$. 
It is also clear that the same arguments can be repeated
with
exchanged particle labels $2\leftrightarrow 3$ and $3\leftrightarrow 1$
and we end up with {\em the same nodal condition}, 
${\bf r}_3\cdot({\bf r}_1\times {\bf r}_2)=0$. This shows that
 the node is encountered
when all three electrons lie on a plane passing
through the origin. 
Now we need to prove that this is the only node 
since there might possibly be
other nodal surfaces not revealed by the parametrization above.
The node given above clearly fulfills the tiling 
property and all symmetries of the state. 
Furthermore, the state is the lowest quartet of 
$S$ symmetry and odd parity (lower quartets such as $1s2s3s$,$1s2s2p$,
and $1s2p^2$ have either different parity or symmetry) and for the
ground state
we expect that the number of nodal cells will be minimal.
This is indeed true 
since the node above specifies only two nodal cells
(one positive, one negative): an electron is either on one or 
the other side
of the nodal plane passing through the remaining two electrons.
Furthermore, any distortion of the node from the plane necessarily leads 
to additional nodal cells (see Fig.~\ref{fig:blob}), which 
can only increase energy by imposing higher curvature 
(kinetic energy) on the wave function.
This is basically the Feynman's argument from the proof 
demonstrating that the energy of fermionic ground state is 
always above the energy of the bosonic 
ground state (and also essentially the same 
argument as used for the proof of the tiling property~\cite{davidnode}).
In fact, all higher excited states of this symmetry 
(HF wave function of $2p$ orbitals, e.g., excited state of B$^{+2}$ ion) 
have additional nodes, as expected (see Fig.~\ref{fig:p3_excit}).
Given all the arguments above we conclude the proof that the plane
is the {\em exact node}.
Note that it is identical to the node of Hartree-Fock 
wave function of $2p$ orbitals given by  
$\Psi_{HF}={\rm det} [\rho(r)x, 
\rho(r)y, \rho(r)z]$ where $\rho(r)$ is a radial function.

The coordinate transformation above is not the only one that can 
be used to analyze this state. 
The high symmetry of the problem
enables us to find an alternative coordinate map 
with definitions of $\cos\beta$ modified to 
$\cos\beta'=[({\bf r}_1\times {\bf r}_2)\times {\bf r}_{12}^+]
\cdot  {\bf r}_{12}/
[|({\bf r}_1\times {\bf r}_2)\times {\bf r}_{12}^+|
|{\bf r}_{12}^+|]$
 and $\gamma$ to
$\gamma'$ by  redefinition of ${\bf e}_z$ to
${\bf e'}_z=[[({\bf r}_1\times {\bf r}_2) \times {\bf r}_{12}]\times {\bf r}_{12}^+]
/|[({\bf r}_1\times {\bf r}_2)\times {\bf r}_{12}]\times {\bf r}_{12}^+ |$
and ${\bf e'}_y={\bf e'}_z\times {\bf e}_x$.
In the redefined coordinates the search for the node
simplifies to an action of $P_{12}$
on $\Psi(r_{12}^+, r_{12}, r_3, \cos\alpha,\cos\beta',\gamma')$ 
\begin{equation}
\Psi(...,-\cos\alpha,\cos\beta',\gamma')=
-\Psi(...,\cos\alpha,\cos\beta',\gamma')
\end{equation}
since the distances and $\cos\beta', \gamma'$ are invariant to $P_{12}$.
Obviously, this leads to the same 
nodal condition as derived above.

It is quite interesting to compare these two coordinate maps
with $\beta,\gamma$ and $\beta', \gamma'$.
Although parity and exchange are independent operators, the analysis above shows 
that in an appropriate coordinate system they imply the same nodal surface.
Both these operators cause an identical sign change
of the wave function indicating thus a special symmetry of the
$^4S(p^3)$ ground state node, which is higher than would be expected
solely from antisymmetry. Similar observation was made in a study 
of fermion node in another case of two electron atomic state
~\cite{andersp,darionew}.

\subsection{Two-Electron Triplet $ ^3P (p^2)$  and  
$^3\Sigma_g(\pi^2)$ States} \label{nodessec:level22}
Apparently, the exact node of this case was derived in a different 
context by Breit in 1930~\cite{Breit30,Bressanininew,darionew}. 
Here we offer an independent
proof which enables us to apply  the analysis 
to some molecular states with the same symmetries.
The exact node for the $ ^3P (p^2)$ state can be found 
in a similar way as in the case of quartet above.
The state has even parity, 
cylindric symmetry, say,  around $z$-axis, and is odd
under rotation by $\pi$ around $x,y$ axes, 
$R(\pi x)$, $R(\pi y)$.
The mapping of  
Cartesian coordinates
which enables to analyze the wave function 
 symmetries is given by  
\begin{equation}
({\bf r}_1, {\bf r}_2) \to
(r_{12}^+,r_{12},\cos\omega,\cos\beta,\varphi,\varphi')
\end{equation}
where 
$\cos\omega=
{\bf z}_0\cdot({\bf r}_1\times {\bf r}_2)/
|{\bf r}_1\times {\bf r}_2|$
with
${\bf z}_0$ being the unit vector in the $z$-direction
and
$\varphi'$ being the azimuthal angle of ${\bf r}_{1}\times  {\bf r}_{2}$;
 $\varphi'$ can be omitted due to the cylindric symmetry.
Further, $\varphi$  is the azimuthal angle of ${\bf r}^+_{12}$ 
in the relative coordinate system with the $x$-axis unit vector
given by a projection of ${\bf z}_{0}$ into the plane
defined by ${\bf r}_1, {\bf r}_2$,
i.e., ${\bf e}_x= {\bf z}_{0p}/|{\bf z}_{0p}|$, 
 ${\bf e}_z=({\bf r}_1\times {\bf r}_2)/|{\bf r}_1\times {\bf r}_2|$ and 
 ${\bf e}_y={\bf e}_z\times {\bf e}_x$.
Action of $P_IP_{12}R(\pi x)$ reveals that the wave function is
invariant in the simultaneous change  $(\cos\beta,\varphi)$ $\to$ 
$(- \cos\beta, -\varphi)$. This property and action of  
$P_{12}$ to the wave function together lead to 
\begin{equation}
\Psi(...,-\cos\omega,...)=
-\Psi(...,\cos\omega,...)
\end{equation}
with the rest of the variables unchanged.
The node is therefore given by $\cos\omega=0$
and is encountered when an electron hits the plane which contains the
$z$-axis and the other electron.
As in the previous case
 the nodal plane fulfills the tilling property and manifestly divides
the space into two nodal cells  so that we can conclude that this
node is exact.
The exact node again agrees with the node of Hartree-Fock wave function 
$\Psi={\rm det}[\rho(r)x,\rho(r)y]$. 

\begin{table}
\caption{Total energies (in Hartrees) of N$^{+}$, N$^{+2}$ and N$^{+3}$ ions with core electrons
eliminated by pseudopotentials~\cite{lester}. The energies are calculated
by variational
(VMC) and fixed-node diffusion (DMC) quantum Monte Carlo and configuration
interaction (CI) methods. The HF energies are given as a reference for estimation of
the correlation energies.}
\begin{center}
\begin{tabular}{l c c c c}
\hline
\hline
State  & HF  & CI  & VMC  & DMC \\
\hline
 $^3P(p^2)$  & -5.58528 &  -5.59491  &    -5.59491(2)  &  -5.59496(3)     \\
 $^4S(p^3)$  & -7.24716 &  -7.27566  &    -7.27577(1)  &  -7.27583(2)     \\
 $^5S(sp^3)$  & -8.98570 & -9.02027  &    -9.01819(4)  &  -9.01962(5)     \\
\hline
\hline
\end{tabular}
\end{center}
\label{tab_xx}
\end{table}

The fixed-node QMC energies for the 
$^4S (p^3)$  and $^3P(p^2)$ cases derived above were calculated for 
a nitrogen cation
 with valence electrons in these states.
The core electrons were eliminated
by pseudopotential~\cite{lester}. The trial wave function was
of the commonly used form with single HF determinant times a Jastrow correlation 
factor~\cite{qmcrev}.
Note that the pseudopotential nonlocal $s-$channel does not couple
to either odd parity $S$ state or even parity $P(p^2)$ state so that
that the nonlocal contribution to the energy vanishes exactly.

In order
to compare the fixed-node QMC calculations with an independent 
method
we have carried out also 
configuration interaction calculations 
with ccpV6Z basis~\cite{dunning} (with up to three
$g$ basis functions), which
generates more than 100 virtual orbitals in total. In the CI method the wave function
is expanded in excited determinants and we have included
all single, double  and triple excitations.
Since the doubles and triples include two- and three-particle correlations
exactly, the accuracy of the CI results is limited only by the size
of the basis set. By comparison with other two- and three-electron
 CI calculations we estimate that the order of magnitude of
the basis set CI bias is  $\approx$ 
0.01 mH for two electrons and  $\approx$ 0.1 mH.
and for three electrons (despite the large number of virtuals 
the CI expansion converges relatively slowly~\cite{kutz} in the maximum
angular momentum of the basis functions, in our case $l_{max}=4$).
The pseudopotentials we used were identical in both QMC and CI 
calculations.

The first two rows of Tab.~\ref{tab_xx} show the total energies of
variational and fixed-node DMC calculations 
with the trial wave functions with HF nodes together with
results from the CI calculations.
For $^3P(p^2)$ the energies agree within a few hundredths of mH
with the CI energy being slightly higher but within two standard
deviations from the fixed-node QMC result.  For  
$^4S (p^3)$ the CI energy is clearly above the fixed-node DMC by about 
0.17 mH as expected due to the limited basis set size. In order to 
illustrate the effect of the fixed-node approximation in the case
when the HF node is {\em not} exact we have also included calculations
for four electron state $^5S(sp^3)$ (for further discussion
of this Hartree-Fock node see Sec~\ref{nodessec:level33} below). For this case, 
we estimate
that the CI energy 
is above the exact one by $\approx 0.3$ mH so that the fixed-node energy 
is significantly {\em higher} than both CI and exact energies.
Using these results we estimate that the  
fixed-node error is  $\approx $ 1 mH, i.e., close to 3\% of the correlation energy. 

Since in the $p^2$ case we have assumed 
cylindric symmetry, the derived node equation is applicable to any
such potential, e.g., equidistant homo-nuclear dimer, trimer, etc,
with one-particle orbitals 
$\pi_x,\pi_y$ which couple into the triplet state
$^3\Sigma_g(\pi_x\pi_y)$.

Note that the parametrization given above 
automatically provides also one of the very few known
exact nodes in atoms so far~\cite{node1s2s_1,node1s2s_2}, i.e.,
the lowest triplet state of He $^3 S (1s2s)$.
The spherical symmetry makes 
 angles $\omega$ and $\varphi$  irrelevant and simplifies
the two-electron wave function dependence to distances $r_1,r_2, r_{12}$ 
or, alternatively, to
 $r_{12},r_{12}^+,\cos\beta$.
Applying $P_{12}$ to wave function 
$\Psi(r_{12},r_{12}^+,\cos\beta)$
leads to
\begin{equation}
-\Psi(r_{12},r_{12}^+,\cos\beta)=
\Psi(r_{12},r_{12}^+,-\cos\beta) 
\end{equation}
so that the node is given by the condition $\cos\beta=0$, i.e., $r_1-r_2=0$.

In addition, the presented analysis sheds some light on the He $^3P(1s2p)$
state node which
was investigated before~\cite{andersp} as having higher symmetry
than implied by the wave function symmetries.
 The symmetry operations reveal that the wave function
depends on $|\cos\omega|$ and that the node is related to the
simultaneous flips such as  $(\cos\beta,\varphi)$ $\to$ 
$(- \cos\beta, -\varphi)$ or angle shifts
$\varphi \to \varphi+\pi$. Since, however, two of the variables are involved,
the node
has a more complicated shape as the previous study illustrates~\cite{andersp}.
In order to test the accuracy of the HF node we have carried out a
fixed-node diffusion Monte Carlo
calculation of the  He $^3P(1s2p)$ state.  The resulting total energy of
 -2.13320(4) H is in an excellent agreement with the estimated exact
value of -2.13316 H~\cite{Schwartz},
which shows that the HF node is very close to
the exact one~\cite{james}.

\section{Approximate Hartree-Fock Nodes}\label{nodessec:level3} 
It is quite instructive to investigate the nodes of half-filled 
subshells of one-particle states with higher angular momentum.

\begin{figure}[!t]
\begin{center}
\includegraphics[width=\columnwidth]{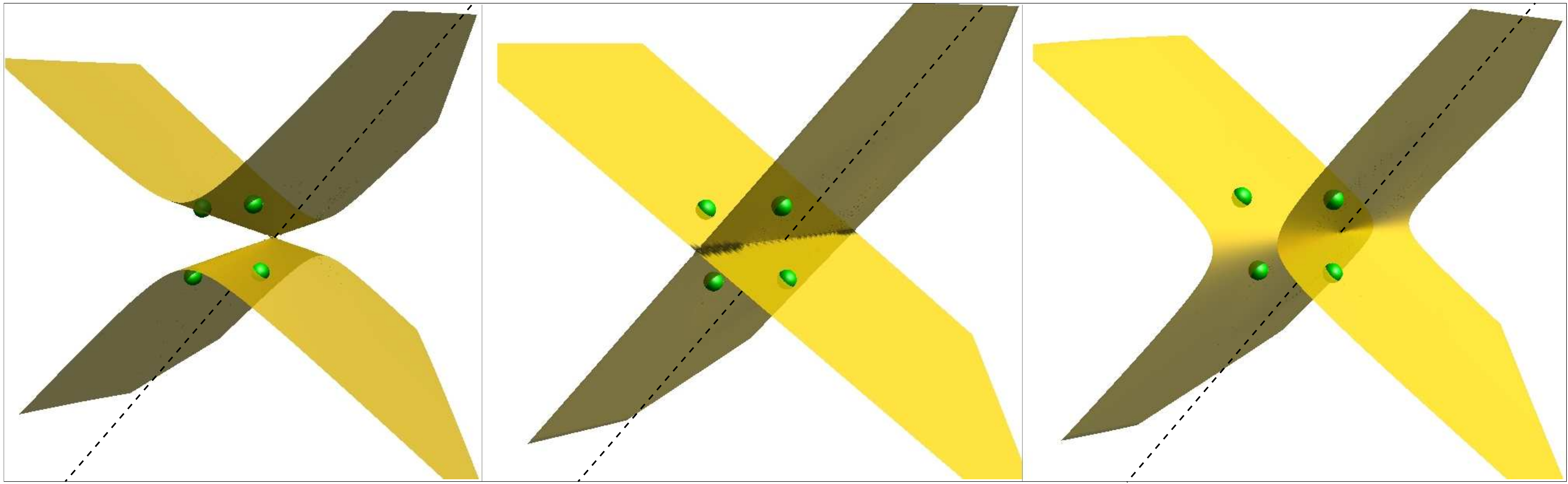}
\end{center}
\caption{ 
The 3D projected  Hartree-Fock node of $^6S(d^5)$ state, which is an elliptic cone
(left and right pictures). 
The middle picture illustrates a
case when two pairs of two electrons lie on orthogonal
planes which pass through the origin. This two-plane node is 
of lower dimension because of the additional condition 
on positions of the electrons. It appears
as a crossover between the cones with different
orientation (left 
and right pictures). The small spheres show the positions
of the four electrons while the line denotes the $z-$axis.}
\label{fig:d5}
\end{figure}

\subsection{Hartree-Fock Node of $^6S(d^5)$ State}\label{nodessec:level31}
The HF determinant wave function for $^6S(d^5)$
is given by
\begin{equation}
\Psi_{HF}= \prod_{i=1}^5 \rho(r_i){\rm det}[2z^2-x^2-y^2,x^2-y^2,xz,yz,xy],
\end{equation}
where $\rho(r_i)$ is the radial part of the $d$-orbital and 
we assume that all the orbitals
are from the same $l=2$ subshell, e.g., $3d$ subshell. Since all radial functions
are the same they factor out from the determinant and for the 
purpose of finding the node they can be omitted. The
$S$ symmetry allows to rotate the system so that, say, electron 1 is
on the $z$-axis, and then the corresponding column in the Slater matrix
becomes $(2z_1^2,0,0,0,0)$. Assuming that $z_1\neq 0$ we 
can then write the nodal condition as
\begin{equation}
{\rm det}[x^2-y^2,xz,yz,xy] =0.
\end{equation}
Using one of the electrons as a 
{\em probe} (i.e., looking at the node from the
perspective of one of the electrons) we can find the projection 
of the node to 3D space.
By denoting the probe electron coordinates
simply as
$(x,y,z)$ and by expanding the determinant 
we get
\begin{equation}\label{eq:5dHFstart}
(x^2-y^2)m_1 +xzm_2+yzm_3+xym_4=0
\end{equation}
where 
$m_i$ are the corresponding cofactors. We divide out
the first cofactor assuming that it is nonzero 
(not a crucial assumption as clarified below).
We get 
\begin{equation}\label{eq:cone5d}
(x^2-y^2)+axz+byz+cxy=0
\end{equation}
where  $\; a=m_2/m_1$,  $b=m_3/m_1$, $c=m_4/m_1$.
By completing the square
this can be further rearranged to 
\begin{equation}\label{eq:cone5d2}
(x-k_1y)(x-k_2y) +z(ax+by) =0
\end{equation}
with
$k_{1,2}=(-c\pm\sqrt{c^2+4})/2$. 
Let us define
rotated and rescaled coordinates 
\begin{align}
u^*&=-(ak_2-b)(x-k_1y)/(k_1-k_2)\\
v^*&=(ak_1-b)(x-k_2y)/(k_1-k_2)\\
w^*&=z[(ak_1-b)(ak_2-b)]/(k_1-k_2)^2
\end{align}
so we can write the Eq. (\ref{eq:cone5d}) as 
\begin{equation}\label{eq:5dHFgeneral}
u^*v^* +w^*u^* +w^*v^*=0.
\end{equation}
Note that this equation has a form which is
 identical to Eq. (\ref{eq:5dHFstart}) with 
$m_1=0$ so this representation is correct for general $m_1$.
After some effort one finds that Eq. (\ref{eq:5dHFgeneral}) is
a cone equation (i.e., $d_{z^2}$ orbital) as can be easily
verified by using the following identity
\begin{equation}
(2u^2-v^2-w^2)/8=u^*v^* +w^*u^* +w^*v^*,
\end{equation}
where  $ u=u^*+v^*+2w^*$, $ v=(-u^*+v^*+2w^*)$,
$w=(u^*-v^*+2w^*)$. 
 The 3D projected node is therefore rotated and rescaled 
 (elliptic) cone. 

At this point it is useful to
clarify how
the derived node projection cone
is related to the complete 14-dimensional node.
Remarkably, the 3D projection
 enables us to understand some of the properties
of the 14-dimensional manifold. 
First, the cone orientation 
and elliptic radii (i.e., rescaling of the 
two axes with respect
to the third one) are determined
by the position of the four electrons in 3D space: 
with the exception of special lower dimensional cases explained below
there always exists a unique cone given by the  Eq. (\ref{eq:5dHFgeneral})
which "fits" the positions of the four electrons. 
Besides the special cases (below)
we can therefore define
a projection of a single point
in $4\times 3=12$-dimensional space of four electrons  
onto a cone. That also implies that the complete
12-dimensional space describes a set (or family) of cones which
are 3D projections of the nodal manifold.
Similar projection strategies are often used in algebraic geometry
to classify or analyze surfaces with complicated
topologies or with high dimensionality.

Since the cone orientation and two radii
 are uniquely defined by the point in 12 dimensions
and the cone itself is a 2D 
 surface in 3D space of the probe electron, the complete
node then has 12+2=14
dimensions.  Therefore the $d^5$ HF node
is a set of cone
surfaces specified by the positions of the electrons . 
This 
particular form
is simply a property of the 
$d^5$  Hartree-Fock determinant.
From the derivation above it is clear that after factoring 
out the radial parts one
obtains
a homogeneous second-order polynomial in
three variables with coefficients determined by the 
positions of the four
electrons. 
In fact, from the theory of quadratic surfaces 
~\cite{rektorys}, one finds
that a general elliptic cone can possibly fit up to 
five 3D points/electrons, 
however, in our case the cone has an additional constraint.
Our system was reoriented so
that one of the electrons lies on the $z$-axis;  that 
implies that the $z$-axis lies on the cone.
Therefore the cone always 
cuts the $xy$  (i.e., $z=0$) plane in two lines, which are orthogonal to each other.
The orthogonality can be verified by imposing
$z=0$ in Eq. (\ref{eq:cone5d2}) and checking that $k_1k_2=-1$. 
In addition, one can find
 "degenerate" configurations with two pairs of two electrons
lying on orthogonal planes (Fig.~\ref{fig:d5}).
This corresponds to the "opening" of the cone 
 with one of the elliptic radii becoming
infinite and the resulting node having a form of
 two orthogonal planes (Fig.~\ref{fig:d5}).
Since in this case there is an additional condition 
on the particle positions, the two-plane node has
lower dimension and is a zero measure 
subnode of the general 14-dimensional node.
The condition is equivalent to
$A_{44}=b^2-a^2-abc=0$,  where
$A_{44}$ is one of the quadratic invariants~\cite{rektorys}.
There are more special cases of lower dimensional nodes:
(a) when two electrons lie on a straight line going through
the origin;
(b) when three electrons lie on a plane
going through the origin; 
(c) when four electrons lie in a single plane.

Remarkably, the analysis above enables us to find the number
of nodal cells. From Fig.~\ref{fig:d5} one can infer
that by appropriate repositioning of the four electrons 
the cone surface smoothly "unwraps" the domains inside the cone, forms
two crossing planes and then "wraps" around the cone domains of the opposite
sign. That implies that an electron inside one of the cone
regions
can get to the region outside of the cone (with the same 
wave function sign) without any node crossing, using only appropriate  
concerted repositioning of the remaining four electrons.
That enables us to understand that a point in the
15-dimensional space (positions of five electrons)
 can continuously scan 
the plus (or minus) domain of the wave function:
 there are only two maximal nodal cells.

\begin{figure}[!t]
\begin{center}
\includegraphics[width=\columnwidth]{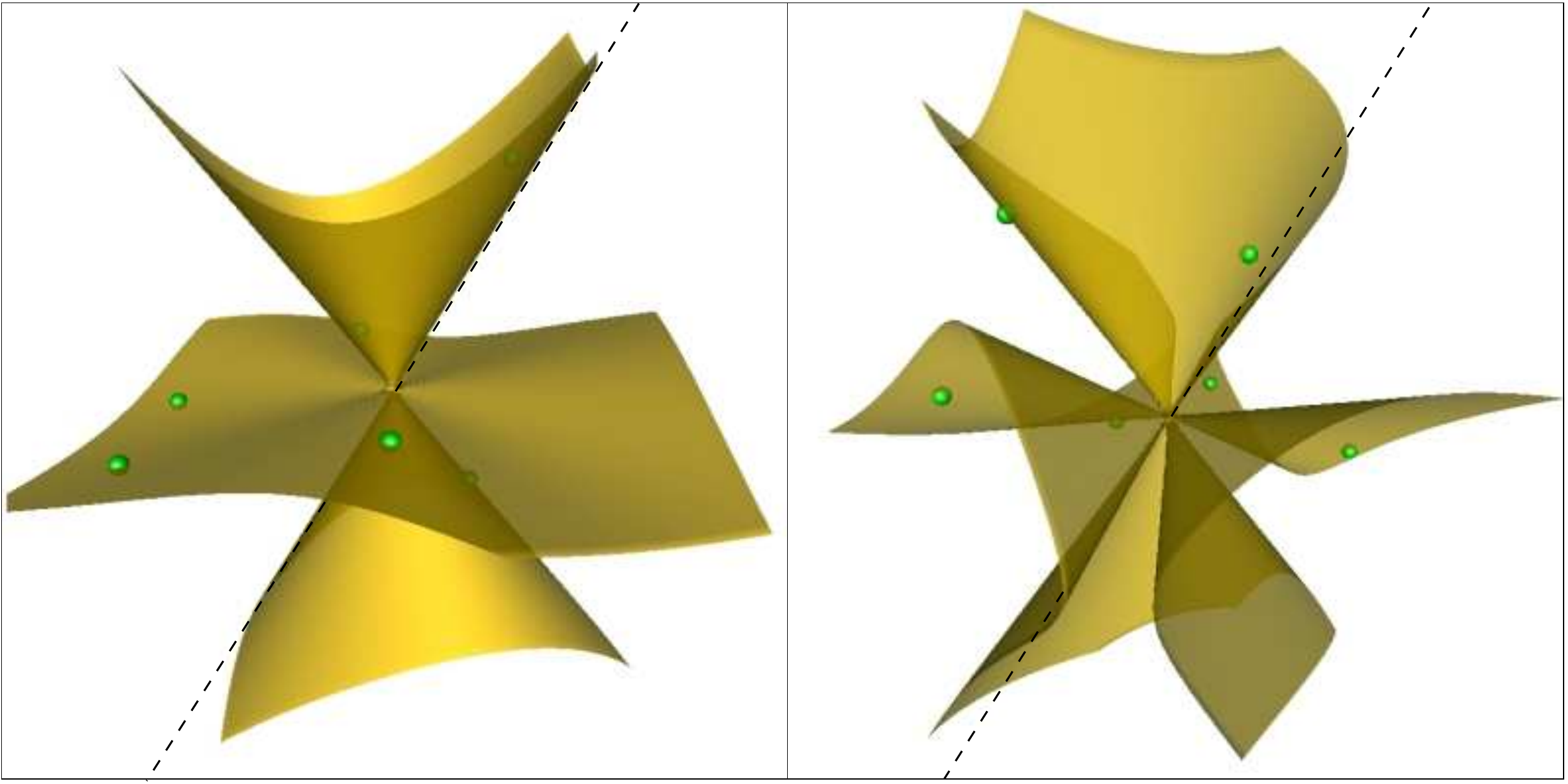}
\end{center}
\caption{Projected Hartree-Fock node of $ ^8S(f^7)$ state. The node has
two topologies: cone times planar surface or a cone "fused"
with planar surface what forms a single sheet surface. There is a smooth transition
between these two forms depending on the positions of six electrons which are
denoted by the small spheres. Note that the node contains the $z-$axis which
is denoted by the dashed line. 
\label{fig:f7}}
\end{figure}

\subsection{Hartree-Fock Nodes of the $^8S(f^7)$ Ion}\label{nodessec:level32}
We will use similar strategy as in the preceding case. 
After rotating one of the electrons to $z$-axis we
expand the determinant in the probe electron column 
and eliminate the radial orbitals which form
an overall prefactor of the Slater determinant since we assume that
all seven $f$-states are from the same $l=3$ subshell (e.g., 4$f$).
We get 
\begin{align}
(m_1x+m_2y)(4z^2-x^2-y^2)&+m_3z(x^2-y^2)+m_4xyz\nonumber \\
+m_5x(x^2-3y^2)&+m_6y(y^2-3x^2)=0.
\end{align}
Note that the node  contains the $z$-axis 
and there are $two$ possible values of $z$ for any $x,y$ since 
the form is quadratic in $z$. This restricts the node 
shapes significantly and by further analysis one can find that
the nodal surface projection into 3D has two topologies 
(Fig.~\ref{fig:f7}). The first one
is  a cone times a planar surface (topologically
equivalent to the $Y_{40}$ spherical harmonic).
Note that, in general, the  planar surface is deformed from
a straight plane since it passes
through the origin and, in addition, it fits three of the electrons.
The second topology is 
a "fused" cone and planar surface, which results in a general single
sheet cubic surface.
The node transforms smoothly between 
these two topologies depending on how the six 
electrons move in space.
These two topologies define the projection of the node 
into the probe 3D space and therefore enable us to capture the 
many-dimensional node for this particular Hartree-Fock state. 
This again enables to describe the complete node using
a theorem from algebraic geometry which states that any cubic 
surface is determined by an appropriate 
mapping of six points in a projective
plane~\cite{pedoe,semple,harris}. To use it we first need to
realize the following property of  
the 3D projected node: The node equation above contains
only a homogeneous polynomial in $x,y,z$
which implies that in spherical coordinates
the radius can be eliminated and the node is dependent 
only on angular variables.
Hence, any line defined by an arbitrary point on the node
and the origin (i.e., a ray) lies on the node.
In other words, 
we see that the surface is ruled, i.e., it can be created by continuous 
sweep(s) of ray(s)
passing through origin. This enables us to project the positions of the six
electrons on an arbitrary plane, which does not contain the origin, and 
the node will cut such a plane in a cubic curve. As we mentioned above,
a theorem from
the algebraic geometry of cubic surfaces and curves says 
that any cubic surface is fully described by 
six points 
in a projective plane (see~\cite{pedoe,semple,harris}). 
For ruled surfaces any plane not passing
through the origin is a projective plane and therefore we can specify
a one to one correspondence
between the $6\times 3=18$ dimensional space and our cubic surface
in 3D. Obviously, there will be a number of lower-dimensional 
nodes which will correspond to positions
of electrons with additional constraints such as when they lie on 
curve with the degree lower than cubic; i.e., a conic.

\begin{figure}[!t]
\begin{center}
\includegraphics[width=\columnwidth]{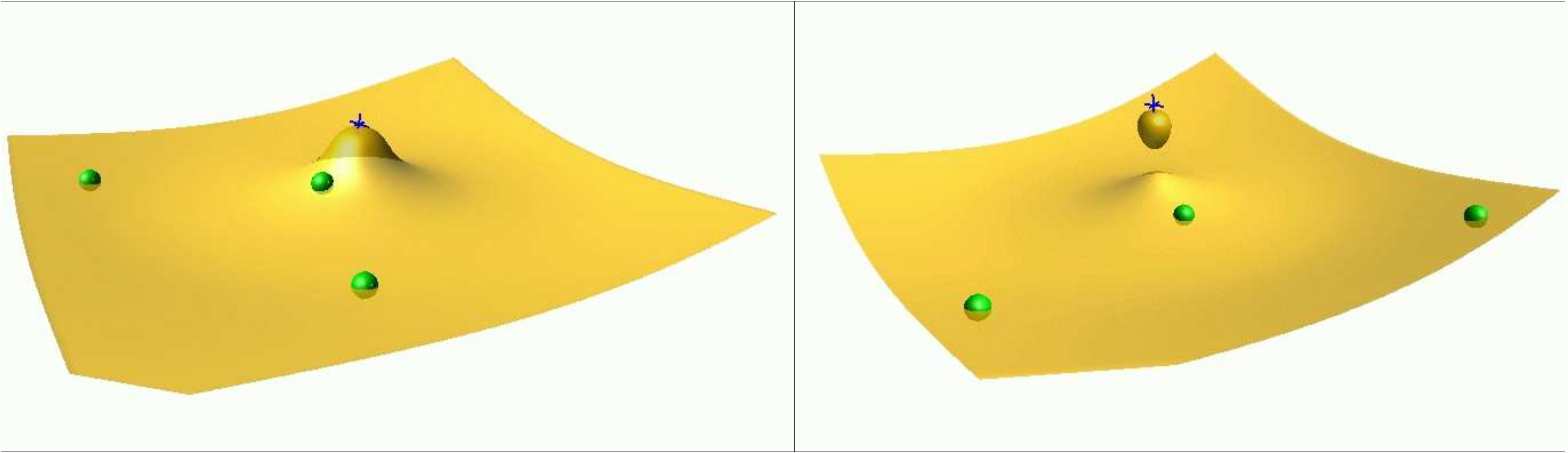}
\end{center}
\caption{
The 3D projection of the nitrogen cation $^5S(sp^3)$
Hartree-Fock node (the core electrons are eliminated by pseudopotentials). 
The projected node exhibits two
topologies. It is either
a planar surface deformed by the radial orbital functions 
at the nucleus or,
in certain configurations, the deformation forms a small bubble
detached form the surface (the picture on the right). 
The small cross is the location of the ion while the small spheres
denote positions of electrons.
}\label{fig:sp3}
\end{figure}

\subsection{Hartree-Fock Nodes of the $^5S(sp^3)$ Ion}\label{nodessec:level33}
The HF node for  this two-shell spin-polarized 
state can be investigated in a similar way as in previous cases
with a new feature that the radial parts will
be present in the expansion of the determinant.
By expanding the determinant in the column
of the probe electron with position $x,y,z$ 
the 3D node projection is simply given by
\begin{equation}
x+b'y+c'z+ d'\eta(r)=0,
\end{equation}
where $b',c',d'$ depend on ratios of cofactors 
and $\eta(r)=\rho_s(r)/\rho_p(r)$
is the ratio of radial parts of $s$ and $p$ orbitals 
and $r=\sqrt{x^2+y^2+z^2}$.
The probe electron will see a plane with a 
approximately bell-shape deformation in the area of the nucleus 
(see Fig.~\ref{fig:sp3}).
The shape of deformation depends on the ratio of $s$ and $p$ radial parts and the 
magnitudes and signs of the cofactors.
For certain 
configurations the deformation is so large that it gets detached
from the surface and forms a separated ellipsoid-like bubble.
The bubble results from the radial dependence of $\eta(r)$ which  
for pseudized core is not a monotonic function and therefore
can create new topologies.
Note that despite the fact that the 3D projection shows a separated
region of space (the bubble) the complete node has again 
the minimal number of nodal cells
property. To understand this, suppose that the probe electron is
located inside the bubble and wave function there has a positive sign.
Let us try to imagine how the electron can get
to the other positive region (the other side of
the planar surface).
Seemingly, the electron would need to cross the nodal
surface twice (the surface of the bubble and the planar surface).
However, the complete node is a collective-coordinate object 
and by moving the other two electrons
in an appropriate way the bubble attaches to the surface and
then fuses into a single surface (Fig.~\ref{fig:sp3}, left)
so that the probe electron can reach 
the positive region without node crossing.

In order to see whether the correlation would change the HF node
we have carried out a limited study of the CI wave function nodes for this
case; we have found some differences but we have not discovered any
dramatic changes 
to the HF nodes.  To quantify 
this further we have calculated the
CI energy (with the same basis and level of correlation as in the previous 
cases) and the result is in the last row of Tab.~\ref{tab_xx}. We estimate that 
the fixed-node bias of the HF node
is of the order of   $\approx$ 0.001 Hartree which is close to $\approx$
3\% of the correlation energy. Obviously,  the DMC energy is above
the exact one and percentage-wise the amount of missing 
correlation energy is
not insignificant. We conjecture that
the HF node is reasonably close to the exact one although the 
fine details of the nodal surface are not captured perfectly.
\subsection{Hartree-Fock Nodes of Spin-Polarized $p^3d^5$  and $sp^3d^5$ Shells with $S$ Symmetry}\label{nodesec:level34}
\begin{figure}
\begin{center}
\includegraphics[width=\columnwidth]{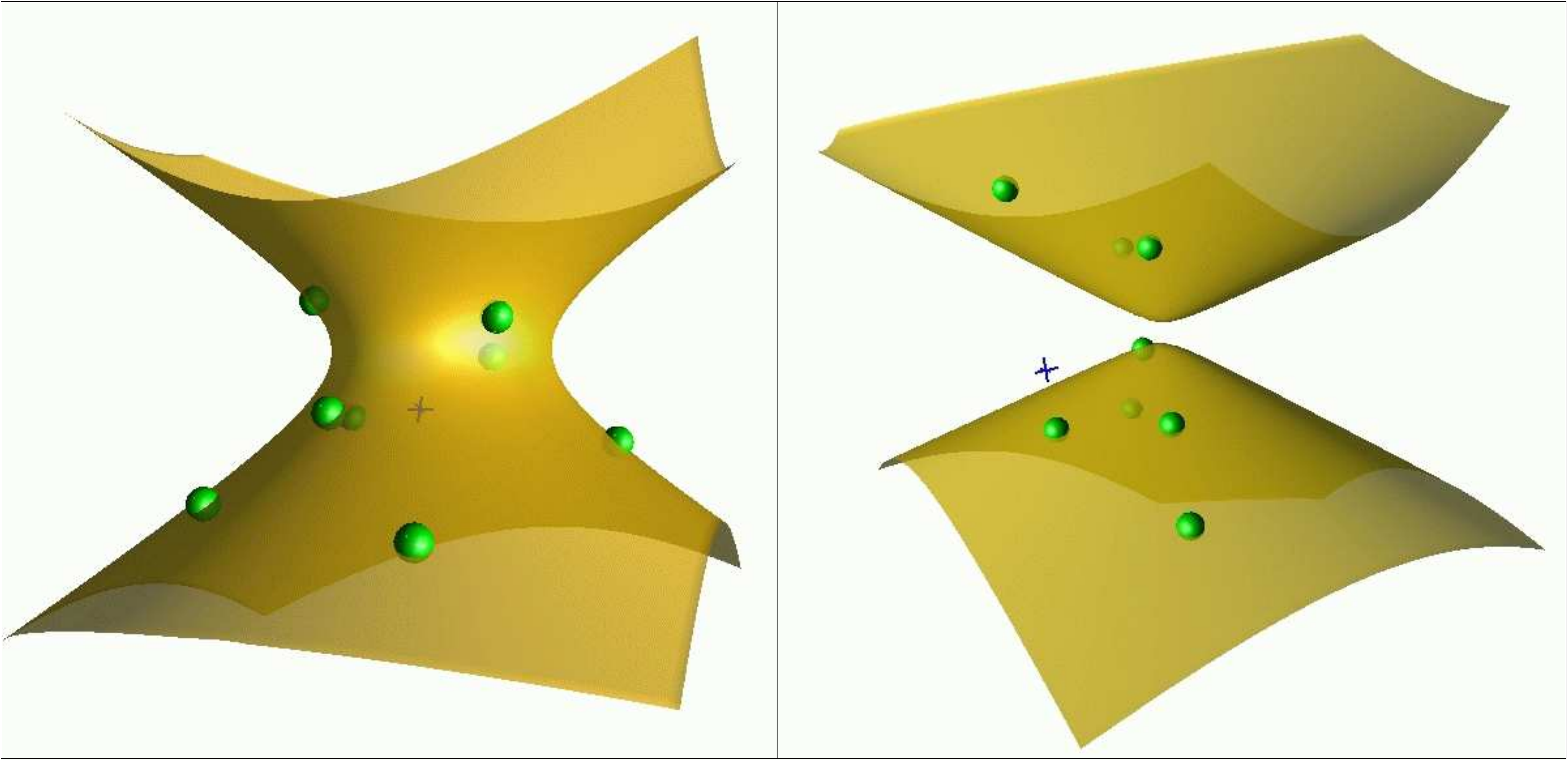}
\end{center}
\caption{
The 3D projection of the angular part of the $^{10}S(sp^3d^5)$ state
 Hartree-Fock node 
(with radial parts of orbitals identical for all $spd$ orbitals).  
The projection has a topology
of a single-sheet or double-sheet hyperboloid. The small cross shows the 
location of the nucleus while the spheres illustrate the electron positions. 
}\label{fig:sp3d5}
\end{figure}

\begin{figure}[!t]
\begin{center}
\includegraphics[width=\columnwidth]{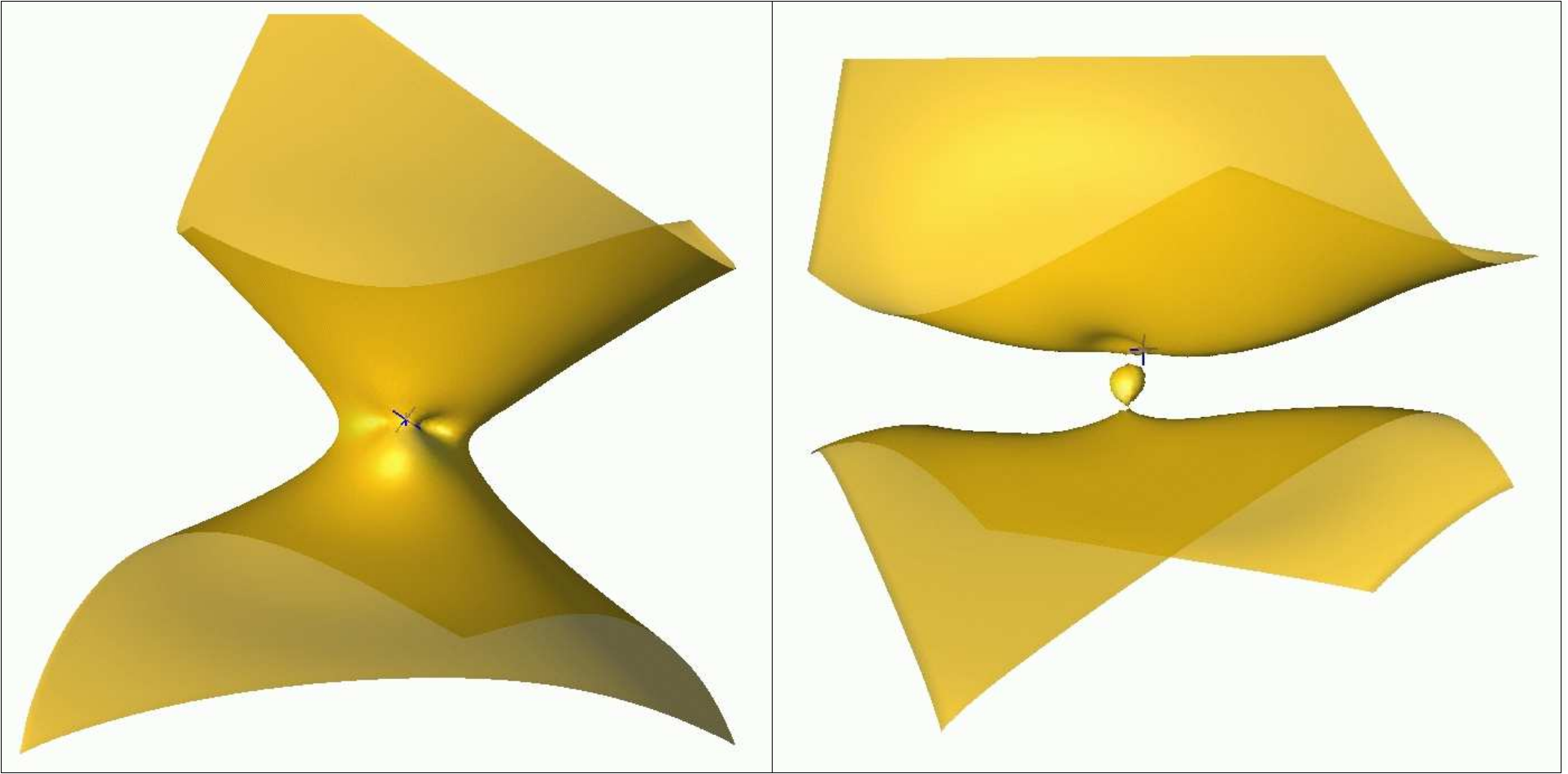}
\end{center}
\caption{ Projected Hartree-Fock node of
$^{10}S(sp^3d^5)$ of the majority spin valence electrons in
$\mathrm{Mn}^{+2}$ ion. The Ne-core electrons are eliminated 
by pseudopotentials. Note the deformations from the radial
parts of orbitals, including a  
small bubble detached from the rest of the surface (the right picture). 
For clarity, 
the positions of electrons 
have been omitted.   
}\label{fig:Mn}
\end{figure}

Let us for a moment assume a model wave function
 in which the radial parts of $s,p,d$
orbitals are identical. Then, using the arrangements similar
to $d^5$ case, we can 
 expand the determinant of $p^3d^5$ in one column and 
for the 3D node projection we then get
\begin{equation}
2u^2-v^2 -w^2 +\alpha u +\beta v +\gamma w=0,
\end{equation}
where $u,v,w$ are appropriate linear combinations of $x,y,z$. 
This can be further
rewritten as 
\begin{equation}
2(u+\alpha/4)^2-(v-\beta/2)^2 -(w-\gamma/2)^2 +\delta_0 =0,
\end{equation}
where
$\delta_0=(-\alpha^2/2+\beta^2+\gamma^2)/4$.
It is clear that the 
 quadratic surface is offset
from the origin (nucleus)
by a vector normal to 
$\alpha u +\beta v +\gamma w =0$
plane.
Using the properties of quadratic surfaces one finds that
 for $(\alpha^2/(\alpha^2+\beta^2+\gamma^2))<2/3$ the node
is
a single-sheet hyperboloid with the radius $\sqrt{\delta_0}$; otherwise
it has a shape of
 a double-sheet hyperboloid. The double-sheet hyperboloid forms when
there is 
an electron located close to the origin.
A special case is a cone which corresponds to ($ \delta_0=0$).
The case of $sp^3d^5$ is similar, but with different $\delta_0$,
which now has a contribution from the $s$-orbital (see Fig.~\ref{fig:sp3d5}). 
Once we include also the correct radial parts of orbitals in the 
$s,p,d$ channels 
the coefficients of the quadratic form
depend on both cofactors and orbital radial functions.
The resulting
 nodal surface is deformed beyond an ideal quadric and shows some
more complicated structure around the nucleus (see Fig.~\ref{fig:Mn})
as illustrated on HF nodes of the majority spin electrons in Mn$^{+2}$ ion
(note that the Ne-core electrons were eliminated by pseudopotentials).

\section{Nodes and Spin Correlations}\label{nodessec:level5}
We conjecture that a non-degenerate ground state of 
any given symmetry possesses only two maximal nodal cells. 
It was demonstrated for the considered fully spin-polarized systems
that the corresponding HF wave functions have the desired topology, i.e., two maximal nodal cells. 
On the other hand, for partially spin-polarized and unpolarized systems 
the corresponding HF wave functions of the form
$\Psi_{HF}={\rm det}[\varphi_{\alpha}^{\uparrow}({\bf r}_i)]
{\rm det} [\varphi_{\beta}^{\downarrow}({\bf r}_j)]$
lead to four nodal cells because there are two nodal cells 
for each independent spin subspace (assuming more than one electron 
in each spin subspace). 
To see how correlations eliminate a redundant nodal structure 
due to the artificial HF spin-up and spin-down separation 
we compare the nodal structure for HF and CI wave functions 
in two specific cases of spin-unpolarized states.

\begin{figure}[!t]
\begin{center}
\includegraphics[width=\columnwidth]{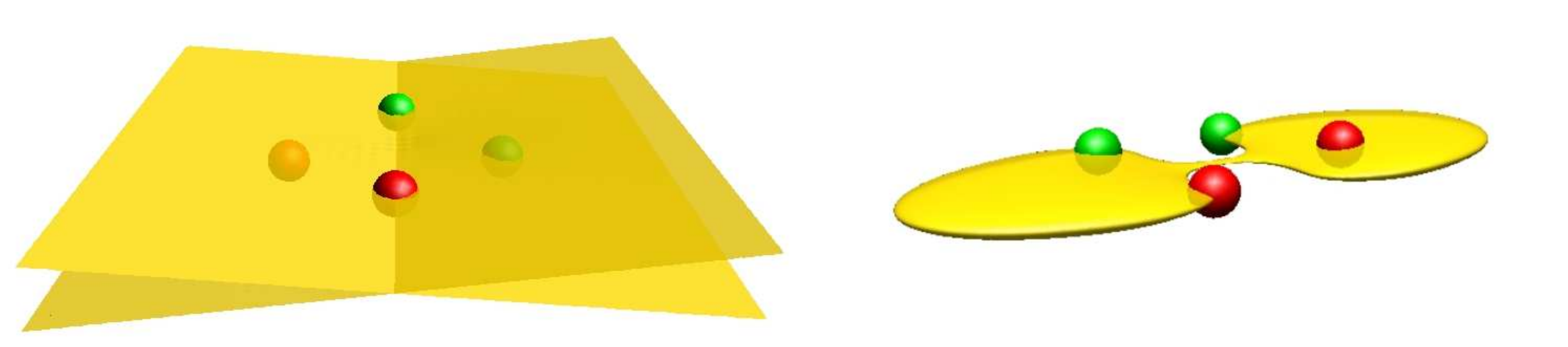}
\end{center}
\caption{The 3D projected HF and CI nodes of unpolarized $^{1}S(p^6)$ state.
The HF node (left) consists of four nodal regions 
(two for each spin channel)
while the CI node (right) exhibits only two nodal cells.
Positions of four electrons (small spheres) are fixed, 
two different colors indicate opposite spins. 
The projected nodal surface is sampled with a pair 
of electrons of opposite spins which are positioned 
close together.}
\label{fig:p6nodes1}
\end{figure}

\begin{figure}
\begin{center}
\includegraphics[width=\columnwidth]{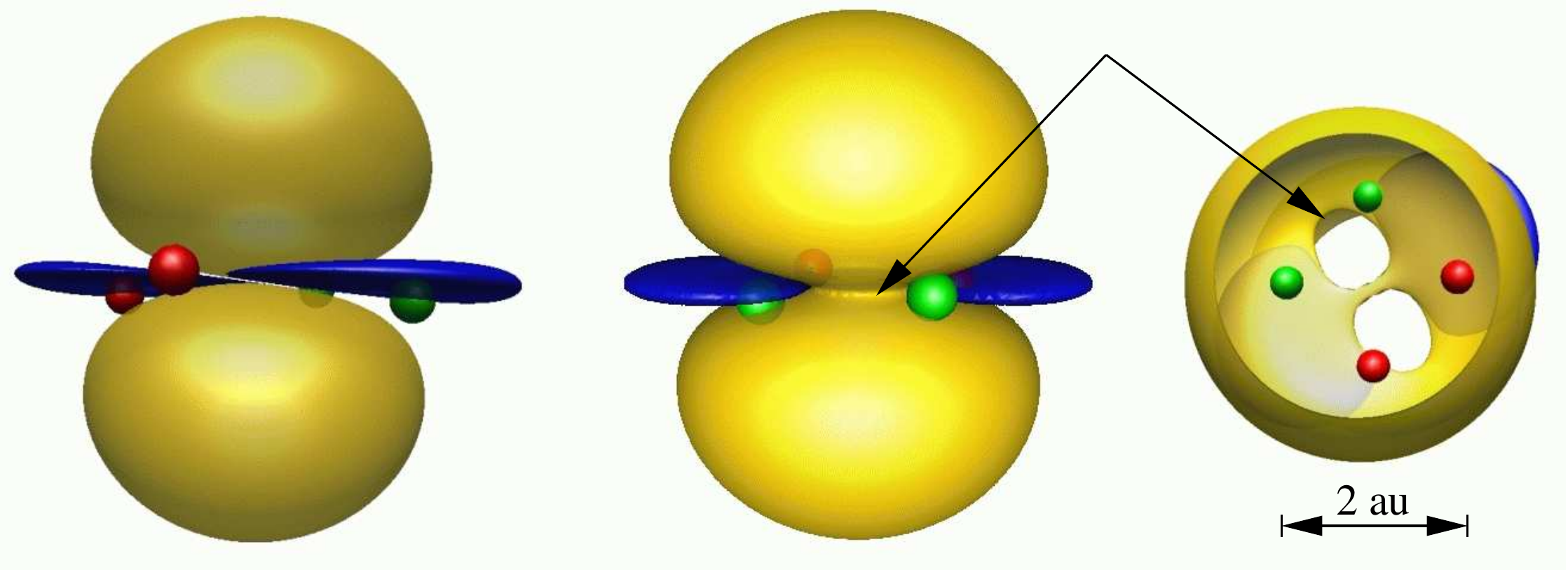}
\end{center}
\caption{The 3D projected isosurfaces of HF and CI wave functions 
for $^{1}S(p^6)$ state. 
Different colors (dark and light) represent opposite signs.
The HF wave function (left) has a discontinuity 
between regions of the same sign. In the CI wave function (right)
the regions of the same sign are connected. A cut through
(a view from top) reveals open channels (indicated by arrows) 
between regions of the same sign.}
\label{fig:p6nodes2}
\end{figure}

\subsection{HF and CI Nodes for Singlet State $^{1}S(p^6)$}
The 3D projected nodal structure of 
$^{1}S(p^6)$ unpolarized state is shown in Fig.~\ref{fig:p6nodes1}. 
We fixed positions of four electrons and sample for the nodal surface
with a pair of electrons of opposite spins which are close to
each other. 
The HF node consists of four nodal regions with a planar node 
for each spin subspace. The planar node is even exact
for the case of three electrons in the same spin subspace, 
i.e., $^{4}S(p^3)$.
On the other hand, the CI wave function, which is close to the exact 
ground state of the given symmetry, leads to only two nodal cells.
If we write $\Psi_{CI}=\sqrt{1-\epsilon^2} \Psi_{HF} 
+ \epsilon \Psi_{excit}$ where $\Psi_{excit}$ includes 
single, double, and triple excitations, we find that 
$\epsilon\approx 0.01$. Introducing a small correlation 
via nonzero $\epsilon$ makes a dramatic change to the topology 
of the HF nodal structure despite of a small total energy gain.
Comparing the corresponding isosurfaces for HF and CI wave functions
(Fig.~\ref{fig:p6nodes2}) the structure of the wave functions seems to be very similar. 
However, a closer inspection reveals 
that regions of the same sign, which are disconnected in the HF wave function, 
are now connected by narrow channels in the CI wave function.

\begin{figure}[!ht]
\begin{center}
\includegraphics[width=\columnwidth]{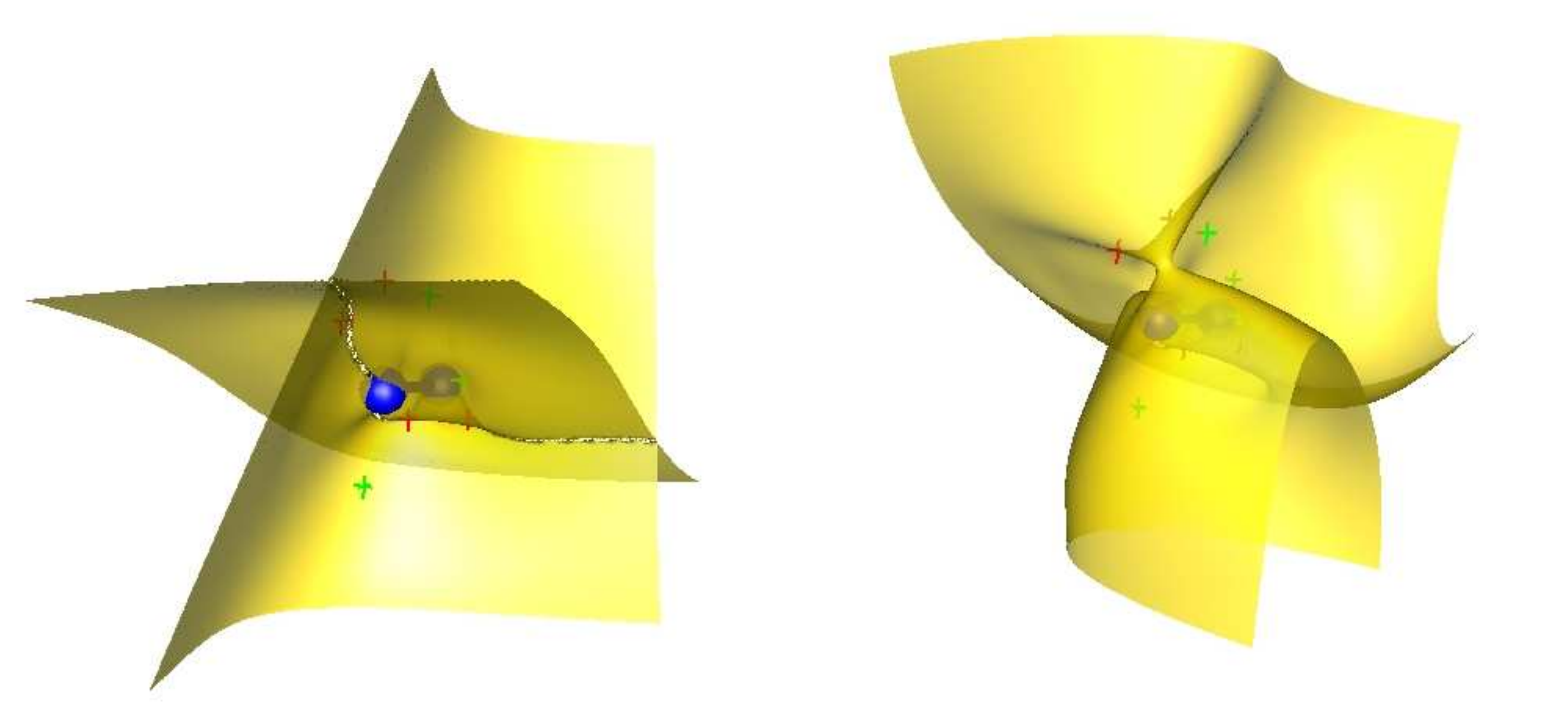}
\end{center}
\caption{The 3D projected nodes for N$_2$ dimer. 
The HF nodes (left) are bended and distorted due to spin correlations
into CI nodal surface (right) with just two nodal cells.
Sampling of nodes is performed with a pair of electrons with opposite spins, 
which are close to each other; positions of other electrons (crosses) 
are fixed; small spheres indicate ions.}
\label{fig:n2nodes}
\end{figure}

\subsection{HF and CI Nodes for N$_2$ Dimer State $^{1}\Sigma_g^+$} 
The projected nodal structure of the spin-unpolarized 
ground state $^{1}\Sigma_g^{+}$ for N$_2$ dimer
is shown in Fig.~\ref{fig:n2nodes}. The He-core electrons are eliminated by 
pseudopotentials. The HF wave function
with the separation of electrons into independent spin-up and spin-down 
subspaces forms fours nodal cells. In the CI case we can see that 
the HF nodes have been distorted and bended to build up channels, 
which connect regions of the same sign, i.e., two maximal nodal cells are obtained.

In order to compare accuracy of the nodes we have carried out 
a fixed-node diffusion Monte Carlo simulations  
for N$_2$ dimer with core electrons eliminated by pseudopotentials~\cite{lester}.
For the HF trial wave function (including the Jastrow factor, 
which does not change nodes) we have obtained the total energy 
-19.8395(7) H, which recovers 94\% of correlation energy.
For the CI trial function with more than 5000 determinants 
from double excitations of 45 orbitals used and ccpV6Z basis
(with up to $f$ basis functions) the total energy 
reaches -19.870(5) H, which recovers 98\% of 
correlation energy. 
The estimated exact energy is \mbox{-19.8822}~H. The Fig.~\ref{fig:n2nodes} illustrates a good quality of the HF nodes except for 
the regions where two nodal surfaces cross each other.

\section{Conclusions}\label{nodessec:level6} 
We summarized the most recent knowledge about the general properties of 
nodes of fermionic wave functions.
Further, we have investigated the nodes of atomic and molecular spin-polarized systems
with one-particle states in $s,p,d$ channels.
We have studied cases with high symmetries, which 
enabled us to find exact nodes for several 
states  with a few electrons ($p^2, p^3, \pi^2$). 
Moreover, the projection of multi-dimensional 
manifolds into 3D space enabled us to 
study and characterize properties of nodes, in particular,
their topologies for the Hartree-Fock wave functions. 
This analysis has provided useful insights and 
enabled us to formulate general transformation of one-particle coordinates 
using coordinate translation (backflow) and metric tensor
to capture inhomogeneities and  rotation symmetries.
We test a special form of a backflow coordinate transformation
in Ch.~\ref{ch:bf} of the thesis. 
Finally, we have demonstrated that the more accurate CI 
wave functions for two specific cases of spin-unpolarized states have 
the nodal structure of only two maximal nodal cells. 

\chapter{Pfaffian Pairing Wave Functions}\label{ch:pfaffians}
\begin{center}
\begin{minipage}{\columnwidth}
Sections of this chapter also appeared in:
\begin{center}
{\bf Pfaffian pairing wave functions in electronic-structure\\ 
quantum Monte Carlo simulations},\\
M.~Bajdich, L.~Mitas, G.~Drobn\'y, L.~K, Wagner, and K.~E. Schmidt,\\
Phys. Rev. Lett \textbf{96}, 130201 (2006).\\
(Received 14 December 2005; published 5 April 2006)\\
{\it \copyright2006 The American Physical Society}
\end{center}

\begin{center}
{\bf Pfaffian pairing wave functions and their properties\\ 
in electronic structure quantum Monte Carlo methods},\\
M.~Bajdich, L.~Mitas, L.~K. Wagner, and K.~E. Schmidt,\\
cond-mat/0610850  (2006).\\ 
(Submitted to Phys. Rev. B)\\
\end{center}
\end{minipage}
\end{center}
\newpage
\section{\label{sec:level1} Introduction}

The key challenge for successful application of fixed-node DMC is 
to develop methods, which can eliminate 
the fixed-node bias or at least make it smaller 
than experimental error bars for the given quantity.
This is a difficult task, once we realize that
the fermion nodes, which are subject of Ch.~\ref{ch:nodes}, 
are complicated high-dimensional manifolds determined 
by the many-body effects. 
So far, improvement in the accuracy of trial wave functions 
has proved to be one realistic approach to finding better approximations 
for the nodes. This approach has an additional benefit in forcing us to 
think about the relevant correlation effects and their 
compact and computationally efficient description.

The commonly used  QMC trial wave functions have the Slater-Jastrow 
form, which can be written as  $ \Psi_{T}=\Psi_{A} \exp[U_{corr}]$,
where $\Psi_A$ is the antisymmetric part while $U_{corr}$
describes the electron-electron and higher-order correlations as described in Sec.~\ref{subsec:twf}. 
The antisymmetric component is typically one or a linear
combination of several Slater determinants of one-particle orbitals
such as a configuration interaction expansion introduced in Sec.~\ref{sec:intro:postHF}.
To overcome the limit of one-particle orbitals the two-particle
or pair orbital has been suggested.  In condensed systems one 
such example is the Bardeen-Cooper-Schrieffer (BCS) wave function, which 
is an antisymmetrized product of singlet pairs. The singlet pair is sometimes 
referred to as geminal and the resulting wave function as the antisymmetrized geminal product (AGP).
It has been recently used  to calculate several atoms and molecules as well as  
superfluid Fermi gases~\cite{carlsonbcs,sorellabcs1,sorellabcs2}.
The results show promising gains when compared to 
the single-determinant Hartree-Fock (HF) wave functions, 
nevertheless, in partially
spin-polarized systems the improvements are
less pronounced due to the lack of pair correlations in
the spin-polarized subspace~\cite{sorellabcs1,sorellabcs2}. 
The spin-polarized 
(triplet) pairing wave functions  lead to Pfaffians (instead of 
determinants) and have been mentioned a few times before
and applied to model systems~\cite{bouchaud2,Bhattacharjee,kevin}. 

In this chapter, we further develop the idea from Sec.~\ref{subsec:twf}, 
in which we have proposed the description of electronic systems
by a novel generalized pairing wave function in the Pfaffian form. 
This chapter is organized as follows:
In Sec.~\ref{sec:pf:algebra}, we present a set of the key mathematical identities
and formulas for Pfaffians, some of them derived for the first time. 
In Sec.~\ref{sec:pf:pairingwf}, we establish the connection of a 
generalized Pfaffian pairing wave function to BCS and HF wave functions.
The resulting Pfaffian wave functions are tested on atomic and molecular
systems in variational and fixed-node diffusion Monte
Carlo methods as described in Sec.~\ref{subsec:pf:spf}.
In Secs.~\ref{subsec:pf:mpf} and~\ref{subsec:pf:aip}, we investigate also a generalizations to linear combinations 
of Pfaffians and to antisymmetrized independent singlet pairs and compare 
the results from the point of view of recovered energies and compactness of
the wave functions. Finally, in Sec~\ref{subsec:pf:nodes}, we analyze the fermion nodes for some of the wave functions
and point out the topological differences between HF, Pfaffian and 
an essentially exact wave functions for a given test example.

\section{Algebra of Pfaffians}\label{sec:pf:algebra}
\subsection{Definitions}
First introduced by Arthur Cayley in 1852~\cite{cayley2},  the Pfaffian is named after 
German mathematician Johann Friedrich Pfaff.
Given a $2n\times 2n$ skew-symmetric matrix $A=\left[a_{i,j}\right]$, the 
Pfaffian of $A$ is defined as antisymmetrized product

\begin{align}\label{eq:pfdef}
{\rm pf}[A]&={\mathcal A}[a_{1,2}a_{3,4}\ldots a_{2n-1,2n}]\nonumber\\
&=\sum_{\alpha} {\rm sgn}(\alpha)\ a_{i_1, j_1}a_{i_2, j_2} \ldots a_{i_n, j_n},
\end{align}  
where the sum runs over all possible $(2n-1)!!$ pair partitions 
$\alpha=\{(i_1, j_1),(i_2, j_2),\ldots ,(i_n, j_n))\}$ of $\{1,2,\ldots,2n\}$ with $i_k < j_k$.
The sign of permutation associated with the partition $\alpha$ is denoted as
${\rm sgn}(\alpha)$. 
The Pfaffian for a matrix of odd order equals to zero. The following example gives Pfaffian
of a  $A(4 \times 4)$ skew-symmetric matrix
\begin{equation}
{\rm pf}\begin{bmatrix}
0 & a_{12} & a_{13} & a_{14} \\
-a_{12} & 0 & a_{23} & a_{24} \\
-a_{13} & -a_{23} & 0 &  a_{34} \\
-a_{14} & -a_{24} & -a_{34} & 0\\
\end{bmatrix}=a_{12}a_{34}-a_{13}a_{24}+a_{14}a_{23}.
\end{equation}   
It can be also evaluated recursively as 
\begin{align}\label{eq:pfrecurr}
{\rm pf}[A]&=\sum_{j=2}^{2n} a_{1,j}\sum_{\alpha_{1,j}}{\rm sgn}(\alpha_{1,j})\ a_{i_1, j_1}a_{i_2, j_2} \ldots a_{i_{n-1}, j_{n-1}}\nonumber\\ 
&\equiv \sum_{j=2}^{2n} a_{1, j} P_c(a_{1, j}) ,
\end{align} 
where $\alpha_{1,j}$ is partition with $i_k, j_k\neq 1, j$ and $P_c(a_{1, j})$ is 
defined as Pfaffian cofactor of $a_{1, j}$.
The cofactor for an element $a_{j,k}$ is given by a formula, 
\begin{equation}\label{eq:pfcof}
P_c(a_{j,k})=(-1)^{j+k+1}{\rm pf}[A(j,k;j,k)]  
\end{equation}
where the matrix $A(j,k;j,k)$ has the rank $2(n-1)\times 2(n-1)$  and is obtained  from  $A$ by eliminating
$j$ and $k$ rows and columns. 
\subsection{Calculation of a Pfaffian}
There exist several identities involving Pfaffians and determinants.
For any  $2n \times 2n$ skew-symmetric matrix $A$ and arbitrary matrices $B(2n \times 2n)$ 
and $M(n \times n)$ we have the following relations:
\begin{subequations}
\begin{align}
{\rm pf}[A^T]&=(-1)^n {\rm pf}[A] \label{eq:pfident1}\\
{\rm pf}[A]^2&={\rm det}[A] \label{eq:pfident2}\\
{\rm pf} \left[\begin{array}{cc}
A_1 & 0 \\
0 & A_2
\end{array} \right]&={\rm pf}[A_1]{\rm pf}[A_2] \label{eq:pfident5}\\
{\rm pf}[BAB^T]&={\rm det}[B]{\rm pf}[A]\label{eq:pfident4} \\
{\rm pf}\left[\begin{array}{cc}
0 & M \\
-M^T & 0
\end{array} \right]
&=(-1)^{{n(n-1)}\over{2}}{\rm det}[M]\label{eq:pfident3}
\end{align}
\end{subequations}
Key ideas of respective proofs:
\begin{itemize}
\item[(\ref{eq:pfident1})] Each permutation contains product of 
$n$ pairs resulting in an overall $(-1)^n$ factor.

\item[(\ref{eq:pfident2})] This is a well-known Cayley's relationship between
the Pfaffian and the determinant of a skew-symmetric
matrix. Since it has been proved many times before in variety of ways~\cite{nakahara,sss,cayley}  
we do not give this proof here. Using this relation we rather prove
 a more general version of Cayley's identity~\cite{cayley}
in App.~\ref{appendix:Cayley}, which we were not able to find 
anywhere else except in the original Cayley's paper~\cite{cayley}.

\item[(\ref{eq:pfident5})] Use the expansion by Pfaffian cofactors.

\item[(\ref{eq:pfident4})] By squaring (4d), using Eq.~(\ref{eq:pfident2}), and taking the
square root one finds
${\rm pf}[BAB^T]=\pm{\rm det}[B]{\rm pf}[A]$. Substituting the identity matrix $I$ for $B$ 
one finds $+$ to be the correct sign.

\item[(\ref{eq:pfident3})] Assume
\begin{align*}
B=\left(\begin{array}{cc}
M & 0\\
0 & I
\end{array}\right)
\quad {\rm and} \quad
A=\left(\begin{array}{cc}
0 & I\\
-I & 0
\end{array}\right)
\end{align*}
in Eq.~(\ref{eq:pfident4}). The overall sign is given by value of ${\rm pf}[A]$.
\end{itemize}

The identities listed above imply several important properties.
First, Eqs.~(\ref{eq:pfident4}) and (\ref{eq:pfident3}) show that every determinant 
can be written as a Pfaffian, but on the contrary, 
only the absolute value of Pfaffian can be given by determinant [Eq.~(\ref{eq:pfident2})].
The Pfaffian is therefore a generalized form of the determinant.
Second, by substituting suitable matrices~\cite{matrices} for $M$
in Eq.~(\ref{eq:pfident4}) one can verify the following three  
properties of Pfaffians~\cite{galbiati}, similar to the well-known
properties of determinant.
\begin{itemize}
\item[(a)] Multiplication of a row and a column by a constant
is equivalent to multiplication of Pfaffian by the same constant.  
\item[(b)] Simultaneous interchange of two different rows and corresponding columns changes the sign of Pfaffian.
\item[(c)] A multiple of a row and corresponding column added to 
to another row and corresponding column does not change the value of Pfaffian.
\end{itemize}
It is also clear that
any skew-symmetric matrix can be brought to block-diagonal form by an orthogonal transformation.
Recursive evaluation [Eq.~(\ref{eq:pfrecurr})] implies that the Pfaffian of block-diagonal matrix is directly given by 
\begin{equation}\label{eq:pf:blockdiag}
{\rm pf}\begin{bmatrix}
0 & \lambda_1  & & & &  \\
-\lambda_1 & 0 & & & & 0 & \\
& & 0 & \lambda_2 & & \\
& & -\lambda_2 & 0 & & \\
& & & &  \ddots & \\
& 0 & & & & 0 & \lambda_n \\
& & & & & -\lambda_n & 0  \\
\end{bmatrix}=\lambda_1 \lambda_2 \ldots \lambda_n.
\end{equation}
Therefore by employing a simple Gaussian elimination technique with row pivoting (see  App.~\ref{appendix:pfaffianroutines}) 
we can transform any skew-symmetric matrix into block-diagonal form and obtain 
its Pfaffian value in $O(n^3)$ time.

However, in QMC applications, one often needs to 
evaluate the wave function after a single electron
update. Since Cayley~\cite{cayley} showed (for proof see App.~\ref{appendix:Cayley})
that  
\begin{align}\label{eq:cayley}
{\rm det}& \left[\begin{array}{ccccc}
0  & b_{12}  & b_{13} &\ldots &  b_{1n}\\
-a_{12}  & 0  & a_{23} & \ldots &  a_{2n}\\
-a_{13}  & -a_{23} & 0  & \ldots &  a_{3n}\\
 \vdots & \vdots & \vdots &  \ddots &  \vdots \\
-a_{1n} & -a_{2n} & -a_{3n} & \ldots &  0\\
\end{array}\right]\\
&={\rm pf} \left[\begin{array}{cccccc}
0  & a_{12}  & a_{13} &\ldots &  a_{1n}\\
-a_{12}  & 0  & a_{23} & \ldots &  a_{2n}\\
-a_{13}  & -a_{23} & 0  & \ldots &  a_{3n}\\
 \vdots & \vdots & \vdots &  \ddots &  \vdots \\
-a_{1n} & -a_{2n} & -a_{3n} & \ldots &  0\\
\end{array}\right]\,
{\rm pf} \left[\begin{array}{ccccc}
0  & b_{12}  & b_{13} &\ldots &  b_{1n}\\
-b_{12}  & 0  & a_{23} & \ldots &  a_{2n}\\
-b_{13}  & -a_{23} & 0  & \ldots &  a_{3n}\\
 \vdots & \vdots & \vdots &  \ddots &  \vdots \\
-b_{1n} & -a_{2n} & -a_{3n} & \ldots &  0\\
\end{array}\right],\nonumber
\end{align}
we can relate the Pfaffian of original matrix ${\rm pf}[A]$ to  
the Pfaffian of a matrix with updated first row and column ${\rm pf}[B]$ using the inverse 
matrix $A^{-1}$ in only $O(n)$ operations by 
\begin{equation}\label{eg:inverseupdate}
{\rm pf}[B]=\frac{{\rm det}[A]\sum_j b_{1j}A^{-1}_{j1}}{{\rm pf}[A]}={\rm pf}[A]\sum_j b_{1j}A^{-1}_{j1}.
\end{equation}
The second part of Eq.~(\ref{eg:inverseupdate}) was obtained by 
taking advantage of the identity in Eq.~(\ref{eq:pfident2}).
Slightly more complicated relation between ${\rm pf}[A]$ and ${\rm pf}[B]$ 
can be derived if one considers simultaneous change of two separate rows and columns,
which represents the two electron update of a wave function.  

\subsection{Gradient and Hessian of Pfaffian}
In the case of linear dependence of matrix elements $A$ on a set of parameters $\{c\}$, 
one can derive the following useful relations:
\begin{equation}\label{eq:gradient}
\frac{1}{{\rm pf}[A]}\frac{\partial {\rm pf}[A]}{\partial c_i}=\frac{1}{2} {\rm tr}
\left[ A^{-1}\frac{\partial A}{\partial c_i}\right]
\end{equation}
and
\begin{align}\label{eq:hessian}
\frac{1}{{\rm pf}[A]}\frac{\partial^2{\rm pf}[A]}{\partial c_i \,\partial c_j}=&\frac{1}{4}
{\rm tr}
\left[A^{-1}\frac{\partial A}{\partial c_i}\right] \, {\rm tr}\left[ A^{-1}\frac{\partial A}{\partial c_j}\right] 
\\ \nonumber
&-\frac{1}{2} {\rm tr} \left[  A^{-1}\frac{\partial A}{\partial c_i}A^{-1}\frac{\partial A}{\partial c_j} \right],
\end{align}
where 
$A^{-1}$ is again the inverse of $A$. 

\section{\label{sec:level2} Pairing Wave Functions}\label{sec:pf:pairingwf}
In order to contrast the properties of pair wave functions with the ones build from
one-particle orbitals we will first recall 
the well-known fact from the Hartree-Fock theory. The simplest antisymmetric
wave function for $N$ electrons constructed from one-particle orbitals is the Slater {\em determinant}
\begin{equation}\label{eq:hfslater}
\Psi_{HF}= {\rm det} [\tilde\varphi_k({\bf r_i},\sigma_i)]={\rm det} [\tilde\varphi_k(i)]; \quad i,k=1,\ldots,N,
\end{equation}
where tilde means that the one-particle states depend on
both space and spin variables. Clearly, for $N$ electrons
this requires $N$ linearly independent spin-orbitals, which form an orthogonal
set in canonical HF formulation. 

Let us now consider the generalization of the one-particle orbital
to a two-particle (or pair) orbital
$\tilde{\phi}(i,j)$, where tilde again denotes dependence on both
spatial and spin variables. The simplest antisymmetric wave function 
for $2N$ electrons constructed from the pair orbital is a
{\em Pfaffian} 
\begin{equation}\label{eq:generalpairingwf}
\Psi={\mathcal A}[\tilde{\phi}(1,2),\tilde{\phi}(3,4) \ldots\tilde{\phi}(2N-1,2N)]={\rm pf}[\tilde{\phi}(i,j)].
\end{equation}
The antisymmetry is guaranteed by the definition (\ref{eq:pfdef}), since the signs 
of pair partitions alternate depending on the parity of the corresponding
permutation. 
The important difference from Slater determinant is that in the simplest case only {\em one} 
pair orbital is necessary. (This can be generalized, of course, as will be shown later.)
If we further restrict our description to systems with collinear spins, 
the pair orbital $\tilde{\phi}({\bf r}_i, \sigma_i; {\bf r}_j, \sigma_j)$
for two electrons in positions ${\bf r}_i$ and ${\bf r}_j$ 
and with spins projections $\sigma_i$ and $\sigma_j$ can be expressed as 
\begin{align}\label{eq:generalpair}
\tilde{\phi}({\bf r}_i, \sigma_i; {\bf r}_j, \sigma_j)&=
\phi(i,j)
\langle \sigma_i\sigma_j|[|\uparrow \downarrow\rangle -|\downarrow\uparrow\rangle]/\sqrt{2}\\ \nonumber
&+\chi^{\uparrow \uparrow}(i,j)\langle \sigma_i\sigma_j|\uparrow\uparrow\rangle\\ \nonumber
&+\chi^{\uparrow \downarrow}(i,j)
\langle \sigma_i\sigma_j|[|\uparrow \downarrow\rangle +|\downarrow\uparrow\rangle]/\sqrt{2}\\ \nonumber
&+\chi^{\downarrow \downarrow}(i,j)\langle \sigma_i\sigma_j|\downarrow\downarrow\rangle.
\end{align}
Here  
$\phi(i,j)=\phi({\bf r}_i,{\bf r}_j)$ is even while $\chi^{\uparrow \uparrow}$, $\chi^{\uparrow \downarrow}$ and $\chi^{\downarrow \downarrow}$ are 
odd functions of spatial coordinates. 
In the rest of this section we will discuss special cases of the wave function~(\ref{eq:generalpairingwf}).

\subsection{Singlet Pairing Wave Function}\label{subsec:singlet}
Let us consider the first $1,2, ...,N$ electrons to be spin-up and the rest $N+1, ...,2N$ 
electrons to be spin-down and allow only $\phi({\bf r}_i,{\bf r}_j)$ in $\tilde{\phi}({\bf r}_i, \sigma_i; {\bf r}_j, \sigma_j)$ to be non-zero. 
Using the Pfaffian identity [Eq.~(\ref{eq:pfident3})] we can write 
the wave function for $N$ singlet pairs, also known as the BCS wave function (or AGP), in the following form
\begin{equation}
\Psi_{BCS}={\rm pf}\begin{bmatrix}
0 & 
{\boldsymbol \Phi}^{\uparrow\downarrow}\\
-{\boldsymbol \Phi}^{\uparrow\downarrow T}&
0 \\
\end{bmatrix}={\rm det}[\boldsymbol \Phi^{\uparrow\downarrow}],
\end{equation}
which is simply a determinant of the $N\times N$ matrix $\boldsymbol \Phi^{\uparrow\downarrow}=\left[\phi (i,j)\right]$ as was shown
previously~\cite{bouchaud1,bouchaud3}.

It is straightforward to show that the BCS wave function contains the 
restricted HF wave function as a special case.
Let us define the Slater matrix $C=\left[\varphi_i(j)\right]$ where $\{\varphi_i\}$ is a set of HF occupied
orbitals. Then we can write 
\begin{equation}
\Psi_{HF}={\rm det}[C]{\rm det}[C]=
{\rm det}[CC^T]={\rm det}[{\boldsymbol \Phi}_{HF}^{\uparrow\downarrow}],
\end{equation}
where 
\begin{equation}
( \boldsymbol \Phi_{HF}^{\uparrow\downarrow})_{i,j}=\phi_{HF}(i,j)=\sum_{k=1}^{N}\varphi_k(i)\varphi_k(j).
\end{equation}
On the other hand, we can think of the BCS wave function as a natural generalization
of the HF one. To do so we write the singlet pair orbital as
\begin{equation}\label{eq:phi}
\phi(i,j)=\sum_{k,l}^{M>N}S_{k,l}\varphi_k(i)\varphi_l(j)=
{\boldsymbol \varphi}(i)\,{\bf S}\,{\boldsymbol \varphi}(j),
\end{equation}
where the sum runs over all $M$ (occupied and virtual) single-particle orbitals and ${\bf S}$ is some symmetric matrix. 
Therefore, we can define one-particle orbitals, which diagonalize this matrix and call them 
\emph{natural orbitals of a singlet pair}.


The BCS wave function is efficient for describing
systems with single-band correlations such
as Cooper pairs in conventional BCS
superconductors, where pairs form from
one-particle states close to the Fermi level.

\subsection{Triplet Pairing Wave Function}\label{subsec:triplet}
Let us assume, in our system of $2N$ electrons, that the first $M_1$ electrons are spin-up 
and remaining $M_2=2N-M_1$ electrons are spin-down. Further, we restrict $M_1$ and $M_2$ to be even numbers.  
Then by allowing only $\chi^{\uparrow \uparrow}(i,j)$ and $\chi^{\downarrow \downarrow}(i,j)$
in (\ref{eq:generalpair}) to be non-zero, we obtain from expression~(\ref{eq:generalpairingwf})
by the use of Eq.~(\ref{eq:pfident5})
\begin{equation}
\Psi_{TP}={\rm pf}\begin{bmatrix}
 {\boldsymbol \xi}^{\uparrow\uparrow} & 
0\\
0 & {\boldsymbol \xi}^{\downarrow\downarrow}\\
\end{bmatrix}={\rm pf}[{\boldsymbol \xi}^{\uparrow\uparrow}]{\rm pf}[{\boldsymbol \xi}^{\downarrow\downarrow}],
\end{equation}
where we have introduced $M_1\times M_1(M_2\times M_2)$ matrices 
${\boldsymbol \xi}^{\uparrow\uparrow(\downarrow\downarrow)}=\left[\chi^{\uparrow\uparrow(\downarrow\downarrow)}(i,j)\right]$.
To our knowledge, this result was never explicitly stated and
only the weaker statement that the square of wave function simplifies to a product of determinants 
has been given~\cite{bouchaud1,bouchaud3}.

The connection to a restricted HF wave function for the above state can be again 
established as follows.
In accord with what we defined above,
 ${\rm det}[(C)^{\uparrow(\downarrow)}]$ are
spin-up(-down) Slater determinants of some HF orbitals $\{\varphi_i\}$.
Then, by taking advantage of  Eq.~(\ref{eq:pfident3}) we can write
\begin{align}
\Psi_{HF}&={\rm det}[C^{\uparrow}]{\rm det}[C^{\downarrow}]\\ \nonumber
&=\frac{{\rm pf}[C^{\uparrow} A_1 {C^{\uparrow}}^T]{\rm pf}[C^{\downarrow} A_2 {C^{\downarrow}}^T]}{{\rm pf}[A_1]{\rm pf}[A_2]},
\end{align}
given $A_1$ and $A_2$ are some skew-symmetric non-singular matrices.
In the simplest case, when $A_1$ and $A_2$ have block-diagonal form~(\ref{eq:pf:blockdiag}) with all values $\lambda_i=1$,
one gets
\begin{equation}
\Psi_{HF}={\rm pf}[\boldsymbol \xi_{HF}^{\uparrow\uparrow}]{\rm pf}[\boldsymbol \xi_{HF}^{\downarrow\downarrow}].
\end{equation}
The pair orbitals can be then expressed as
\begin{align}
(\boldsymbol \xi_{HF}^{\uparrow\uparrow(\downarrow\downarrow)})_{i,j}&=\chi_{HF}^{\uparrow\uparrow(\downarrow\downarrow)}(i,j)\\ \nonumber
&=\sum_{k=1}^{M_1(M_2)/2}(\varphi_{2k-1}(i)\varphi_{2k}(j)-\varphi_{2k-1}(j)\varphi_{2k}(i)).
\end{align} 
Similarly to the singlet pairing case, one can also think of the triplet pairing wave function 
as a natural generalization of the HF one. To do so we write the triplet pair orbitals as
\begin{align}\label{eq:chi}
\chi(i,j)^{\uparrow\uparrow(\downarrow\downarrow)}&=\sum_{k,l}^{M>M_1(M_2)}A^{\uparrow\uparrow(\downarrow\downarrow)}_{k,l}\varphi_k(i)\varphi_l(j) \nonumber \\
&={\boldsymbol \varphi}(i)\,{\bf A}^{\uparrow\uparrow(\downarrow\downarrow)}\,{\boldsymbol \varphi}(j),
\end{align}
where again the sum runs over all $M$ (occupied and virtual) single-particle orbitals and ${\bf A}^{\uparrow\uparrow(\downarrow\downarrow)}$ are some skew-symmetric
matrices. Therefore, we can define one-particle orbitals, which block-diagonalize these matrices and call them 
\emph{natural orbitals of a triplet spin-up-up(down-down) pair}. 

\subsection{Generalized Pairing Wave Function}
Let us now consider a partially spin-polarized system with
unpaired electrons. In order to introduce both types 
of pairing we allow $\chi^{\uparrow \uparrow}(i,j)$, $\chi^{\downarrow \downarrow}(i,j)$
and $\phi(i,j)$ in (\ref{eq:generalpair}) to be non-zero. 
However, we omit the $\chi^{\uparrow \downarrow}(i,j)$ term. 
Then our usual ordered choice of electrons labels with all spin-up electrons first 
and remaining electrons spin-down enables us to directly write from (\ref{eq:generalpairingwf}) the
singlet-triplet-unpaired (STU) orbital Pfaffian wave function~\cite{pfaffianprl}
\begin{equation}
\Psi_{STU}=
{\rm pf}\begin{bmatrix}
{\boldsymbol \xi}^{\uparrow\uparrow} & 
{\boldsymbol \Phi}^{\uparrow\downarrow} & 
{\boldsymbol\varphi}^{\uparrow} \\
-{\boldsymbol \Phi}^{\uparrow\downarrow T} &
{\boldsymbol \xi}^{\downarrow\downarrow} &
{\boldsymbol \varphi}^{\downarrow} \\
-{\boldsymbol\varphi}^{\uparrow T} &
-{\boldsymbol\varphi}^{\downarrow T} &
0 \;\; \\
\end{bmatrix},
\end{equation} 
where the bold symbols are block matrices or vectors of
corresponding orbitals as defined in Sections~\ref{subsec:singlet} and \ref{subsec:triplet} and $T$ denotes
transposition. For a spin-restricted STU wave function the
pair and one-particle orbitals of spin-up and -down channels
would be identical. 

The Pfaffian form can accommodate
both singlet and triplet pairs as well as one-particle
unpaired orbitals into a single, compact wave function.
The correspondence of STU Pfaffian wave function to HF wave function can 
be established in a similar way to the pure singlet and triplet pairing cases.

\section{Results}\label{sec:pf:results}
The Pfaffian wave functions were used 
in QMC calculations
by variational and fixed-node diffusion Monte Carlo.
As we have mentioned earlier,
the VMC trial  wave function is a product
of an antisymmetric part $\Psi_A$
times the Jastrow correlation factor
\begin{equation}
\Psi_{VMC} ({\bf R}) = \Psi_{A}({\bf R}) \exp[U_{corr}(\{r_{ij}\},\{r_{iI}\},\{r_{jI}\})],
\end{equation}
where  $U_{corr}$ depends on
electron-electron, electron-ion and, possibly, on 
electron-electron-ion combinations of distances as described in Sec.~\ref{subsec:twf}.
For the antisymmetric part we
have used $\Psi_A=\Psi_{HF}$ and $\Psi_A=\Psi_{STU}$. 
Some tests were also done with $\Psi_A=\Psi_{BCS}$ to compare with recent 
results~\cite{sorellabcs1,sorellabcs2}.
The pair orbitals were expanded  
in products of a one-particle orbital
basis~\cite{sorellabcs1} according to Eqs.~(\ref{eq:phi}) and (\ref{eq:chi}).
The expansions include both occupied and unoccupied
(virtual) one-particle orbitals.
The one-particle atomic and molecular
orbitals used were either Hartree-Fock orbitals or natural orbitals~\cite{szabo} from 
CI correlated calculations. 
Typically, we used about 10
virtual orbitals. The natural orbitals produced better and more systematic
results than the HF ones.
The pair orbital expansion coefficients were then optimized
in VMC by minimizations of energy, variance or a combination
of energy and variance (for details, see Sec.~\ref{subsec:opt3}). The optimization procedure 
requires the calculation of gradient and the Hessian of the wave function 
according to Eqs.~(\ref{eq:gradient}) and (\ref{eq:hessian}). 

\subsection{Single Pfaffian Calculations}\label{subsec:pf:spf}

\begin{figure}[!ht]
\centering
\includegraphics[width=\columnwidth]{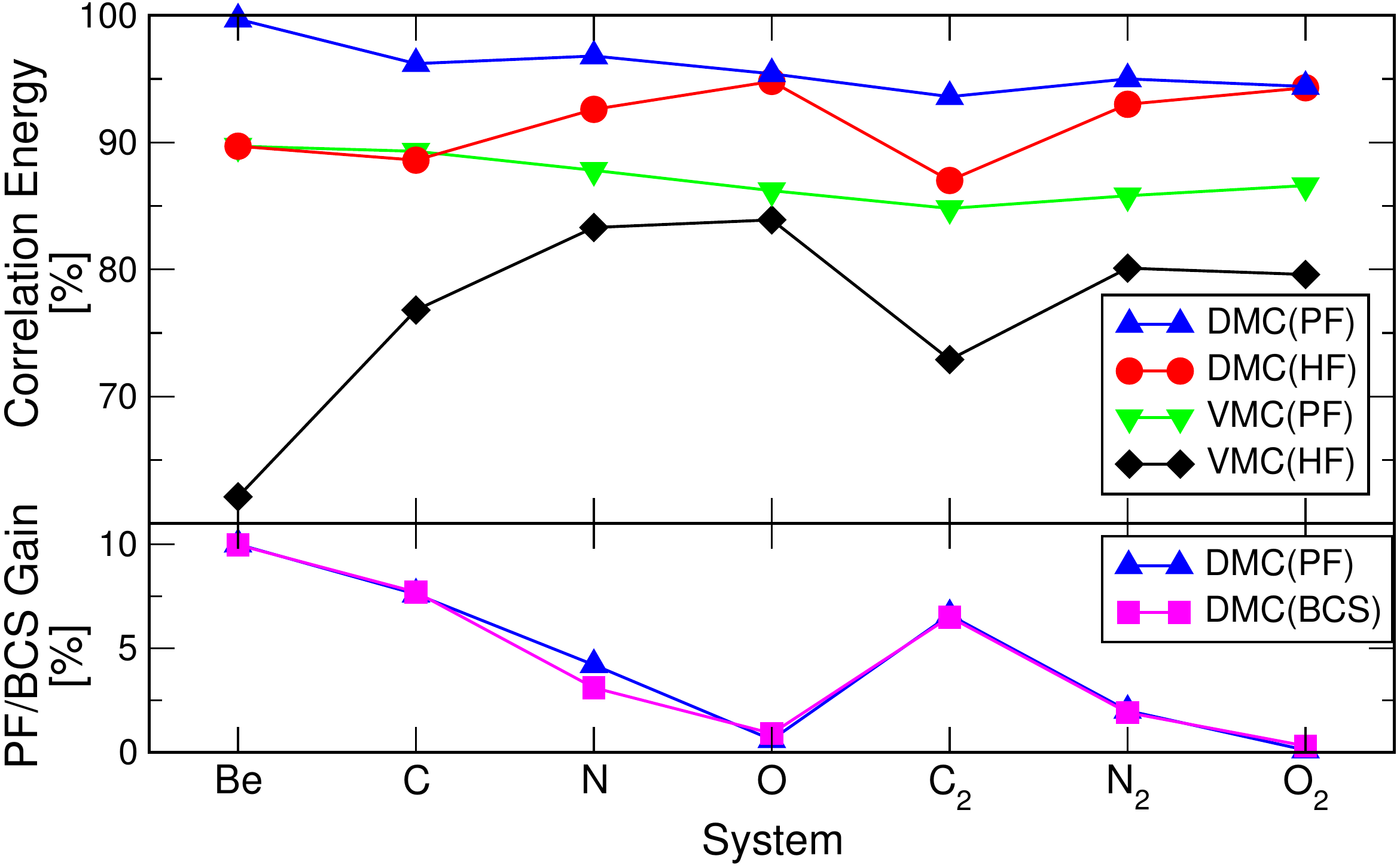}
\caption{
Correlation energies obtained by QMC methods
with the different trial wave functions: VMC
and fixed-node DMC with HF nodes (HF) and STU Pfaffian nodes (PF).
The lower plot shows the fixed-node DMC correlation energy gains
over HF nodes for BCS and STU Pfaffian
wave functions.
The statistical error bars are of the symbol sizes or smaller.
Except for the Be atom all the calculations used the same pseudopotentials~\cite{lester,stevens}.
\label{fig:singlepfresults}}
\end{figure}

\begin{table}[!t]
\caption{ Total energies for C, N and O atoms and their dimers with 
amounts of the correlation energy recovered in VMC and DMC methods with wave functions 
as discussed in the text. Unless noted otherwise, 
the numbers in parentheses are the statistical errors in the last digit from corresponding QMC calculation.
Energies are in Hartree atomic units.
For C, N, O atoms we used the correlation energies by Dolg~\cite{dolgcpl}(0.1031, 0.1303, 0.1937 H).
For the estimation of correlation energies of dimers we needed accurate HF energies at experimental distances~\cite{NIST-JANAF}
and the estimated exact total energies. 
Each exact total energy was estimated as a sum of total energies of constituent atoms minus experimental binding energy~\cite{NIST-JANAF,UBJ,HH}
adjusted for experimental zero-point energy~\cite{HH}.} 
\begin{minipage}{\columnwidth}
\renewcommand{\thefootnote}{\alph{footnote}}
\renewcommand{\thempfootnote}{\alph{mpfootnote}}
\centering
\begin{tabular}{l c c c c c c }
\hline
\hline
 \multicolumn{1}{l}{Method/WF}&\multicolumn{1}{c}{C}&\multicolumn{1}{c}{E$_{corr}$[\%]} & \multicolumn{1}{c}{N} &\multicolumn{1}{c}{E$_{corr}$[\%]}&\multicolumn{1}{c}{O}& \multicolumn{1}{c}{E$_{corr}$[\%]}\\
\hline 
HF                & -5.31471   &   0     & -9.62892   & 0       & -15.65851   &  0      \\[-0.6em]
VMC/HF            & -5.3939(4) & 76.8(4) & -9.7375(1) & 83.3(1) & -15.8210(6) & 83.9(3) \\[-0.6em]
VMC/BCS           & -5.4061(2) & 88.6(2) & -9.7427(3) & 87.3(2) & -15.8250(3) & 86.0(2) \\[-0.6em]
VMC/STU           & -5.4068(2) & 89.3(2) & -9.7433(1) & 87.8(1) & -15.8255(3) & 86.2(2) \\[-0.6em]
DMC/HF            & -5.4061(3) & 88.6(2) & -9.7496(2) & 92.6(2) & -15.8421(2) & 94.8(1) \\[-0.6em]
DMC/BCS           & -5.4140(2) & 96.3(2) & -9.7536(2) & 95.7(2) & -15.8439(4) & 95.7(2) \\[-0.6em]
DMC/STU           & -5.4139(2) & 96.2(2) & -9.7551(2) & 96.8(1) & -15.8433(3) & 95.4(2) \\[-0.6em]
Est./Exact        & -5.417806  & 100     & -9.759215  & 100     & -15.85216   & 100     \\
\hline 
\multicolumn{1}{l}{Method/WF} &  \multicolumn{1}{c}{C$_2$} &  \multicolumn{1}{c}{E$_{corr}$[\%]} &  \multicolumn{1}{c}{N$_2$} &  \multicolumn{1}{c}{E$_{corr}$[\%]} &  \multicolumn{1}{c}{O$_2$} &  \multicolumn{1}{c}{E$_{corr}$[\%]} \\
\hline 
HF                & -10.6604    & 0       & -19.4504   &  0       & -31.3580    & 0      \\[-0.6em]
VMC/HF            & -10.9579(4) & 72.9(1) & -19.7958(5)&  80.0(1) & -31.7858(6) & 79.6(1)\\[-0.6em]
VMC/BCS           & -11.0059(4) & 84.7(1) & -19.8179(6)&  85.0(1) & -31.8237(4) & 86.7(1)\\[-0.6em]
VMC/STU           & -11.0062(3) & 84.8(1) & -19.821(1) &  85.8(2) & -31.8234(4) & 86.6(1)\\[-0.6em]
DMC/HF            & -11.0153(4) & 87.0(1) & -19.8521(3)&  93.0(1) & -31.8649(5) & 94.3(1)\\[-0.6em]
DMC/BCS           & -11.0416(3) & 93.5(1) & -19.8605(6)&  94.9(1) & -31.8664(5) & 94.6(1)\\[-0.6em]
DMC/STU           & -11.0421(5) & 93.6(1) & -19.8607(4)&  95.0(1) & -31.8654(5) & 94.4(1)\\[-0.6em]
Est./Exact\footnotemark[3]  & -11.068(5)\footnotemark[1]  & 100.0(10)  & -19.8825(6)\footnotemark[2]   &  100.0(1)     & -31.8954(1)\footnotemark[2]    & 100.0(1)    \\
\hline
\hline
\end{tabular}
\footnotetext[1] { 
There is rather large discrepancy in the experimental values of C$_2$ binding energy
($141.8(9)$~\cite{NIST-JANAF}, $143(3)$~\cite{HH} and $145.2(5)$ kcal/mol~\cite{UBJ}). 
For the estimation of exact energy we have taken the average value of $143(3)$ kcal/mol.}
\footnotetext[2] {Experimental binding energies taken from ref.~\cite{NIST-JANAF}.}
\footnotetext[3] {The error bars on estimated exact total energies are due to experiment.}
\end{minipage}
\label{energies:1}
\end{table}
We have applied these developments to several first row 
atoms and dimers (see Table \ref{energies:1} and Fig.~\ref{fig:singlepfresults}). We used pseudopotentials 
to eliminate the atomic cores~\cite{lester,stevens}, while 
the previous all-electron calculations with BCS wave functions~\cite{sorellabcs1,sorellabcs2} produced percentages 
of the correlation energies in accordance with our BCS wave functions calculations.

Perhaps the most striking result is a systematic percentage of recovered
correlation energy on the level of 94-97\% in DMC method for the STU wave functions
(see Table \ref{energies:1} and Fig.~\ref{fig:singlepfresults}).
Another significant result is that in general the triplet contribution for
these single Pfaffian STU wave functions are small, with the only
exception being nitrogen atom, where we see a
gain of  additional  1\% in correlation energy when compared
to a trial wave function without triplet pairs.
We believe, this is due to the fact, that  
ground state of nitrogen atom has a quartet spin state and 
therefore the highest spin polarization from all studied cases. 
However, only future tests with the presence of non-zero 
$\chi^{\uparrow \downarrow}(i,j)$ in (\ref{eq:generalpair}) 
will determine the full extent of the triplet contribution to the correlation effects. 
Given the pair orbitals have been optimized using VMC method, it is 
natural that the relative gains in correlation energy 
with respect to the HF wave functions are larger on the level of VMC calculations
than in DMC calculations. 
Overall, the single Pfaffian form is 
capable of capturing near-degeneracies
and mixing of excited states for both
spin-polarized and unpolarized systems.

\subsection{Multi-Pfaffian Calculations}\label{subsec:pf:mpf}
\begin{table}
\caption{Percentages of correlation energies recovered 
for C, N and O atoms by VMC and DMC methods 
with wave functions as discussed in the text.
The corresponding number of Pfaffians or determinants $n$ for each
wave function is also shown. For details, see caption of Table~\ref{energies:1}.}
\begin{minipage}{\columnwidth}
\renewcommand{\thefootnote}{\alph{footnote}}
\renewcommand{\thempfootnote}{\alph{mpfootnote}}
\centering
\begin{tabular}{l c c c c c c}
\hline
\hline
\multicolumn{1}{l}{Method/WF}&
\multicolumn{1}{c}{$n$}&\multicolumn{1}{c}{C} &\multicolumn{1}{c}{$n$}&\multicolumn{1}{c}{N}& \multicolumn{1}{c}{$n$}&\multicolumn{1}{c}{O}\\
\hline 
VMC/MPF & 3   & 92.3(1) & 5   & 90.6(1) &  11  & 92.6(3) \\ 
VMC/CI\footnotemark[1]  & 98  & 89.7(4) & 85  & 91.9(2) & 136  & 89.7(4) \\ 
DMC/MPF & 3   & 98.9(2) & 5   & 98.4(1) &  11 &  97.2(1) \\ 
DMC/CI\footnotemark[1]  & 98  & 99.3(3) & 85  & 98.9(2) & 136  & 98.4(2) \\
\hline
\end{tabular}
\footnotetext[1]{The determinantal weights were taken directly from CI calculation without re-optimization in VMC.}
\end{minipage}
\label{energies2}
\end{table}

\begin{table}
\caption{Total energies for C$_2$ and N$_2$ dimers with 
amounts of correlation energy recovered in VMC and DMC methods
with wave functions as discussed in the text. Energies are in Hartree atomic units.
The corresponding number of Pfaffians or determinants $n$ for each
wave function is also shown. For details, see caption of Table~\ref{energies:1}.}
\begin{minipage}{\columnwidth}
\renewcommand{\thefootnote}{\alph{footnote}}
\renewcommand{\thempfootnote}{\alph{mpfootnote}}
\centering
\begin{tabular}{l c c c c c c}
\hline 
\hline
\multicolumn{1}{l}{Method/WF} & \multicolumn{1}{c}{$n$} & \multicolumn{1}{c}{C$_2$} &  \multicolumn{1}{c}{E$_{corr}$[\%]} &  \multicolumn{1}{c}{$n$} &  
\multicolumn{1}{c}{N$_2$} & \multicolumn{1}{c}{E$_{corr}$[\%]}\\
\hline  
VMC/MPF                 &  5   &  -11.0187(2) & 87.8(1) &  5   &  -19.8357(3)   &  89.2(1)\\
VMC/AIP                 &  4!  &  -11.0205(4) & 88.3(1)  &  5!  &  -19.8350(3)   &  89.0(1)\\
VMC/CI\footnotemark[1]  & 148  &  -11.0427(1) & 93.7(1) & 143  &  -19.8463(9)   &  91.6(2)\\
DMC/MPF                 &  5   &  -11.0437(4) & 94.0(1)  &  5   &  -19.8623(5)   &  95.3(1)\\
DMC/AIP                 &  4!  &  -11.0435(7) & 94.0(2)  &  5!  &  -19.8611(3)   &  95.0(1)\\
DMC/CI\footnotemark[1]  & 148  &  -11.0573(2)\footnotemark[2]  & 97.3(1) & 143  &  -19.875(2)    &  98.3(5)\\
\hline
\end{tabular}
\footnotetext[1]{The determinantal weights were re-optimized in the VMC method.}
\footnotetext[2]{Recently, Umrigar~{\it et.al.}~\cite{umrigarC2} published very accurate 
DMC result for fully optimized CI wave function with up to 500 determinants for C$_2$ molecule. 
The resulting well-depth of his calculation is $6.33(1)$ eV, which is only $0.03$ eV form estimated exact value of Ref.~\cite{bytautas}.
The well-depth resulting from our DMC/CI energy of $-11.0573(2)$ H equals to $6.03(1)$ eV.}
\end{minipage}
\label{energies3}
\end{table}


To test the limits of the Pfaffian functional form,
we have proposed a simple extension: the multi-Pfaffian (MPF)
wave function of the form
\begin{align}\label{eq:mpf}
\Psi_{MPF}&=
w_1{\rm pf}[\chi^{\uparrow\uparrow}_1,\chi^{\downarrow\downarrow}_1,\phi_1,\varphi_1]+
w_2{\rm pf}[\chi^{\uparrow\uparrow}_2,\chi^{\downarrow\downarrow}_2,\phi_2,\varphi_2]+\ldots,
\end{align}
where $w_i$ denotes the weight of $i$th Pfaffian.
In order to improve upon the wave function with single STU Pfaffian, the 
additional terms in wave function (\ref{eq:mpf}) have to contain some new excitations not previously present.
As an example of this form, we apply it to the carbon pseudo-atom. 
The pair orbitals, Eqs.~(\ref{eq:phi}) and (\ref{eq:chi}), for this system are  
expanded in the basis of HF occupied orbitals $2s$, $2p_x$ and $2p_y$. 
The choice of singlet
$\phi_1(1,2)=2s(1)2s(2) \equiv \phi_1[2s,2s]$ and spin-up spin-up triplet
$\chi^{\uparrow\uparrow}_1(1,2)=2p_x(1)2p_y(2)-2p_y(1)2p_x(2) \equiv \chi^{\uparrow\uparrow}_1[2p_x,2p_y]$ 
pair orbitals (the other functions are taken to be zero) 
then gives ${\rm pf}[\chi^{\uparrow\uparrow}_1,\phi_1]=\Psi_{HF}[2s^{\uparrow\downarrow},2p_x^{\uparrow},2p_y^{\uparrow}]$. 
However, one can construct the equivalent combinations of pairs as 
$\phi_2[2s,2p_x]$, $\chi^{\uparrow\uparrow}_2[2s,2p_y]$ and  $\phi_3[2s,2p_y]$, $\chi^{\uparrow\uparrow}_3[2s,2p_x]$.
We can therefore include all three Pfaffians into our $\Psi_{MPF}$ and further optimize all the pairing functions 
in VMC method on the space of occupied and virtual orbitals.

Since each pair orbital in $\Psi_{MPF}$ 
contains on the order of $M^2$ pairing coefficients, $M$ being the total number of
one particle orbitals involved,
we limit our expansions to only few Pfaffians. 
However, this can be improved by factor $M$, if we diagonalize the coefficient matrices.
In practice, given the optimization routine in VMC method can safely minimize 
on the order of a few hundred coefficients at the same time, we 
end up doing several partial optimizations. To minimally disrupt 
the weights in partially expanded MPF wave function, any new Pfaffians 
are added in pairs, each initially set to the HF wave function with opposite sign for zero net contribution. 
Then the Pfaffian pair is re-optimized on the set of all single-particle orbitals. Besides
already optimized STU Pfaffian we have an even number of additional 
Pfaffians, which explains the overall odd number of Pfaffians in our MPF expansions.

The results in Table~\ref{energies2} show
that for the atomic systems our MPF wave functions are able to
recover close to  99\% of correlation energy.
Furthermore, comparison with the CI results demonstrates
it is possible to obtain similar quality wave functions with corresponding
improvements of the fermion nodes
at much smaller calculational cost.
However, for the diatomic cases (see Table~\ref{energies3}), 
only very limited gain over single STU Pfaffian wave function correlation 
energies were achieved for MPF wave functions with few Pfaffians. 
We therefore conclude that for obtaining significantly larger gains in correlation
the molecular wave functions require much larger expansions. 


\subsection{Antisymmetric Independent Pairs Wave Function}\label{subsec:pf:aip}
We have also tested the fully antisymmetric independent pairs (AIP) wave function
which introduces one pair orbital per each electron pair. 
For system of $2N$ fermions in singlet state the AIP wave function can be written as
\begin{align}
\Psi_{AIP}&={\mathcal A} [\tilde{\phi}_1 (1,2), \tilde{\phi}_2(3,4), \ldots,\tilde{\phi}_N (2N-1,2N)]\\ \nonumber
&=\sum_{P}{\rm pf}[\tilde{\phi}_{i_1}(1,2), \tilde{\phi}_{i_2}(3,4), \ldots,\tilde{\phi}_{i_N} (2N-1,2N)], 
\end{align}
where the last equation corresponds to the sum over all $N!$ possible permutations 
of $N$ different pair orbitals $\tilde{\phi}_{i}$ for each Pfaffian. 
This wave function is closely related to the wave function 
of an antisymmetrized product of strongly orthogonal geminals~\cite{rassolov}.
The results for C$_2$ and N$_2$ dimers using AIP wave function are given in Table~\ref{energies3}. 

Consideration of independent pairs results in an exponential increase of a number of Pfaffians.
However, captured correlation energy is on the level of small MPF expansion, and 
significantly less than CI with re-optimized weights using the same one-particle orbitals.  
This suggests that to achieve more correlation energy in larger systems 
we have to go beyond double pairing. 

\subsection{Nodal Properties}\label{subsec:pf:nodes}
\begin{figure}[!t]
   \mbox{
       {\resizebox{2in}{!}{\includegraphics{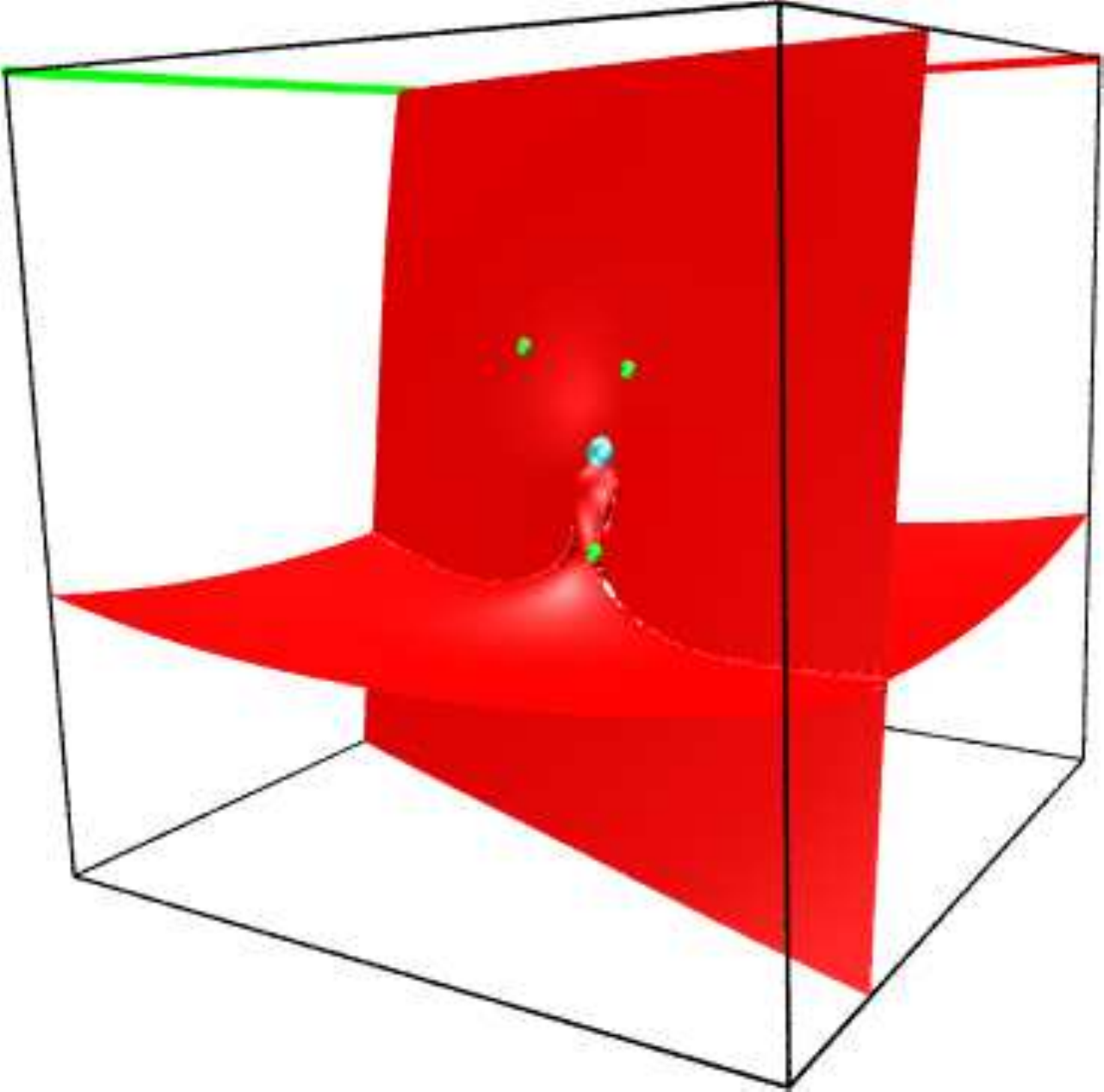}}}
       {\resizebox{2in}{!}{\includegraphics{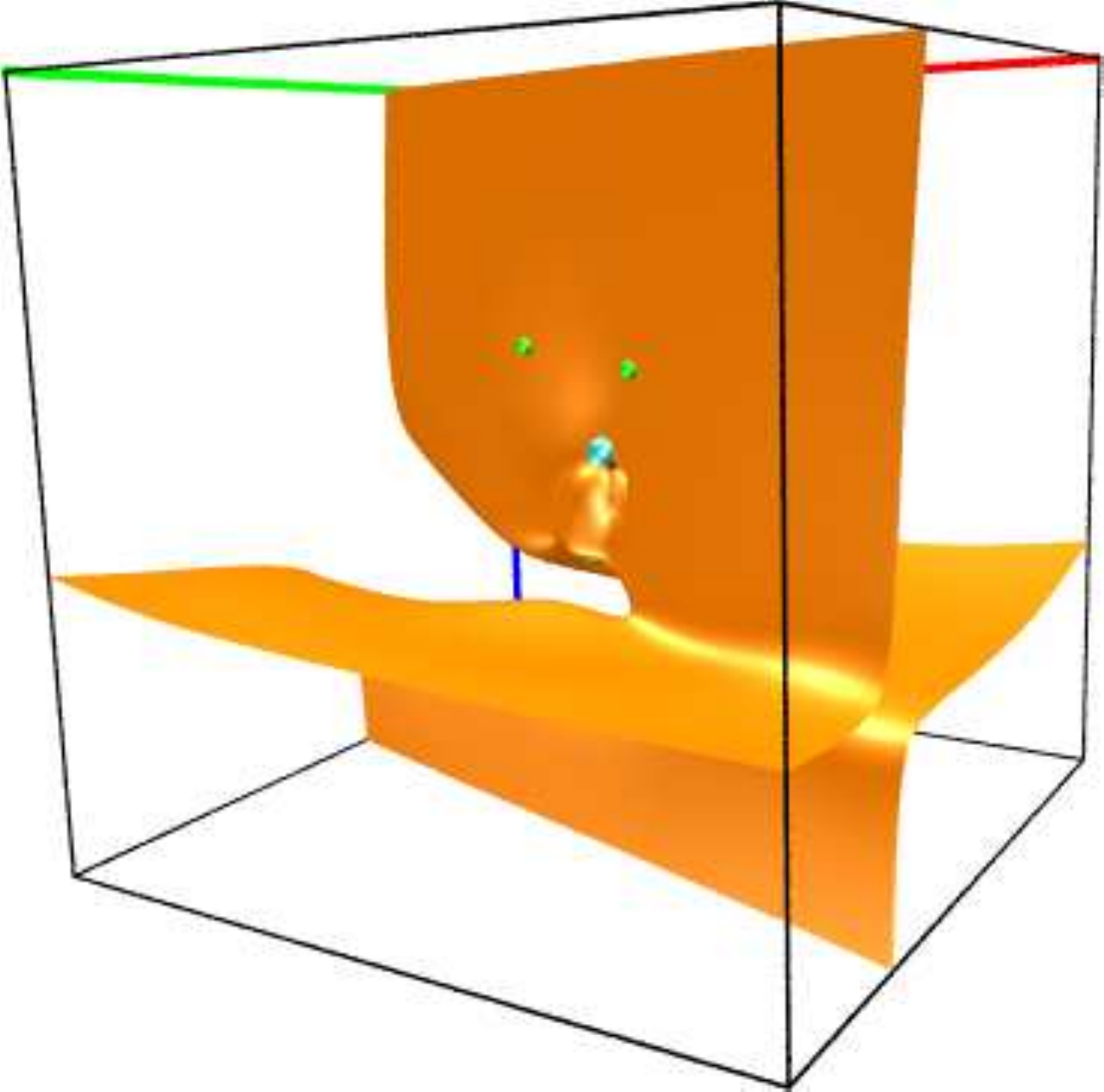}}} 
       {\resizebox{2in}{!}{\includegraphics{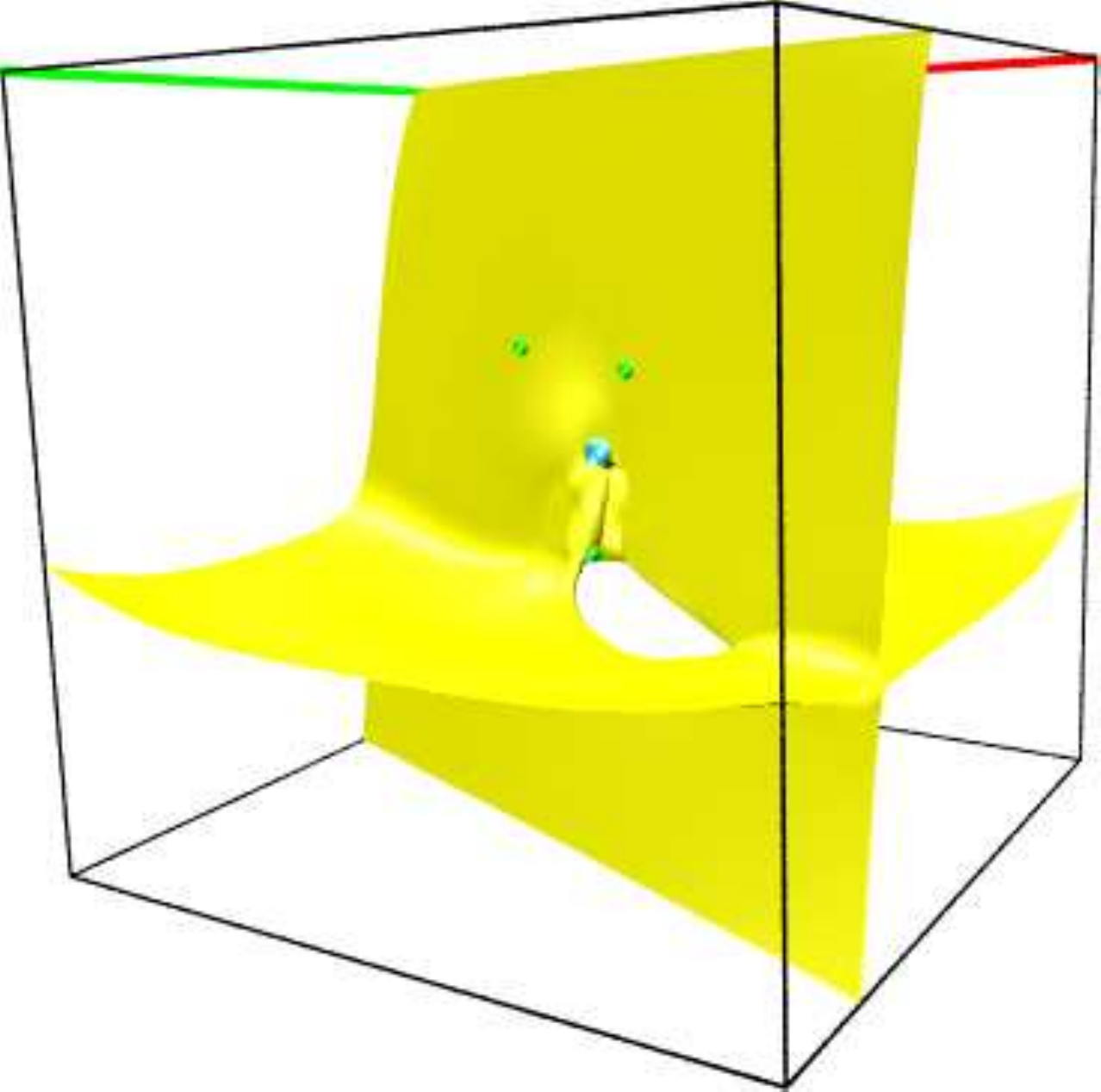}}}
      }
    \mbox{
       {\resizebox{2in}{!}{\includegraphics{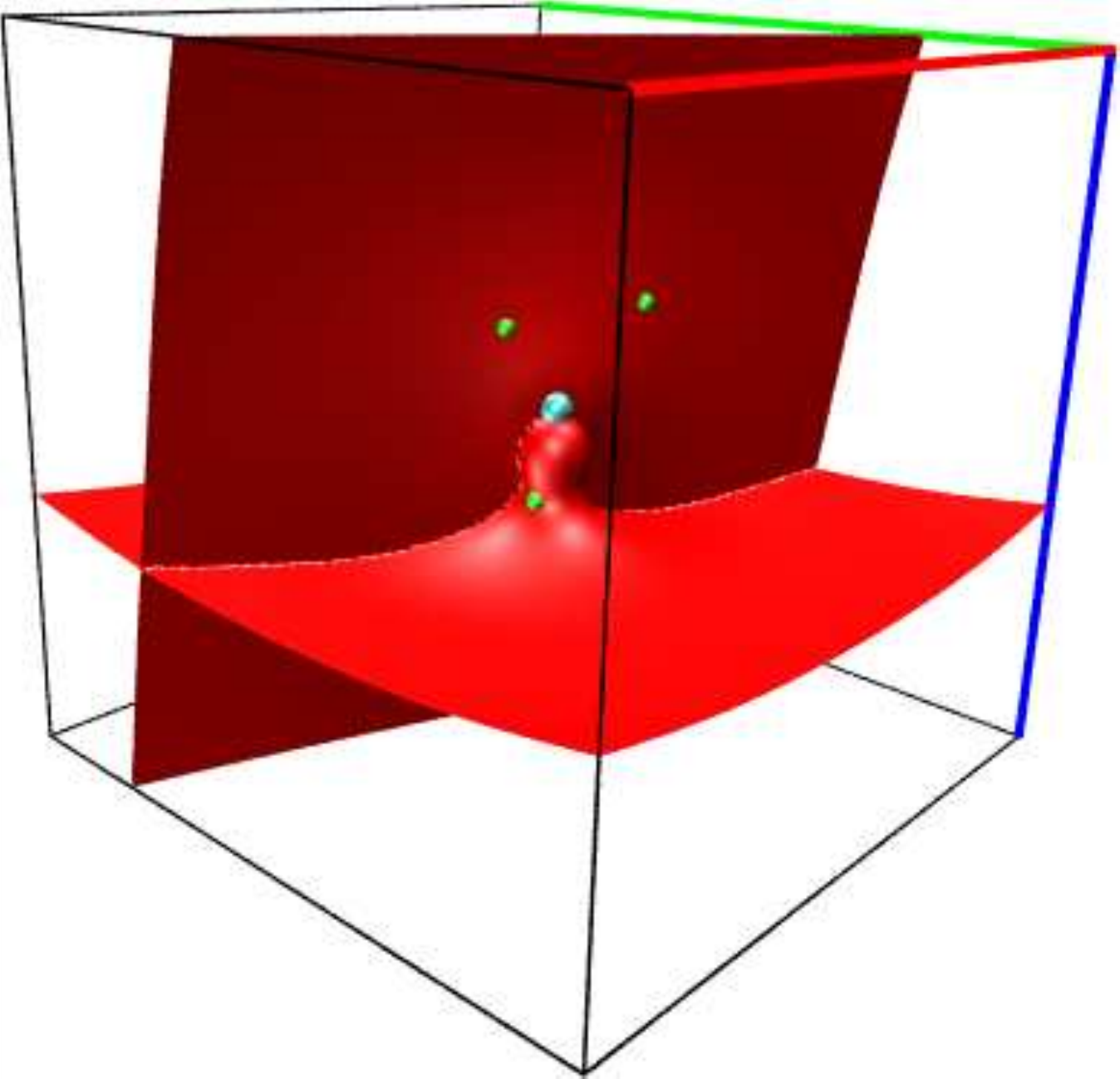}}}
       {\resizebox{2in}{!}{\includegraphics{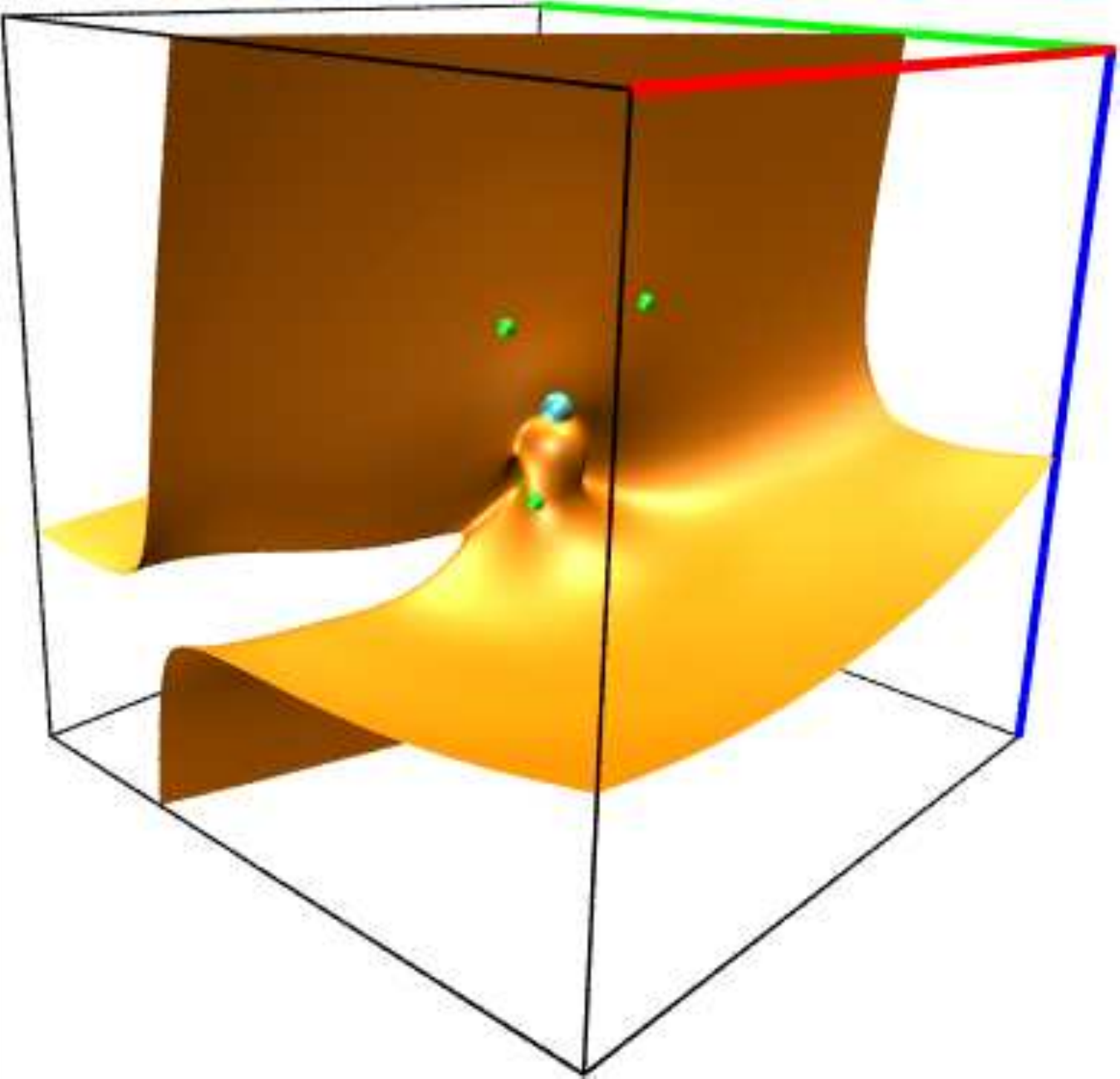}}}
       {\resizebox{2in}{!}{\includegraphics{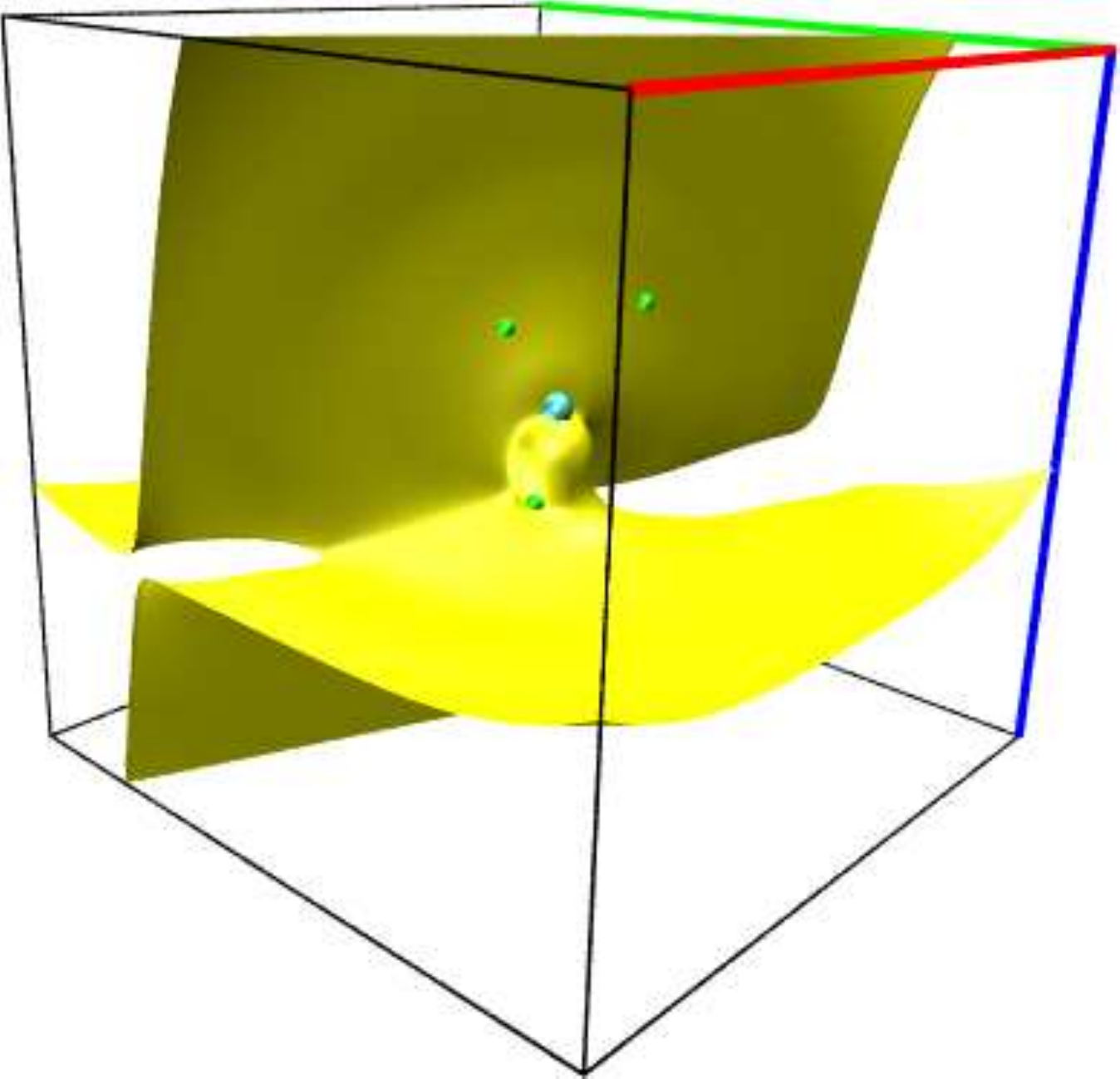}}}
     }
\caption{
A three-dimensional cut through the fermion node hypersurface of oxygen atom 
obtained by scanning the wave function with a spin-up and -down (singlet)
pair of electrons at the equal positions, while keeping the 
rest of electrons at a given VMC snapshot positions (small green spheres). 
Nucleus is depicted in the center of the cube by the blue 
sphere. The three colors (from left to right) show nodes of: Hartree-Fock (red/dark gray);
Multi-Pfaffian nodes (orange/medium gray); and the nodes of the CI
wave function (yellow/light gray) in two different views (upper and lower rows).
The CI nodal surface is very close to the exact one (see text).
The HF node clearly divides the space into four
nodal cells while Pfaffian and CI wave functions
partitioning leads to the minimal number of two nodal cells.
The changes in the nodal topology occur on the appreciable spatial 
scale of the order of 1 a.u.}
\label{fig:nodes2}
\end{figure}
As we have already mentioned in Ch.~\ref{ch:nodes}, the fermion node manifold is defined by an implicit equation
$\Psi(R)=0$ and for $N$ electrons it is a \mbox{$(3N-1)$-dimensional} hypersurface.
With exception of a few exact cases, the nodes of trial wave functions introduce bias into the fixed-node DMC energies. 
Recently a number of authors have reported improvement in nodal structure of trial wave functions 
~\cite{Bressanininew,sorellabcs1,sorellabcs2,pfaffianprl,umrigarC2,drummond_bf,rios_bf}.

The effect of pairing correlations on nodes can be highlighted by direct comparison.
The Fig.~\ref{fig:nodes2} shows the example of nodal structure of oxygen atom. 
Here we compare the nodal surfaces of HF (no pairing), MPF Pfaffian (STU pairing) and a
high accuracy CI wave function with more than 3000 determinants, which gives 
essentially exact fermion nodes (i.e., $99.8(3)\%$ of correlation energy in fixed-node DMC).

It is clear that the changes in the nodal surfaces are significant, 
the most important one being the elimination of artificial 
four nodal cells resulting from the independence of spin-up and -down
channels in HF. The Pfaffian smooths-out the crossings and fuses
the compartments of the same sign into the single ones. These topology changes
therefore lead to the minimal number of two nodal cells, an effect observed
in correlated context previously~\cite{davidnode,dariobe,darionew,lubos_nodeprl,lubos_nodeprb}.
However, the nodes of the Pfaffian wave functions could be further improved if
the scheme for direct optimization of nodes of trial 
wave functions were used~\cite{carlsonbcs,chang2005}. 
Another interesting result from our work is that despite such a substantial change
in the nodal structure the amount of missing correlation energy is still non-negligible. 

\section{\label{sec:level5} Conclusions}
To summarize, we have proposed Pfaffians with
singlet pair, triplet pair and unpaired
orbitals as variationally rich and compact wave functions.
They offer significant and systematic improvements
over commonly used Slater determinant-based wave functions. 
We have included a set of key mathematical 
identities with proofs, which are needed 
for the evaluation and update of the Pfaffians.
We have also shown connections of HF and BCS (AGP) wave functions to more general 
Pfaffian wave function.
We have further demonstrated that 
Pfaffian pairing wave functions are able to capture a large fraction of
missing correlation energy with consistent treatment 
of both spin-polarized and unpolarized pairs.
We have explored multi-Pfaffian wave functions
which enabled us to capture additional correlation.
While for atomic systems the results are comparable to large-scale CI wave functions, 
molecular systems most probably require much larger multi-Pfaffian expansions
than we have explored. As another test of the variational potential of
pairing we have employed the fully-antisymmetrized independent pairs wave function 
in Pfaffian form and we have found that it does not lead to additional gains 
in correlation energy. 
We therefore conclude that more general functional forms 
together with more robust large-scale optimization methods might be necessary 
in order to obtain further improvements. 
The gains in correlation energy for Pfaffians come from
improved fermion nodes, which are significantly closer to the exact ones than the HF nodes, and exhibit 
the correct topology with the minimal number of two nodal cells.

\chapter{Backflow Correlations in Slater and Pfaffian Wave Functions}\label{ch:bf}
\section{Introduction}
Another approach for improvement of the nodal accuracy of variational trial wave functions 
is to employ backflow correlations first introduced by Feynman and Cohen~\cite{feynman}
for liquid $^4$He. Since then, a number of authors~\cite{schmidt_bf,panoff,moskowitz,kwon1,kwon2,kwon3}
showed that the backflow correlations are helpful also in fermionic systems. 
Recently, the backflow transformation of electron coordinates has been applied to 
chemical (inhomogeneous) systems~\cite{drummond_bf,rios_bf}. 
In this chapter, we report on application of backflow transformation, already introduced in the Sec.~\ref{subsec:twf},
to two simple but nontrivial testing cases: carbon atom and its dimer. 
It is the first application of backflow correlations to the multi-determinantal and Pfaffian wave functions.

\section{Backflow Wave Function Form}
As already discussed in Sec.~\ref{subsec:twf}, our trail wave function has the form
\begin{align}
\Psi_T({\bf R})=\Psi_A({\bf X})\times \exp[U_{corr}({\bf R})],
\end{align}
where ${\bf X}=({\bf x}_1,\ldots,{\bf x}_N)$ represents some quasi-particle coordinates 
dependent on all $N$ electron positions ${\bf R}$. Further, $\Psi_A$
is either (multi)-determinantal or Pfaffian wave function and $U_{corr}$
is the Jastrow correlation factor both defined in the previous chapters. 

The quasi-coordinate of $i$th electron at position ${\bf r}_i$ is given as 
\begin{align}
{\bf x}_i&={\bf r}_i+{\boldsymbol \xi}_i({\bf R}) \nonumber \\
&={\bf r}_i+{\boldsymbol \xi}_i^{en}({\bf R})+{\boldsymbol \xi}_i^{ee}({\bf R})+{\boldsymbol \xi}_i^{een}({\bf R}),
\end{align}
where ${\boldsymbol \xi}_i$ is the $i$th electron's backflow displacement 
divided to the contributions from one-body (electron-nucleus), two-body (electron-electron) 
and three-body (electron-electron-nucleus) terms. 
They can be further expressed as 
\begin{align}\label{eg:bfterms}
{\boldsymbol \xi}_i^{en}({\bf R})&=\sum_I \chi(r_{iI}) {\bf r}_{iI} \\
{\boldsymbol \xi}_i^{ee}({\bf R})&=\sum_{j\ne i} u(r_{ij}) {\bf r}_{ij} \\
{\boldsymbol \xi}_i^{een}({\bf R})&=\sum_I \sum_{j\ne i} [w_1(r_{ij},r_{iI},r_{jI}) {\bf r}_{ij} + w_2(r_{ij},r_{iI},r_{jI}) {\bf r}_{iI}],
\end{align}
where ${\bf r}_{ij}={\bf r}_i-{\bf r}_j$ and ${\bf r}_{iI}={\bf r}_i-{\bf r}_I$. The $\chi$, $u$ and $w_1$ with $w_2$
terms are similar to one, two and three-body Jastrow terms and are further expanded 
as 
\begin{align}
\chi(r)&=\sum_k c_k a_k(r), \\
u(r)&=\sum_k d_k b_k(r), \\
w_{1,2}(r_{ij},r_{iI},r_{jI})&=\sum_{klm} g_{klm} a_k(r_{iI})a_l(r_{jI})b_m(r_{ij}). 
\end{align}
The one dimensional basis functions $\{a\}$ and $\{b\}$ are chosen as Gaussians with the center in origin to 
preserve the electron-electron [Eqs.~(\ref{eq:cupsupup}) and~(\ref{eq:cupsupdown})] and electron-nucleus [Eq.~(\ref{eq:necusp})] cusp conditions.
The set of variational parameters $\{c\}$, $\{d\}$ and $\{g\}$ is minimized with respect to 
mixture of energy and variance in NEWTON\_OPT method from Sec.~\ref{subsec:opt3}. In addition, all electron-electron 
coefficients ($\{d_k\}$ and $\{g_{klm}\}$ with fixed $k$ and $l$) are allowed to be different for spin-like and for spin-unlike electron pairs. 

\section{Results}
\begin{figure}[!t]
\begin{center}
\begin{minipage}{\columnwidth}
\includegraphics[width=0.55\columnwidth]{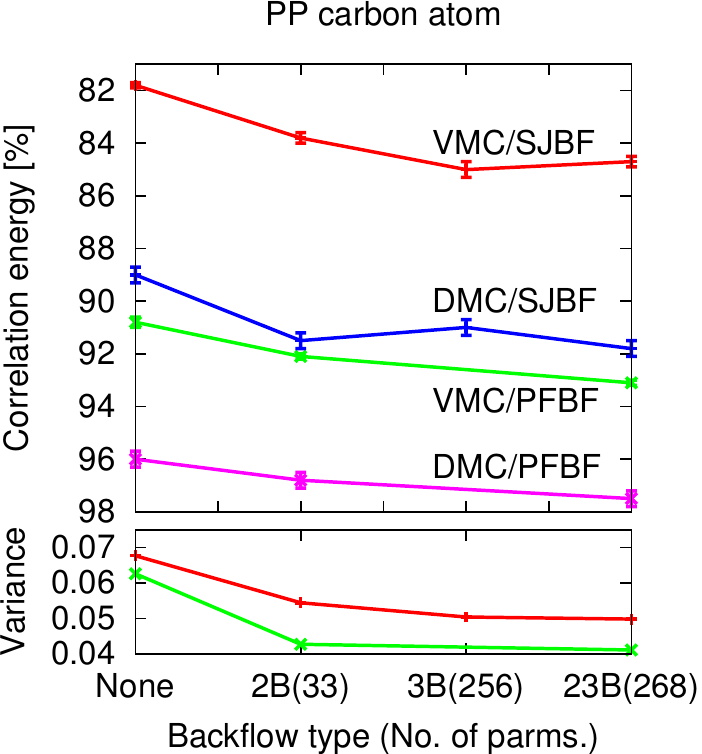}
\includegraphics[width=0.45\columnwidth]{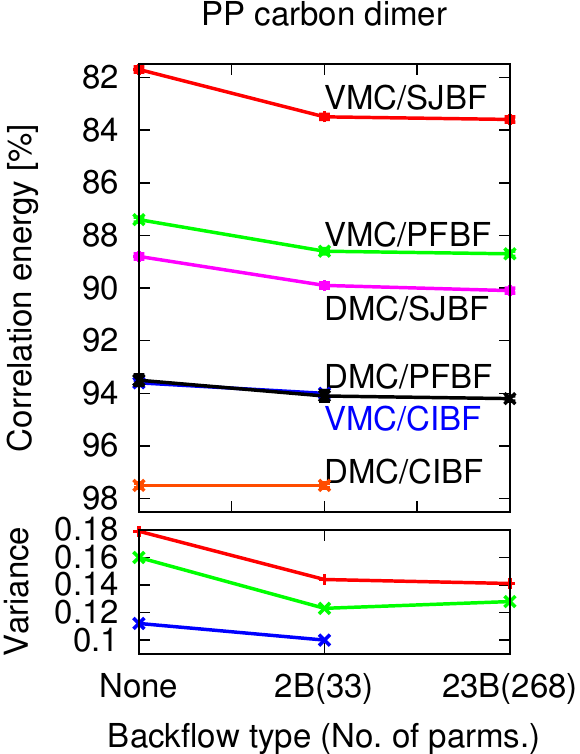}
\end{minipage}
\end{center}
\caption{Slater-Jastrow (SJ), Pfaffian-Jastrow (PF) and CI-Jastrow (CI) wave functions with backflow (BF) 
correlations for carbon pseudopotential (He core) atom (left) and dimer (right) tested in VMC and DMC methods.
Upper figure: Percentages of correlation energy versus a number of backflow parameters. Notation: 2B for all electron-nucleus and electron-electron terms,
3B for all electron-electron-nucleus terms only and 23B for all terms together.
Lower figure: Variance of the local energy versus a number of terms.}
\label{fig:C_bf}
\end{figure}

We test the above backflow correlation function on carbon pseudopotential (He core) atom and its dimer. 
The main results are summarized in Fig.~\ref{fig:C_bf} and can also be found in full detail in Tables~\ref{table:bf:C}
and~\ref{table:bf:C2}. 
The backflow correlations are able to capture additional few percent of correlation energy 
for both Slater-Jastrow and Pfaffian-Jastrow wave functions. Another important feature of backflow is 
20-30\% decrase in variances of local energy with respect to the wave functions without backflow correlations. 
The gains are systematic with increasing number of parameters, however we do not 
find the three-body terms as important as previous study~\cite{rios_bf}. This can be attributed to the fact that 
their main testing case was an all-electron system where the three-body correlations are known to be more significant.

\begin{figure}[!t]
\begin{center}
\begin{minipage}{\columnwidth}
\includegraphics[width=0.5\columnwidth]{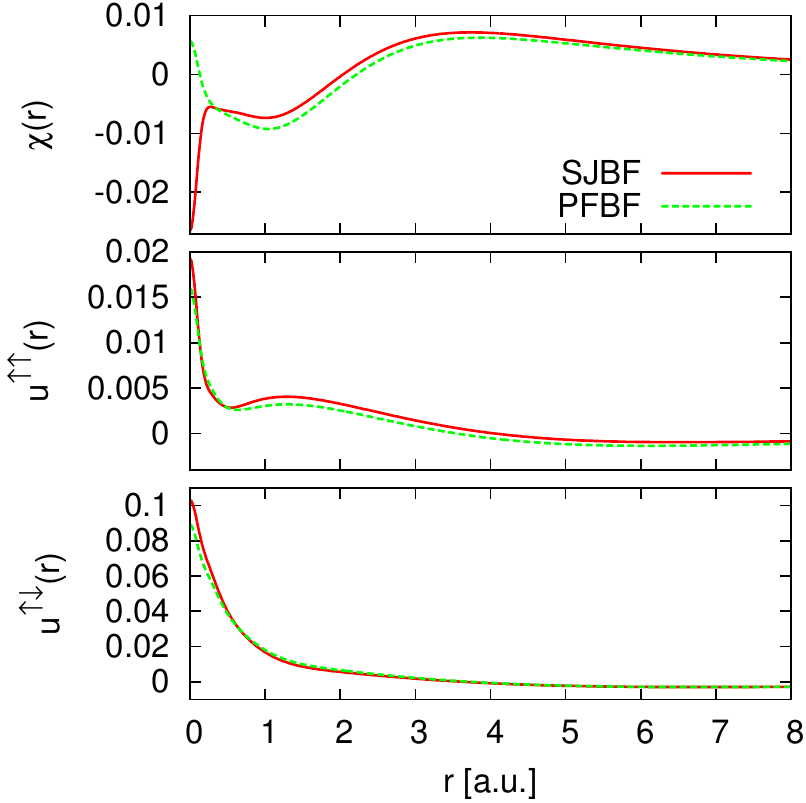}
\includegraphics[width=0.5\columnwidth]{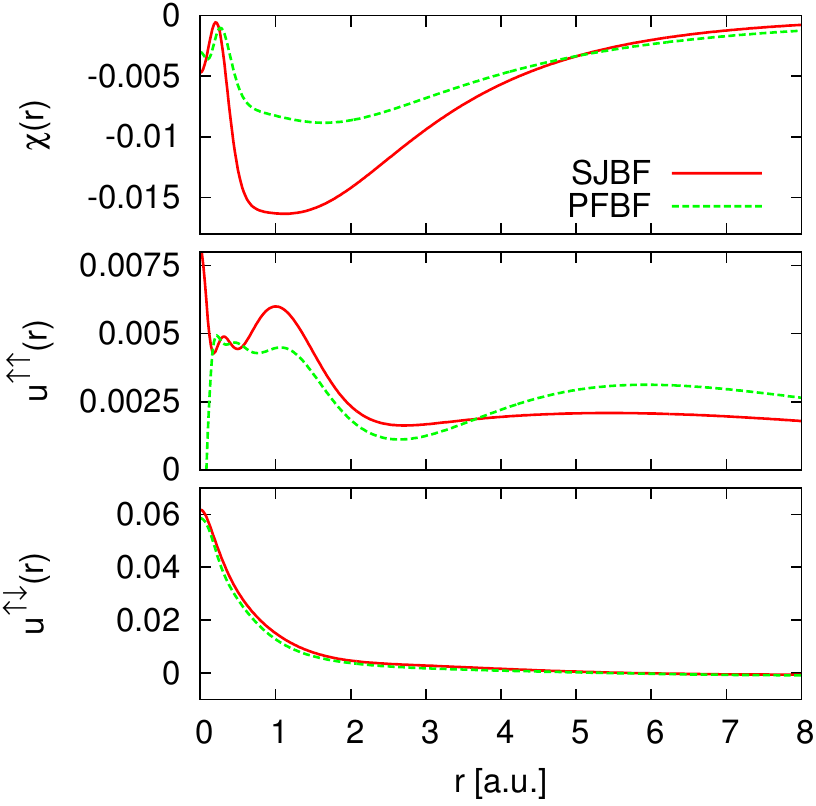}
\end{minipage}
\end{center}
\caption{Optimized electron-nucleus $\chi(r)$ and electron-electron $u(r)$ two-body backflow functions of carbon PP atom (left) 
and dimer (right) without the presence of three body term versus distance $r$. The functions shown are 
for Slater-Jastrow (SJ) and Pfaffian-Jastrow (PF) wave functions with backflow (BF) correlations.
Note that $u^{\uparrow\uparrow}=u^{\downarrow\downarrow}$, i.e., this function is the same for all spin-like electron-electron pairs and
similarly $u^{\uparrow\downarrow}=u^{\downarrow\uparrow}$ is the same for all  spin-unlike electron-electron pairs. }
\label{fig:C_bffunctions}
\end{figure}

In order to gain further insight into the action of backflow transformation on electron coordinates we 
plot the electron-nucleus and electron-electron two-body backflow functions optimized in VMC method without the 
presence of three-body terms. As it is immediately visible from Fig.~\ref{fig:C_bffunctions}, the spin-unlike electron-electron 
functions are almost order of magnitude larger than electron-nucleus and spin-like electron-electron functions. 
They are also characterized with well defined shape, which is almost independent of type of trial wave function. 

\begin{table}
\caption{Slater-Jastrow (SJ), Pfaffian-Jastrow (PF) wave functions with backflow (BF) correlations for carbon pseudo atom
are tested in VMC and DMC methods. Notation is the same as in Fig.~\ref{fig:C_bf}}
\begin{tabular}{l c c c c c  c c c c c  }
\hline
\hline
\multicolumn{1}{l}{Method} &\multicolumn{1}{c}{WF} & \multicolumn{1}{c}{$N_\chi$}  & \multicolumn{1}{c}{$N_u$} & \multicolumn{1}{c}{$N_{w_1}$}  & \multicolumn{1}{c}{$N_{w_2}$} &
 \multicolumn{1}{c}{N$_p$} & \multicolumn{1}{c}{E [H]} &  \multicolumn{1}{c}{$\sigma^2$ [H$^2$]} & \multicolumn{1}{c}{E$_{corr}$[\%]} \\
\hline 
HF  & S & -& - & - & -& -& -5.31471 & - & 0.0 \\
\hline 
VMC & SJ      & -  & -  & -   & -   & -   & -5.3990(1) & 0.0677 & 81.8(1)\\
    & SJBF2B  & 11 & 22 & -   & -   & 33  & -5.4011(2) & 0.0544 & 83.8(2)\\ 
    & SJBF3B  & -  & -  & 128 & 128 & 256 & -5.4023(3) & 0.0504 & 85.0(3)\\
    & SJBF23B & 4  & 8  & 128 & 128 & 268 & -5.4020(2) & 0.0498 & 84.7(2)\\
    & PF      & -  & -  & -   & -   & -   & -5.4083(2) & 0.0626 & 90.8(2)\\
    & PFBF2B  & 11 & 22 & -   & -   & 33  & -5.4097(1) & 0.0427 & 92.1(1)\\
    & PFBF23B & 4  & 8  & 128 & 128 & 268 & -5.4107(1) & 0.0411 & 93.1(1)\\
\hline 
DMC & SJ      & -  & -  & -   & -   & -   & -5.4065(3) & - & 89.0(3)\\
    & SJBF2B  & 11 & 22 & -   & -   & 33  & -5.4090(3) & - & 91.5(3)\\
    & SJBF3B  & -  & -  & 128 & 128 & 256 & -5.4085(3) & - & 91.0(3)\\
    & SJBF23B & 4  & 8  & 128 & 128 & 268 & -5.4094(3) & - & 91.8(3)\\
    & PF      & -  & -  & -   & -   & -   & -5.4137(3) & - & 96.0(3)\\
    & PFBF2B  & 11 & 22 & -   & -   & 33  & -5.4145(3) & - & 96.8(3)\\
    & PFBF23B & 4  & 8  & 128 & 128 & 268 & -5.4152(3) & - & 97.5(3)\\
\hline
Est.& Exact & - & -& - & - & - & -5.417806 & - & 100.0 \\
\hline
\hline
\end{tabular}
\label{table:bf:C}
\end{table}

\begin{table}
\caption{Slater-Jastrow (SJ), Pfaffian-Jastrow (PF)  and CI-Jastrow (CI) wave functions with backflow (BF) correlations for carbon dimer.
The notation is the same as in Table~\ref{table:bf:C}.}
\begin{minipage}{\columnwidth}
\renewcommand{\thefootnote}{\alph{footnote}}
\renewcommand{\thempfootnote}{\alph{mpfootnote}}
\centering
\begin{tabular}{l c c c c c  c c c c c  }
\hline
\hline
\multicolumn{1}{l}{Method} &\multicolumn{1}{c}{WF} & \multicolumn{1}{c}{$N_\chi$}  & \multicolumn{1}{c}{$N_u$} & \multicolumn{1}{c}{$N_{w_1}$}  & \multicolumn{1}{c}{$N_{w_2}$} &
 \multicolumn{1}{c}{N$_p$} & \multicolumn{1}{c}{E [H]} &  \multicolumn{1}{c}{$\sigma^2$ [H$^2$]} & \multicolumn{1}{c}{E$_{corr}$[\%]} \\
\hline 
HF & S & -& - & - & -& -& -10.6604 & - & 0.0 \\
\hline 
VMC & SJ\footnotemark[1]     & -  & -  & -   & -   & -   & -10.9936(4) & 0.179 & 81.7(1)\\
    & SJBF2B                 & 11 & 22 & -   & -   & 33  & -11.0012(3) & 0.144 & 83.5(1)\\
    & SJBF23B                & 4  & 8  & 128 & 128 & 268 & -11.0014(2) & 0.141 & 83.6(1)\\
    & PF\footnotemark[2]     & -  & -  & -   & -   & -   & -11.0171(2) & 0.160 & 87.4(1)\\
    & PFBF2B                 & 11 & 22 & -   & -   & 33  & -11.0223(3) & 0.123 & 88.7(1)\\
    & PFBF23B                & 4  & 8  & 128 & 128 & 268 & -11.0223(2) & 0.128 & 88.7(1)\\
    & CI\footnotemark[3]     &  - & -  & -   & -   & -   & -11.0420(4) & 0.112 & 93.6(1)\\
    & CIBF2B                 & 11 & 22 & -   & -   & 33  & -11.0440(3) & 0.100 & 94.0(1)\\
\hline 
DMC & SJ      & -  & -  & -   & -   & -   & -11.0227(2) & - & 88.8(1)\\
    & SJBF2B  & 11 & 22 & -   & -   & 33  & -11.0269(4) & - & 89.9(1)\\
    & SJBF23B & 4  & 8  & 128 & 128 & 268 & -11.0280(3) & - & 90.1(1)\\
    & PF      & -  & -  & -   & -   & -   & -11.0419(9) & - & 93.5(2)\\
    & PFBF2B  & 11 & 22 & -   & -   & 33  & -11.0443(6) & - & 94.1(2)\\
    & PFBF23B & 4  & 8  & 128 & 128 & 268 & -11.0447(3) & - & 94.2(1)\\
    & CI      &  - & -  & -   & -   & -   & -11.0579(5) & - & 97.5(1)\\
    & CIBF2B  & 11 & 22 & -   & -   & 33  & -11.0580(4) & - & 97.5(1)\\
\hline
Est.& Exact & - & -& - & - & - & -11.068(5) & - & 100.0 \\
\hline
\hline
\end{tabular}
\footnotetext[1] {Slater determinant contains PBE DFT orbitals.}
\footnotetext[2] {Same PBE DFT orbitals are used also in PF wave function.}
\footnotetext[3] {Uses natural orbitals with weights of determinants re-optimized in VMC.}
\end{minipage}
\label{table:bf:C2}
\end{table}

\section{Conclusions}
We have presented the first application of Pfaffian and multi-determinantal wave functions 
with backflow correlations to chemical systems. Results for two testing cases of carbon pseudo atom and its dimer 
show promising gains in correlations energies, decreases in variances and improvements in the nodal structure.
However, it all comes at the additional computational cost of calculation of the backflow displacement  
and the simultaneous update of all electron positions. In the future, it will be therefore necessary to 
perform the scaling tests to larger systems and compare the overall gain from backflow 
correlations to its overhead cost. 

\chapter{Summary}\label{ch:last}
The QMC methodology has proved to be a powerful technique for obtaining 
the ground state properties of fermionic systems. 
Its only insufficiency comes from the necessity to circumvent the 
fermion-sign problem by the fixed-node approximation. 
The resulting fixed-node errors account for small, but important, 
fraction of the correlation energies. In this dissertation, 
we have presented developments with direct relevance to 
elimination or at least to improvement of these fixed-node errors. 

In the first part of this dissertation, presented in the Ch.~\ref{ch:nodes},
we have analyzed the structure and properties of nodes 
of spin-polarized atomic and molecular fermionic wave functions constructed 
from one-particle orbitals of  $s$, $p$, $d$ and $f$ symmetries.
The study of the cases with high symmetries enabled us to find 
exact nodes for several states  with a few electrons ($p^2, \pi^2$, and $p^3$). 
Furthermore, we have used the projection of multi-dimensional 
manifolds into 3D space to study the topologies of the nodes of Hartree-Fock wave functions. 
Finally, we illustrate how the correlation in the accurate CI wave functions, 
for two specific cases of spin-unpolarized states, 
manifests itself in reducing the nodal structure to only two maximal nodal cells.

In the next chapter, we have proposed a generalized pairing wave function 
based on the Pfaffian functional form. The tests of the Pfaffian pairing 
wave functions on the set of first row atoms and dimers revealed that 
the Pfaffians were able to capture a large fraction of missing correlation
energies. We have also explored extensions to linear combinations of Pfaffians 
with good results for atomic systems, but with limited gains in the correlation energies for molecular systems.
Further, we have also employed a wave function in the form of fully-antisymmetrized independent pairs.
We have found that it does not lead to additional gains in correlation energy.
We conclude that the Pfaffian pairing wave functions offer better description of the fermion nodes than 
the wave functions employing Slater determinant (i.e., Hartree-Fock wave functions), 
and exhibit the correct topology with the minimal number of two nodal cells.

Finally, we have teamed up the Slater, Pfaffian and CI based wave functions
with inhomogeneous backflow transformation. Our preliminary tests
for two chemical systems indicate that the backflow correlations
reduce the variances of local energy and results in 
some improvement of VMC and DMC energies as well as in better description of the fermion nodes. 
However, these gains come at the price of an additional computational cost, justification of which   
will have to be determined in the near future.

\bibliographystyle{uqthesis}
\bibliography{thesismb}

\newpage
\appendix
\chapter*{}
\addcontentsline{toc}{chapter}{Appendices}
\thispagestyle{plain}
\begin{center}
\begin{large}
\textbf{Appendices}
\end{large}
\end{center}
\chapter{Cutoff-Cusp and Polynomial Pad\'e Functions}\label{appendix:functions}
\begin{figure}[!t]
\begin{center}
\includegraphics[width=0.8\columnwidth]{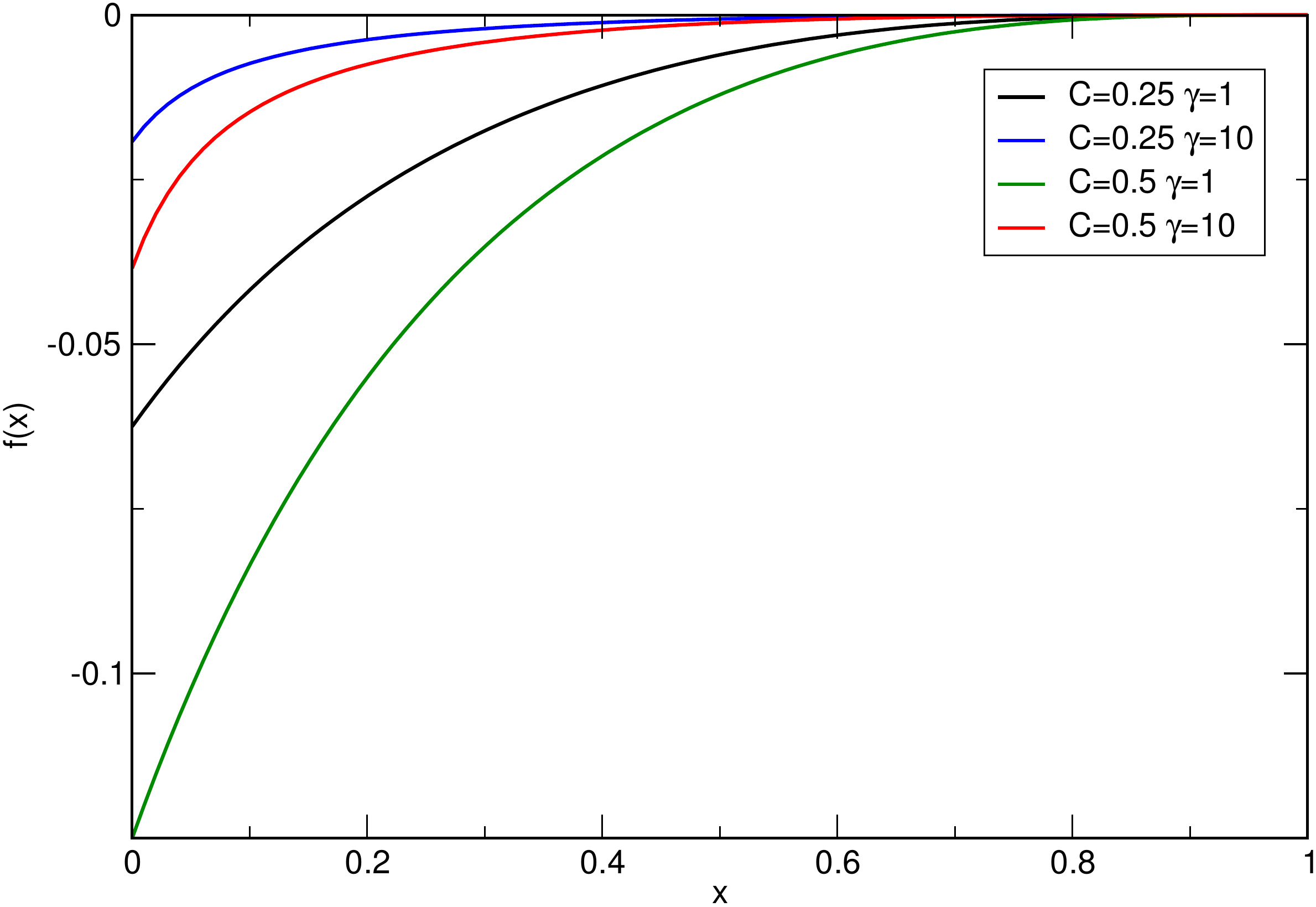}
\end{center}
\caption{Cutoff-Cusp functions with two different curvatures ($\gamma=1$ and $\gamma=10$) for like ($C=\frac{1}{4}$) and 
unlike ($C=\frac{1}{2}$) spins.}
\label{fig:cutofffunc}
\end{figure}

The exact electron-electron cusp conditions are satisfied by the choice of following cusp-function [see Fig.~(\ref{fig:cutofffunc})]
with variable $x=r/r_{cut}$,  where $r_{cut}$ is some cutoff radius and $\gamma$ is the curvature as  
\begin{align}\label{eq:appendix:functions:1}
f_{cusp}(x,\gamma)=C\left(\frac{x-x^2+x^3/3}{1+\gamma(x-x^2+x^3/3)}-\frac{1}{\gamma+3}\right)
\end{align}
The cusp constant is $C=\frac{1}{4}$ for electrons with like and $C=\frac{1}{2}$ for electrons with unlike spins.  

\begin{figure}[!b]
\begin{center}
\includegraphics[width=0.8\columnwidth]{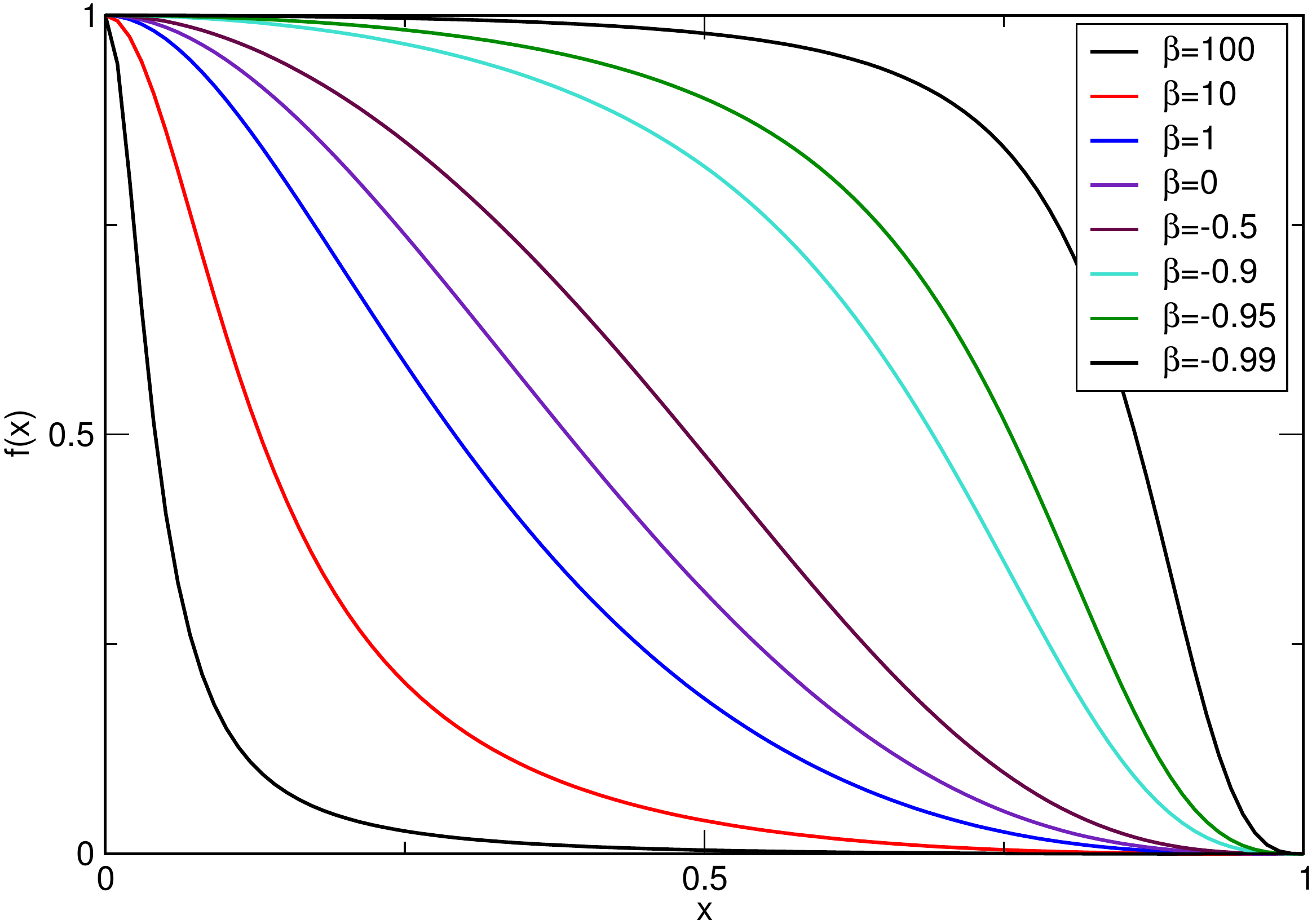}
\end{center}
\caption{Polynomial Pad\'e functions with curvatures $\beta$ ranging from -0.99 to 100.}
\label{fig:bessel}
\end{figure}

Polynomial Pad\'e functions for the same variable $x=r/r_{cut}$ and curvature $\beta$ had proved to 
be excellent choice for describing the electron-electron and electron-nucleus correlation. In calculations, we use 
the form 
\begin{align}
f_{poly-Pade}(x,\beta)=\frac{1-x^2(6-8x+3x^2)}{1+\beta x^2(6-8x+3x^2)}
\end{align}
The $f_{poly-Pade}(0)=1$ with derivative $f'_{poly-Pade}(0)=0$ and also goes smoothly to zero as $r \to r_{cut}$ [see Fig.~(\ref{fig:bessel})].
These conditions are necessary for preserving cusp conditions already fixed by cusp-functions~(\ref{eq:appendix:functions:1}) and choice of orbitals or pseudopotentials. 

\chapter{Proof of Cayley's Identity}\label{appendix:Cayley}
In order to prove the statement in Eq.~(\ref{eq:cayley}) we will proceed by induction. For $n=2$ it is true that
\begin{equation*}
{\rm det} \left[\begin{array}{cc}
0  & b_{12} \\
-a_{12}  & 0 \end{array}\right]
={\rm pf} \left[\begin{array}{cc}
0  & b_{12}\\
-b_{12} & 0\end{array}\right]
{\rm pf} \left[\begin{array}{cc}
0  & a_{12} \\
-a_{12} & 0\end{array}\right].
\end{equation*}
For even $n$ greater than $2$, determinant of our matrix of interest 
can be expanded through its cofactors as
\begin{align}\label{eq:a:expand}
 {\rm det}& \left[\begin{array}{ccccc}
0  & b_{12}  & b_{13} &\ldots &  b_{1,n}\\
-a_{12}  & 0  & a_{23} & \ldots &  a_{2,n}\\
-a_{13}  & -a_{23} & 0  & \ldots &  a_{3,n}\\
 \vdots & \vdots & \vdots &  \ddots &  \vdots \\
-a_{1,n} & -a_{2,n} & -a_{3,n} & \ldots &  0\\
\end{array}\right]=\sum_k -a_{1,k} C(k,1) \nonumber \\ 
&=\sum_k  \sum_l  -a_{1,k} b_{1,l} C(k,1;1,l)
\end{align}
The cofactor can be written as
\begin{equation}
C(k,1;1,l)=(-1)^{k+l+1}{\rm det} \left[A(k,1;1,l)\right],
\end{equation}
where the cofactor matrix  is given by
\begin{equation}
\small
A(k,1;1,l)=
\left[
\begin{array}{ccccccc}
0 & a_{23}  & \ldots & a_{2,k} & \ldots &  a_{2,n}\\
-a_{23}  & 0  & \ldots &  a_{3,k} & \ldots & a_{3,n}\\
 \vdots & \vdots & \ddots & \vdots & \ddots & \vdots  \\
-a_{2,l} & -a_{3,l} & \ldots & -a_{k,l} &\ldots & a_{l,n}\\
\vdots & \vdots & \ddots & \vdots & \ddots & \vdots  \\
-a_{2,n} & -a_{3,n} & \ldots & -a_{k,n} &\ldots & 0
\end{array}\right].
\end{equation}
At this point we would like to use the induction step and rewrite
the determinant cofactor as a product of two pfaffians [Cayley's identity Eq.~(\ref{eq:cayley})].
This would allow us to demonstrate that the expansion is identical to 
the expansion of pfaffians in minors.
In order to do so, however, we have
to shift the $k$-th column by pair column exchanges, so it becomes
the {\em last} column and, similarly, we have to shift the $l$-th row by
pair exchanges, so it becomes the last row.
This involves $k$ pair exchanges of columns and $l$ pair exchanges or rows and 
can be represented by unitary matrices $U_k$ and $U_l$.
It is necessary to invoke these operations so that the 
matrix gets into a form directly amenable for the Cayley's identity, i.e.,
the matrix has to be in a manifestly skew-symmetric form. 
(The sign change from the row/columns exchanges will prove irrelevant as we will 
show below.)
The transformed matrix is given by 
\begin{equation}
A'(k,1;1,l)=U_kA(k,1;1,l)U_l
\end{equation}
and has all zeros on the diagonal with the exception of the last element which is equal to $-a_{k,l}$. 
The last row is given by
\begin{align} 
{\bf v}_r=&(-a_{2,l},\ldots, -a_{k-1,l},-a_{k+1,l},\ldots \nonumber \\
&\ldots,-a_{l-1,l},a_{l,l+1},\ldots, a_{l,n},-a_{k,l}), 
\end{align}
while the last column is given as following
\begin{align}
{\bf v}_c^T=&(a_{2,k},\ldots,a_{k-1,k},-a_{k,k+1},\ldots \nonumber \\
&\ldots,-a_{k,l-1},-a_{k,l+1},\ldots,-a_{k,n},-a_{k,l})^T.
\end{align}
The only non-zero diagonal element $-a_{k,l}$ can be eliminated, 
once we realize that its cofactor contains a determinant of
a skew-symmetric matrix of odd degree,
which always vanishes (proof by Jacob~\cite{Jacobi}).

Now we are ready to perform the induction step, namely to use the 
property that the determinant of a $2(n-1)\times 2(n-1)$ matrix
can be written as given by the Cayley's identity, Eq.~(\ref{eq:cayley}).
We obtain
\begin{align}
{\rm det}[U_kA(k,1;1,l)U_l]&={\rm det}[A'(k,1;1,l)] \\
 &={\rm pf}[A'(1,k;1,k)]\,{\rm pf}[A'(1,l;1,l)]. \nonumber 
\end{align}
We can now apply the inverse unitary transformations and
shift back the columns (and by the skew-symmetry the corresponding rows) in the
first pfaffian and, similarly, the rows (and corresponding columns) 
in the second. This enables us to write
\begin{align}
{\rm pf}&[A'(1,k;1,k)]\,{\rm pf}[A'(1,l;1,l)]\nonumber \\
&={\rm pf}[U_l^{-1}A(1,k;1,k)U_l]\,{\rm pf}[U_kA(1,l;1,l)U_k^{-1}]\nonumber \\
&={\rm pf}[A(1,k;1,k)]\,{\rm pf}[A(1,l;1,l)],
\end{align}
where we have used the identity given by Eq.~(\ref{eq:pfident4}).
We can therefore finally write
\begin{align}
C(k,1;1,l)&=(-1)^{k+l+1}{\rm pf}[A(1,k;1,k)]{\rm pf}[A(1,l;1,l)] \nonumber \\
&=-P_c(a_{1,k})P_c(a_{1,l}),
\end{align}
where $P_c$ denotes a pfaffian cofactor as defined in (\ref{eq:pfcof}). 
Therefore, the determinant expansion in Eq.~(\ref{eq:a:expand}) equals to
\begin{align}
\sum_{k,l} -a_{1,k} b_{1,l} C(k,1;1,l)&=\sum_{k,l} a_{1,k} b_{1,l} P_c(a_{1,k})P_c(a_{1,l}) \nonumber \\
&={\rm pf}[A]{\rm pf}[B]
\end{align}
with matrices $A$ and $B$ defined as in Eq.~(\ref{eg:inverseupdate}).
This concludes the proof of the more general form of the Cayley's identity. Note, if $B=A$,
we trivially obtain well-known formula for the square of
pfaffian [Eq.~(\ref{eq:pfident2})].

\chapter{Core Pfaffian Algorithms}\label{appendix:pfaffianroutines}
\section{Gaussian Elimination with Row-Pivotting Algorithm for Pfaffian Value}
\begin{verbatim}
int RowPivoting(Array2 <doublevar> & tmp, int i, int n){
  //row pivoting algorithm used by Pfaffian_partialpivot
  doublevar big;
  doublevar temp;
  doublevar TINY=1e-20;
  Array1 <doublevar> backup(2*n) ;
  int d=1;
  int k=0;
  big=0.0;
  //find the largest value
  for (int j=i+1;j<2*n;j++){
    temp=fabs(tmp(i,j));
    if(temp > big){
      big=temp;
      k=j;
    }
  }
  if (big<TINY){
    cout <<"Singular row in matrix!!! "<<endl;
    tmp(i,i+1)=TINY;
  }

  if (k!=i+1){
    //exchange k-th column with 2-nd column;
     for (int j=i;j<2*n;j++){
       backup(j)=tmp(j,i+1);
       tmp(j,i+1)=tmp(j,k);
       tmp(j,k)=backup(j);
     }
     //exchange k-th row with 2-nd row;
     for (int j=i;j<2*n;j++){
       backup(j)=tmp(i+1,j);
       tmp(i+1,j)=tmp(k,j);
       tmp(k,j)=backup(j);
     }
     
     d*=-1; //sign change of pfaffian
  }
  return d;
}

doublevar Pfaffian_partialpivot(const Array2 <doublevar> & IN){
  //  returns the pfaffian of skew-symmetric matrix  IN
  //  with partial pivoting
  if (IN.dim[0]%2!=0) return 0.0;
  int n=IN.dim[0]/2;
  Array2 <doublevar> tmp(2*n,2*n);
  doublevar PF=1.0;
  doublevar fac;
  for (int i=0;i<2*n;i++)
    for (int l=0;l<2*n;l++)
      tmp(i,l)=IN(i,l);
 
  int d=1;
  for (int i=0;i<2*n;i=i+2){
    //for given row look for pivoting element
    //exchange if needed
    d*=RowPivoting(tmp, i, n);

    for (int j=i+2;j<2*n;j++){
      fac=-tmp(i,j)/tmp(i,i+1);
      for (int k=i+1;k<2*n;k++){
        tmp(k,j)=tmp(k,j)+fac*tmp(k,i+1);
        tmp(j,k)=tmp(j,k)+fac*tmp(i+1,k);
      }
    }
    PF=PF*tmp(i,i+1);
  }
  return PF*d;
}
\end{verbatim}

\section{Algoritm for the Update of Inverse of Pfaffian Matrix}
\begin{verbatim}
doublevar UpdateInversePfaffianMatrix(Array2 <doublevar> & IN, 
                                      Array1 <doublevar> & row, 
                                      Array1 <doublevar> & column, 
                                      int e)
{
  //update row and column of skew-symmetric inverse matrix IN
  //the ratio of new/old pfaffians is returned by Column(e)
  int n=in.dim[0]/2;
  for (int i=0;i<2*n;i++){
    column(i)=0.0;
    for (int j=0;j<2*n;j++)
      column(i)+=row(j)*IN(j,i);
  }
  
  //to avoid the catastrophe in later division
  if (column(e)==0)
    column(e)=1e-20;

  //rest is just the new inverse matrix IN
  for(int i=0;i<2*n;i++){
    if (i==e){
      IN(i,i)=0.0;
    }
    else {
      IN(e,i)=+IN(e,i)/column(e);
      IN(i,e)=-IN(e,i);
    }
  }

  for(int j=0;j<2*n;j++){
    if (j!=e){
      for (int k=0;k<2*n;k++){
        if (k==j) {
          IN(k,k)=0.0;
        }
        else {
          IN(k,j)-=column(j)*IN(k,e);
          IN(j,k)=-IN(k,j);
        }
      }
    }
  }
  //return the ratio of pfaffians
  return  column(e);
}
\end{verbatim}

\end{document}